\documentclass[journal=aesccq,manuscript=article]{achemso}

\usepackage[utf8]{inputenc}

\title{The ice composition close to the surface of comet 67P/Churyumov-Gerasimenko}

\author{Matthias L{\"a}uter}
\affiliation[Zuse Institute Berlin]
{Zuse Institute Berlin, 14195 Berlin, Germany}
\email{laeuter@zib.de}
\author{Tobias Kramer}
\affiliation{Institut f\"ur Theoretische Physik, Johannes Kepler Universität, 4040 Linz, Austria}
\alsoaffiliation{Department of Physics, Harvard University, Cambridge, MA 02138, USA}
\author{Martin Rubin}
\author{Kathrin Altwegg}
\affiliation{University of Bern, Physikalisches Institut, 3012 Bern, Switzerland}

\begin{document}

\maketitle

\begin{abstract}
The relation between ice composition in the nucleus of comet 67P/Churyumov-Gerasimenko on the one hand and relative abundances of volatiles in the coma on the other hand is important for the interpretation of density measurements in the environment of the cometary nucleus.
For the 2015 apparition, in situ measurements from the two ROSINA (Rosetta Orbiter Spectrometer for Ion and Neutral Analysis) sensors COPS (COmet Pressure Sensor) and DFMS (Double Focusing Mass Spectrometer) determined gas densities at the spacecraft position for the 14 gas species H$_2$O, CO$_2$, CO, H$_2$S, O$_2$, C$_2$H$_6$, CH$_3$OH, H$_2$CO, CH$_4$, NH$_3$, HCN, C$_2$H$_5$OH, OCS, and CS$_2$.
We derive the spatial distribution of the gas emissions on the complex shape of the nucleus separately for 50 subintervals of the two-year mission time.
The most active patches of gas emission are identified on the surface.
We retrieve the relation between solar irradiation and observed emissions from these patches.
The emission rates are compared to a minimal thermophysical model to infer the surface active fraction of H$_2$O and CO$_2$.
We obtain characteristic differences in the ice composition close to the surface between the two hemispheres with a reduced abundance of CO$_2$ ice on the northern hemisphere
(locations with positive latitude).
We do not see significant differences for the ice composition on the two lobes of 67P/C-G.
\end{abstract}


\section{Introduction}

The formation process of comets in the solar nebula at large distances from the sun resulted in cometary nuclei consisting of a mixture of solid ices and non-volatile dust components, including a variety of organic compounds \cite{Geiss1987,Mumma2011}.
The molecular inventory of the volatile species in the nucleus is reflected in the composition of the cometary coma, which can be observed from Earth and during spacecraft missions \cite{Altwegg2019}.
The Rosetta mission \cite{Schulz2009} of the European Space Agency accompanied and investigated the Jupiter-family comet 67P/Churyumov-Gerasimenko (67P/C-G) during its 2015 apparition \cite{Keller2020}.
Rosetta's ROSINA instrument \cite{Balsiger2007} (Rosetta Orbiter Spectrometer for Ion and Neutral Analysis) collected millions of in situ measurements in the coma.
ROSINA consisted of three sensors, the COmet Pressure Sensor (COPS), the Double Focusing Mass Spectrometer (DFMS), and the Reflectron-type Time Of Flight (RTOF) mass spectrometer for the determination of neutral gas densities and of relative abundances.
The observed gas species range from inorganic molecules \cite{Rubin2019}, which include the most abundant volatiles H$_2$O, CO$_2$, and CO, to more complex organic molecules \cite{Rubin2019a}.
Here, we focus our detailed, spatially resolved analysis on 14 volatiles which have been globally characterized in terms of gas production in Läuter et al. \cite{Lauter2020}.
The Rosetta mission additionally contained the remote sensing instruments MIRO \cite{Biver2019} and VIRTIS \cite{Bockelee-morvan2016}, which performed an independent analysis of the coma composition with respect to the ROSINA data.

The ROSINA data were obtained over a wide range of varying observational conditions.
Coma measurements around the cometary nucleus began at a heliocentric distance of 3.6~au in August 2014, continued while approaching perihelion on August 13th 2015 at $1.24~\mathrm{au}$, and finally concluded after two years in September 2016 at a heliocentric distance of $3.7~\mathrm{au}$.
The increasing solar irradiation toward perihelion is linked to a steep increase of gas and dust emissions.
The operational orbit of the Rosetta spacecraft covered the surface of the rotating nucleus (rotation period about 12~hours) by a fast sampling (on the scale of hours) along the longitudinal coordinates and by a slower sampling (on the scale of days) along the latitudinal coordinates of the cometocentric system.
The direct correlation of the spacecraft trajectory to peaks in density yields a localization of gas emissions on the surface \cite{Hassig2015}.
High resolution OSIRIS images \cite{Keller2007,Sierks2015,Preusker2017} of comet 67P/C-G revealed the complicated non-convex shape of the nucleus with a volume of $\approx 19~\mathrm{km}^3$, which requires moving beyond spherical approximations of the nucleus geometry.
The spacecraft distance to the nucleus ranged from a few kilometers to more than $1000\,\mathrm{km}$ which affects the spatial resolution of the instruments on the spacecraft.
The Rosetta mission provided the unique opportunity to monitor the evolution of the cometary coma with in situ measurements over two years.
Cometary flyby missions collect in situ data as well, but are limited to much shorter time periods \cite{Vincent2019,Keller2020}.
The number of comets accessible via Earth bound observations is more comprehensive and covers, besides comets in the solar system, \cite{Cordiner2014,Farnham2021,Bonev2021} also interstellar objects \cite{Cordiner2020,Yang2021}.

One goal of the observations is to determine the composition of the cometary material \cite{Marschall2020}.
We derive maps of best-fit emission rates on the cometary surface which reproduce the measured density variations at the spacecraft position within the inner coma.
The emission maps are obtained by adjusting the emission rates placed on a complex shape of the nucleus within a forward/inverse modeling approach.
This approach was introduced in Kramer et al. \cite{Kramer2017} and Läuter et al. \cite{Lauter2019,Lauter2020}, it is briefly reviewed in Section~\ref{sec:model}, and it does away with the assumption of a spherical symmetric gas expansion of Haser-type models \cite{Haser1957}.
Kramer et al. \cite{Kramer2017} retrieved surface maps showing the emission rates of neutral gas for three different time intervals and found a strong correlation between enhanced gas activity months before perihelion and locations of dust outbursts imaged around perihelion \cite{Vincent2016a}.
Läuter et al. \cite{Lauter2019} extended this analysis to the major species H$_2$O, CO$_2$, CO and O$_2$ and to the entire duration of the Rosetta mission of two years.
The data set was later enlarged to determine the temporal changes of the global production rate of 14 species analyzed by ROSINA \cite{Lauter2020}.

Here, we construct and analyze the surface emission rate of the 14 species.
In Sections~3 and 5 we compare our results with independent observations, which directly image the surface using optical and spectroscopic measurements \cite{Vincent2019,Filacchione2019}.
The detection and analysis of small surface areas based on OSIRIS images has been limited to specific regions and observation windows \cite{Vincent2019}. 
An automated approach to map surface deformations and changes seen by optical instruments across the entire nucleus will improve this comparative approach in the future \cite{Vincent2021}.

Section~4 introduces a minimal thermophysical model.
The nucleus material close to the surface consists of a mixture of various icy components available for sublimation processes and non-volatile dust grains, partly lifted up by the stream of gas.
The modeling of the energy and mass fluxes within the nucleus requires a thermophysical model to relate the ice composition to the localized gas emission \cite{Huebner1999,Huebner2006,Skorov2012,Marboeuf2014,Groussin2019,Prialnik2020,Hoang2020}.
One class of thermophysical models considers the sublimation in response to the irradiation via a steady-state process (the models are discussed e.g. in Keller et al. \cite{Keller2015}).
This allows correlating the sublimation rate with solar irradiation and determining the fraction of the surface that is active.
The assumption of a completely homogeneous ice distribution in these models does not explain the observed non-gravitational accelerations \cite{Kramer2019a,Attree2019,Mottola2020a} and changes of the rotational state of the nucleus \cite{Jorda2016}, in particular the small changes of the orientation of the rotation axis\cite{Kramer2019}. 
In Section~5 we discuss the relation between the surface production and the composition of the near-surface ices.

\section{Data and gas model in the coma}
\label{sec:model}

The data processing and analysis of the measured coma densities and abundances follows our forward/inverse model approach described in previous papers \cite{Kramer2017,Lauter2019,Lauter2020} and is reviewed briefly in this Section.
The starting point is the combination of two data sets, one from the COPS sensor, the other one from the DFMS sensor of the ROSINA instrument.

\begin{figure}
\includegraphics[height=0.35\textwidth]{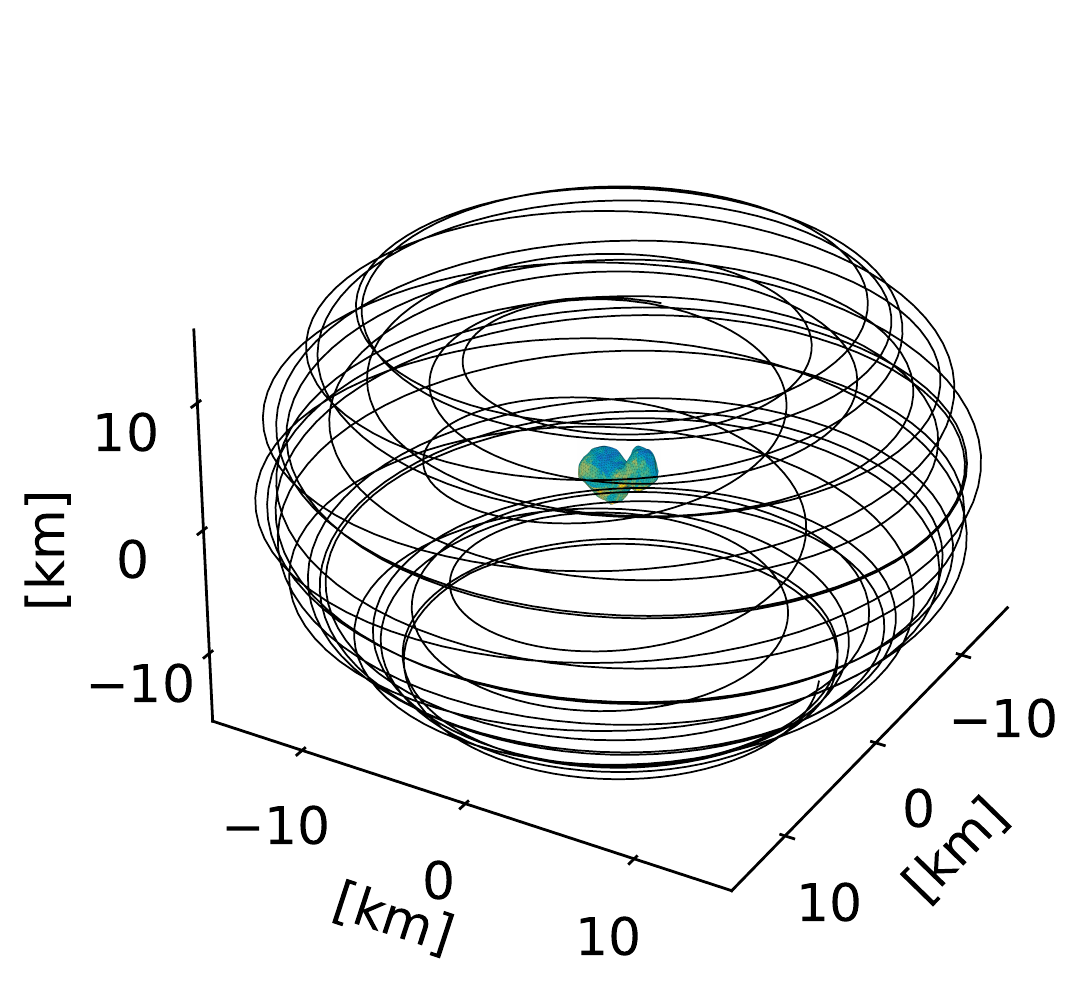}
\caption{Spacecraft trajectory around comet 67P/C-G between 10~km to 20~km for subinterval $I_4$, 317~days to 302~days before perihelion.}
\label{fig:bienenkorb}
\end{figure}

DFMS collected the relative abundances of the gas species $s=$ H$_2$O, CO$_2$, CO, H$_2$S, O$_2$, C$_2$H$_6$, CH$_3$OH, H$_2$CO, CH$_4$, NH$_3$, HCN, C$_2$H$_5$OH, OCS, and CS$_2$ at numerous spacecraft positions within the coma of 67P/C-G.
In combination with the absolute density of the neutral gas (from COPS), the relative abundances for each gas species yield the absolute densities of these species \cite{Rubin2019,Gasc2017a}.
In the following, we consider measurements between August 1st 2014 and September 5th 2016.
This period spans 377~days before the perihelion passage to 390~days afterwards.
We subdivide this interval for the analysis into 50 separated time subintervals $I_1$, ..., $I_{50}$, varying in duration between 7~days and 29~days.
The duration of the subintervals is taken as short as possible to satisfy the model assumption of constant solar irradiation conditions with respect to the heliocentric distance $r_\mathrm{h}$ within one subinterval.
The minimal duration of any subinterval is dictated by two conditions.
First, the sub-spacecraft position should have overflown almost the complete surface area of the nucleus.
This condition is met for example, as shown in Figure~\ref{fig:bienenkorb}, for the spacecraft trajectory within the subinterval $I_4$ around 310~days before perihelion.
Rosetta's orbit varied considerably over the mission and for the used data set the spacecraft distance to the nucleus changed from 8~km to $\approx{}500~\mathrm{km}$.
The second condition is that we require at least as many gas density measurements as there are undetermined surface emitters (3996 in our model).
The number of available measurements differs across subintervals and species and varies from 4443 points for CO to 19212 points for H$_2$O \cite{Lauter2020}.
For a small number of subintervals, some species are excluded from the analysis due to missing data points.  
For the species $s$ in the subinterval $I_j$ we denote the set of data points by  $D_{s,j}=\{(t_\mathrm{data},\rho_\mathrm{data})\}$, the measurement times by $t_\mathrm{data}\in I_j$  and the measured density by $\rho_\mathrm{data}$ at the spacecraft position $\vec{x}_\mathrm{sc}(t_\mathrm{data})$.
To reduce noise effects, our definition of $D_{s,j}$ excludes some outlying data points selected by a $2\sigma$ criterion described in Läuter et al. \cite{Lauter2020}.

\begin{figure}
\includegraphics[height=0.35\textwidth]{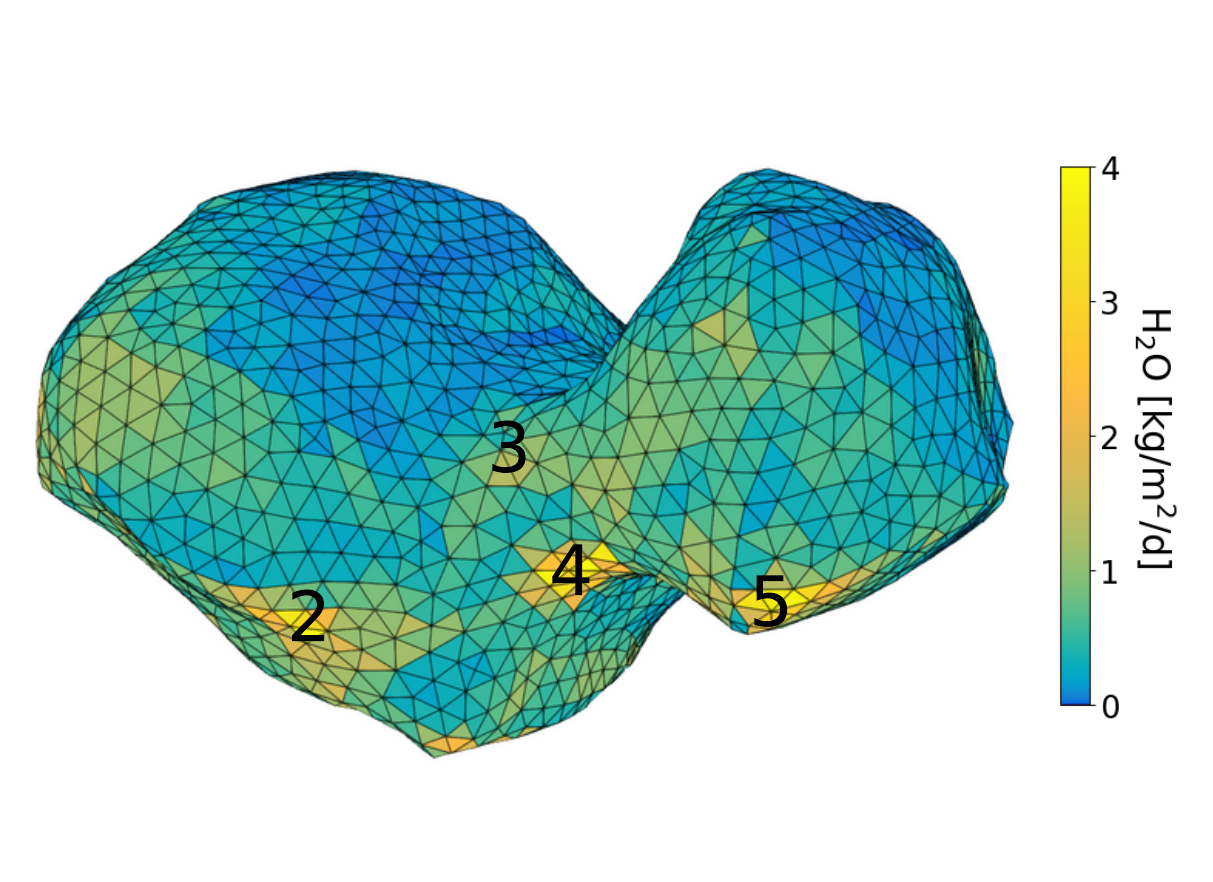}%
\caption{%
Shape of the 67P/C-G nucleus with 3996 triangular elements.
The colors represent surface emission rates for H$_2$O (averaged from 50~days before to 50~days after perihelion), the numbers refer to surface patches.
}
\label{fig:shape}
\end{figure}
For each gas species $s$ within the subinterval $I_j$ we construct a \textit{forward coma model}.
The model considers a steady-state flow of collisionless neutral gas into the vacuum domain around the nucleus up to a distance of $\approx{}500\,\mathrm{km}$ \cite{Kramer2017}.
The coma model makes several assumptions to reduce the number of open parameters and to facilitate the processing and assimilation of the complete ROSINA data set.
For instance, we neglect collisions and changes of the gas velocity, because most of the time the spacecraft trajectory was outside the region of gas acceleration.
These effects could be incorporated by solving the gas equations for rarefied gas based on the direct simulation Monte Carlo method \cite{Tenishev2008}, which has been applied to comet 67P/C-G  \cite{Hansen2016,Combi2020}.
However, the increased computational effort limits the possible spatial resolution required for the determination of gas sources on the surface \cite{Fougere2016,Fougere2016a}.
A fixed number of equally spaced gas sources on the cometary surface serve as emitters for the gas flow in space.
The shape of the nucleus is approximated by a surface mesh with $N_\mathrm{E}=3996$ triangular elements $E_i$.
The shape in Figure~\ref{fig:shape} is derived from the high resolution shape of Preusker et al. \cite{Preusker2017} and features an average size, taken as the diameter of a circle with the same area, of 120~m for the triangular elements $E_i$.
Each element $E_i$ contains one gas source with the surface emission rate $\dot\rho_{s,i,j}$ (see Eq.~(4) in Kramer et al. \cite{Kramer2017}) which represents the time averaged gas production in the subinterval $I_j$.
Narasimha \cite{Narasimha1962} provides an expression for the gas cloud originating from a single gas source with the density $\rho_{i}(\dot\rho_{s,i,j},\vec{x}_\mathrm{sc})$ at the spacecraft position $\vec{x}_\mathrm{sc}$.
$\rho_{i}$ contains two adjustable parameters, the gas velocity $u_{0}$ into the vertical outflow direction and the cone parameter $U_0 = 3$ for the density distribution. 
Here, we take $u_{0}$ to be the function of $r_\mathrm{h}$ given by Hansen et al. \cite{Hansen2016} and assume the same value of $u_{0}$ for all gas species \cite{Biver2019}.
Recapitulating equation (5) in Kramer et al. \cite{Kramer2017}, the steady-state gas density at the spacecraft position
\begin{equation}
\rho_{s,j}(\vec{x}_\mathrm{sc}) = \sum_{i=1}^{N_\mathrm{E}}
\mathrm{occ}_i(\vec{x}_\mathrm{sc}) \rho_{i}(\dot\rho_{s,i,j},\vec{x}_\mathrm{sc})
\end{equation}
is the superposition of element-wise contributions with the occlusion function $\mathrm{occ}_i(\vec{x}_\mathrm{sc})$.
For each subinterval $I_j$  we seek the surface emission rates $\dot\rho_{s,i,j}$ giving best agreement between model densities and observed coma densities $\rho_{s,i}(\vec{x}_\mathrm{sc}(t_\mathrm{data})) = \rho_\mathrm{data}$ for all data points $(t_\mathrm{data},\rho_\mathrm{data})\in D_{s,j}$.
This requires us to solve the least-squares optimization problem given by equation (7) of Kramer et al. \cite{Kramer2017}.
We refer to this step as applying the \textit{inverse model}.

\section{Gas emission rates and spatial distribution}
\label{sec:globalgas}

\begin{table}
\begin{tabular}{l|ccc}
name & time interval & $r_\mathrm{h}~[\mathrm{au}]$  & sub solar lat.\ \\ \hline
perihelion &  August 13th 2015 & 1.24 & $-52^\circ$ \\
$I_\mathrm{inbound}$ & 330 to 270~days before perihelion & 3.0 -- 3.4 & $+37^\circ$ -- $+42^\circ$ \\
$I_\mathrm{ph}$ & 50 before to 50~days after perihelion & 1.24 -- 1.4 & $-52^\circ$ -- $-18^\circ$\\
$I_\mathrm{outbound}$ & 340 to 390~days after perihelion & 3.4 -- 3.7 & $+14^\circ$ -- $+19^\circ$
\end{tabular}
\caption{Definition of time intervals considered in the analysis of the 2015 apparition of comet 67P/C-G.}
\label{tab:intervals}
\end{table}
\begin{figure}
\includegraphics[height=0.30\textwidth]{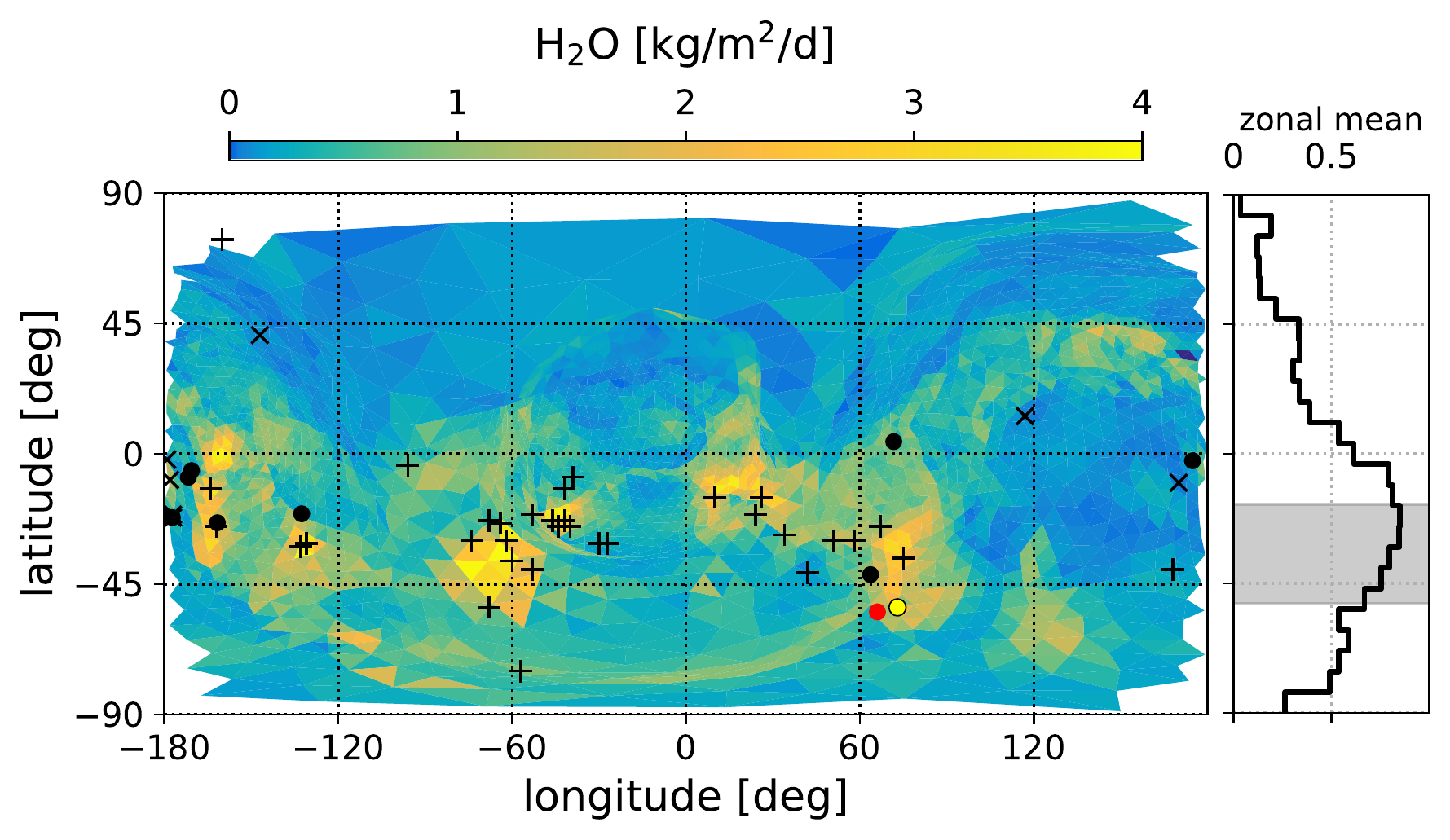}%
\hfill%
\includegraphics[height=0.30\textwidth]{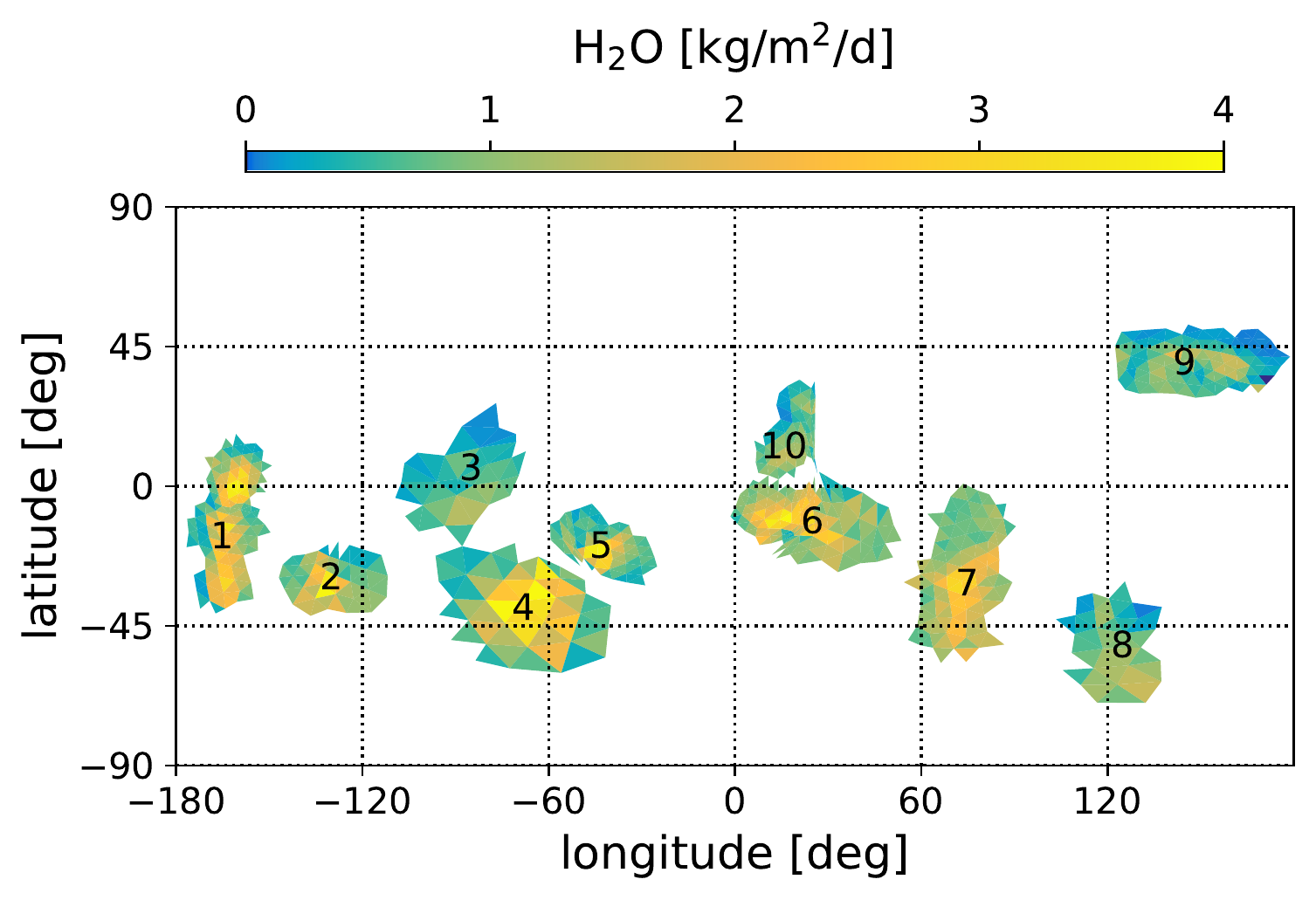}
\caption{%
Surface emission rates for H$_2$O, averaged from 50~days before to 50~days after perihelion (see $I_\mathrm{ph}$ in Table~\ref{tab:intervals}), cometocentric Cheops coordinate system \cite{Preusker2015}.
Left panel: surface map and corresponding zonal mean.
$+$ markers \cite{Vincent2016a} denote outburst locations;
$\times$ markers \cite{Oklay2017}, black dots \cite{Barucci2016}, and
yellow dot \cite{Fornasier2016} denote water-ice patches;
red dot \cite{Filacchione2016} denotes a CO$_2$-ice patch. 
The gray bar denotes the range of subsolar latitudes within $I_\mathrm{ph}$.
Right panel: definition of the patches $A_1$, ..., $A_{10}$ with the highest emission rates (dominated by H$_2$O near perihelion).}
\label{fig:h2operihel}
\end{figure}
\begin{figure}
\includegraphics[height=0.30\textwidth]{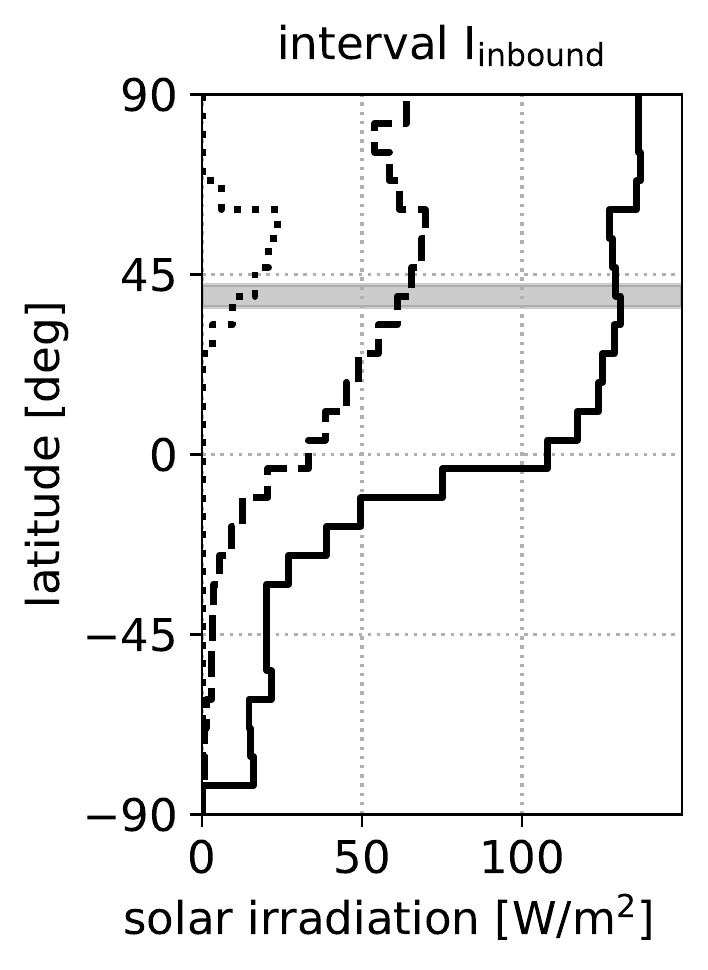}%
\hfill%
\includegraphics[height=0.30\textwidth]{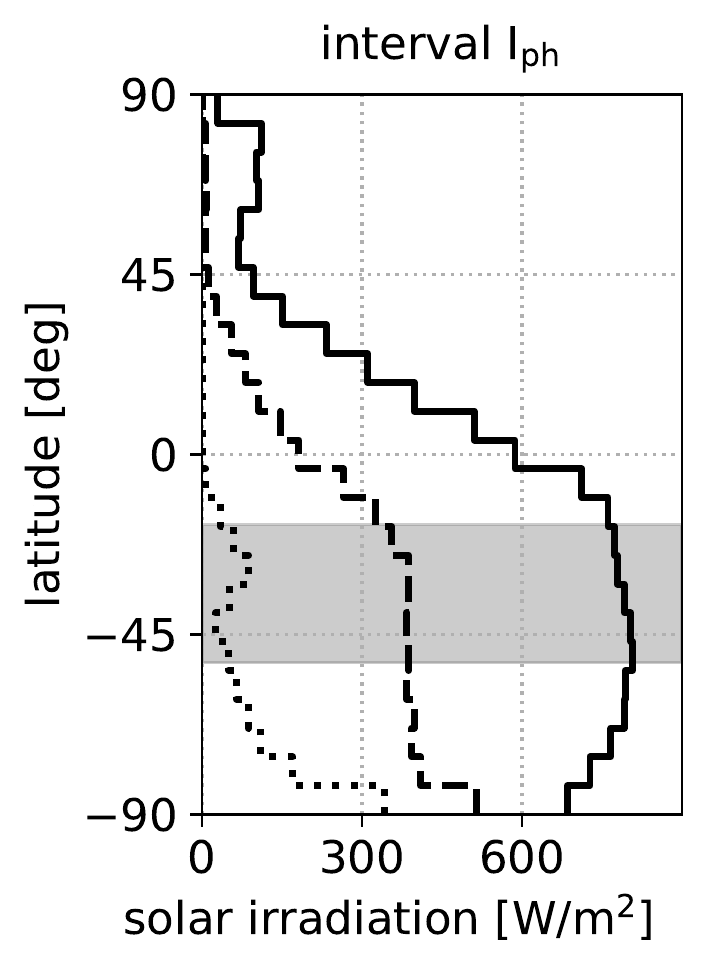}%
\hfill%
\includegraphics[height=0.30\textwidth]{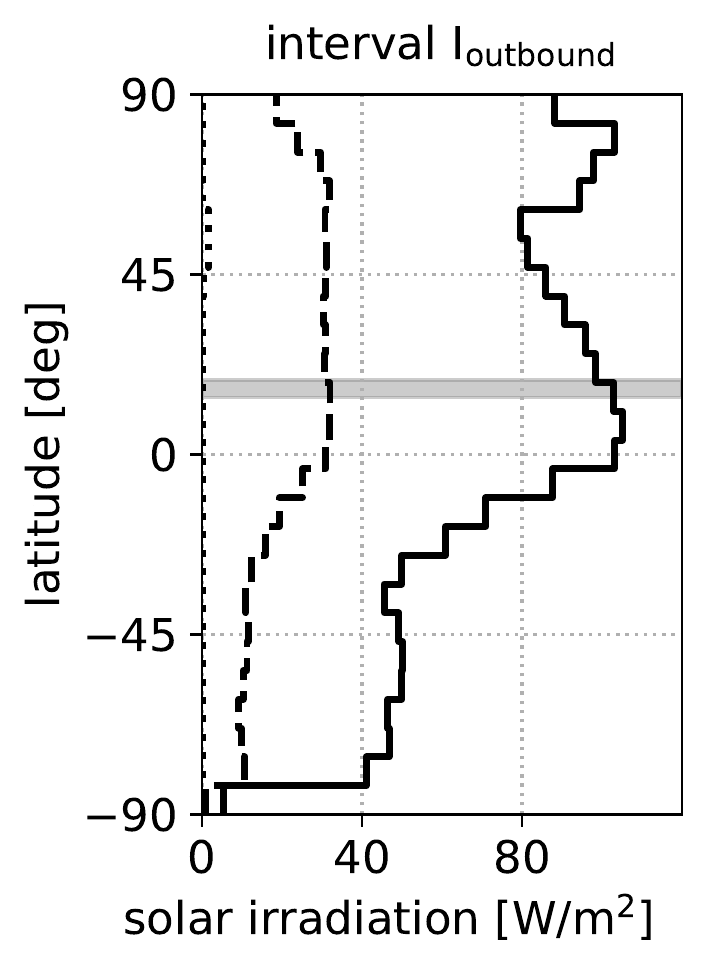}%
\caption{%
Solar irradiation onto the surface of 67P/C-G, diurnal average and zonal mean.
The gray bar denotes the range of subsolar latitudes during the time interval.
Left panel: time interval $I_\mathrm{inbound}$ (see Table~\ref{tab:intervals}),
mid panel: time interval $I_\mathrm{ph}$,
right panel: time interval $I_\mathrm{outbound}$.
Dotted line: minimum irradiation,
dashed line: mean irradiation,
continuous line: maximum irradiation.%
}
\label{fig:irradiation}
\end{figure}
The inverse model determines the surface emission rates, which can be spatially integrated to obtain the global production rates $Q_s$ \cite{Lauter2020}.
The production rates increase toward perihelion, peak at 16 to 27 days after perihelion, and decrease as the solar irradiation diminishes for larger heliocentric distances \cite{Hansen2016,Marshall2017,Shinnaka2017}.
Within the previous papers \cite{Lauter2019,Lauter2020} we have described the temporal evolution of the production rates as well as the surface localization of sources for the major species H$_2$O, CO$_2$, CO, and O$_2$.
The production rates around perihelion (within the time interval $I_\mathrm{ph}$, see Table~\ref{tab:intervals}) represent the major part of the global gas production within a complete apparition.
Analyzing the long term trend of the production rate 190~days after perihelion by a power law fit, we distinguished two groups of species.
The CO$_2$ group features a slower production decay with decreasing solar irradiation (with exponents $-3\leq\alpha$), while the H$_2$O group displays a faster production decay with exponents exceeding $\alpha\leq -4.5$ \cite{Gasc2017,Lauter2020}.

The surface emission rates $\dot \rho_{\mathrm{H}_2\mathrm{O}}$ across the entire surface are represented as a surface map in Figures~\ref{fig:shape} and \ref{fig:h2operihel} for the predominant gas species water on 67P/C-G in the interval $I_\mathrm{ph}$.
To relate these emission rates to solar irradiation qualitatively, the middle panel in Figure~\ref{fig:irradiation} shows the zonal mean of the incoming solar irradiation during this time interval.
The southern hemisphere receives the largest part of the solar irradiation at subsolar latitudes between $-52^\circ$ and $-18^\circ$ within the interval $I_\mathrm{ph}$.
We define the hemispheres as being separated by the equatorial plane through the center of the nucleus.
Thus the southern and northern hemispheres ($A_\mathrm{SH}$ and $A_\mathrm{NH}$) are the surface locations with negative and positive latitudes in the cometocentric Cheops coordinate system \cite{Preusker2015}, respectively.
The zonal mean of the water production and the irradiation are strongly correlated.
Both, water production and maximum solar irradiation peak around latitudes between $-25^\circ$ and $-50^\circ$.
Outside the perihelion period, the relation between sub solar latitude and the latitudes of dominant emission is more involved. 
For instance active CO$_2$ and CO regions continue to emit in the south 300 days after perihelion \cite{Kramer2017}, while the H$_2$O emission regions shift to the North along with the sub solar latitude \cite{Lauter2019}.

\begin{table}
\begin{tabular}{ccl}
region & center (long,lat) & description \\ \hline
$A_\mathrm{SH}$ & & southern hemisphere \\
$A_\mathrm{NH}$ & & northern hemisphere \\
$A_\mathrm{SL}$ & & small lobe \\
$A_\mathrm{BL}$ & & big lobe \\ \hline
$A_1$ & $(-165^\circ,-15^\circ)$ & Imhotep, Apis, Khonsu, $A_\mathrm{BL}$ \\
$A_2$ & $(-130^\circ,-30^\circ)$ & Khonsu, Atum, $A_\mathrm{BL}$ \\
$A_3$ & $(-85^\circ,5^\circ)$ & Anuket, Hapi \\
$A_4$ & $(-70^\circ,-40^\circ)$ & Anuket, Sobek \\
$A_5$ & $(-45^\circ,-20^\circ)$ & Wosret, $A_\mathrm{SL}$ \\
$A_6$ & $(25^\circ,-10^\circ)$ & Bastet, Wosret, $A_\mathrm{SL}$ \\
$A_7$ & $(75^\circ,-30^\circ)$ & Anhur, Bes, Khepry, $A_\mathrm{BL}$ \\
$A_8$ & $(125^\circ,-50^\circ)$ & Bes, Imhotep, $A_\mathrm{BL}$ \\
$A_9$ & $(145^\circ,40^\circ)$ & Ash, $A_\mathrm{BL}$ \\
$A_{10}$ & $(15^\circ,15^\circ)$ & Maat, Bastet, $A_\mathrm{SL}$
\end{tabular}
\caption{
Surface patches and their coordinates of the center, placement with respect to the lobes and to geomorphological features \cite{El-maarry2019} on 67P/C-G.
}
\label{tab:geomorph}
\end{table}
The water emission map shown in Figure~\ref{fig:h2operihel}, left panel, allows us to identify the most active patches\cite{Lauter2019} $A_1$--$A_{10}$, highlighted in the right panel of Figure~\ref{fig:h2operihel}.
These patches overlap partly with the geomorphological features defined by El-Maarry et al. \cite{El-maarry2019}. 
The patches are located on the southern and northern hemispheres $A_\mathrm{SH}$, $A_\mathrm{NH}$, and on the two lobes $A_\mathrm{SL}$, and $A_\mathrm{BL}$, see Table~\ref{tab:geomorph}.
The patches $A_2$--$A_5$ are marked on the three-dimensional shape model in Figure~\ref{fig:shape}.
Although the area of all 10 patches covers only 1/5 of the entire surface area, their combined water emission represents almost half of the global water emission.

The local emission depends on two independent properties of a patch, the composition of the nucleus material and the solar irradiation condition which will be further discussed in Section~\ref{sec:sublimation}.
To discuss the localization of the gas emissions, for a patch $A_p$ we define the mean production
by the average value of the surface emission rate over the patch $A_p$.
At perihelion for water, mean production varies between $0.7\,\mathrm{kg}/\mathrm{m}^2/\mathrm{d}$ on patch $A_3$ and $3.4\,\mathrm{kg}/\mathrm{m}^2/\mathrm{d}$ on the most active patch $A_4$.
To quantify just the activity of a patch $A_p$, we introduce the activity ratio $a_p$ as the mean production divided by the global production rate per area $Q_s/|A_{\mathrm{67P}}|$.
$Q_s/|A_{\mathrm{67P}}|$ is evaluated based on $Q_s$ in Table 2 of Läuter et al.\cite{Lauter2020} and the entire surface area $A_{\mathrm{67P}}=44.31\times 10^6\,\mathrm{m}^2$ of 67P/C-G which yields the value $1.1\,\mathrm{kg}/\mathrm{m}^2/\mathrm{d}$ for water.
Around perihelion this implies activity ratios for the patches $A_3$ and $A_4$
of $a_3=0.7$ and $a_4=3.2$, respectively.
The surface $A_{\mathrm{67P}}$ is related to our smoothed shape model in Section~\ref{sec:model} and thus it is smaller compared to reported areas \cite{Jorda2016,Preusker2017} for high-resolution shape models which vary between $[46.9\pm2.5] \times 10^6\,\mathrm{m}^2$ and $[51.7\pm0.1] \times 10^6\,\mathrm{m}^2$.
This effects the production rates in the order of 10\% and can be taken into account by rescaling the surface emissions reported here accordingly.

In the Refs.~\cite{Kramer2017,Lauter2019} we show the high correlation of the most active gas emitting patches to short-lived outbursts \cite{Vincent2016a} ($+$ markers in Figure~\ref{fig:h2operihel}).
In particular the activity on the patches $A_4$, $A_5$ (regions Sobek and Wosret) displays a strong correlation with the outbursts.
Morphological changes indicating gas activity in $A_5$ have been reported \cite{Fornasier2019a,Fornasier2021} around perihelion and later during the outbound Rosetta mission phase.
For the duration of two years, Oklay et al. \cite{Oklay2017} repeatedly observed water-ice-rich areas close to patch $A_1$ up to a size of $15000~\mathrm{m}^2$ ($\times$ markers in Figure~\ref{fig:h2operihel}).
In the time around 300~days before and around perihelion, Barucci et al. \cite{Barucci2016} find 8 icy patches (black dots in Figure~\ref{fig:h2operihel}) consisting of water ice.
Six of these icy patches are located close to patches $A_1$ (Imhotep), $A_2$ (Khonsu, Atum), and $A_7$ (Anhur).
One hundred days before perihelion Fornasier et al. \cite{Fornasier2016} reported two $1500~\mathrm{m}^2$ areas in the Anhur-Bes region with H$_2$O (yellow dot in Figure~\ref{fig:h2operihel}) and CO$_2$ (red dot in Figure~\ref{fig:h2operihel}) reservoirs \cite{Filacchione2016}.
Both are located in patch $A_7$.

\begin{figure}
\includegraphics[width=0.5\textwidth]{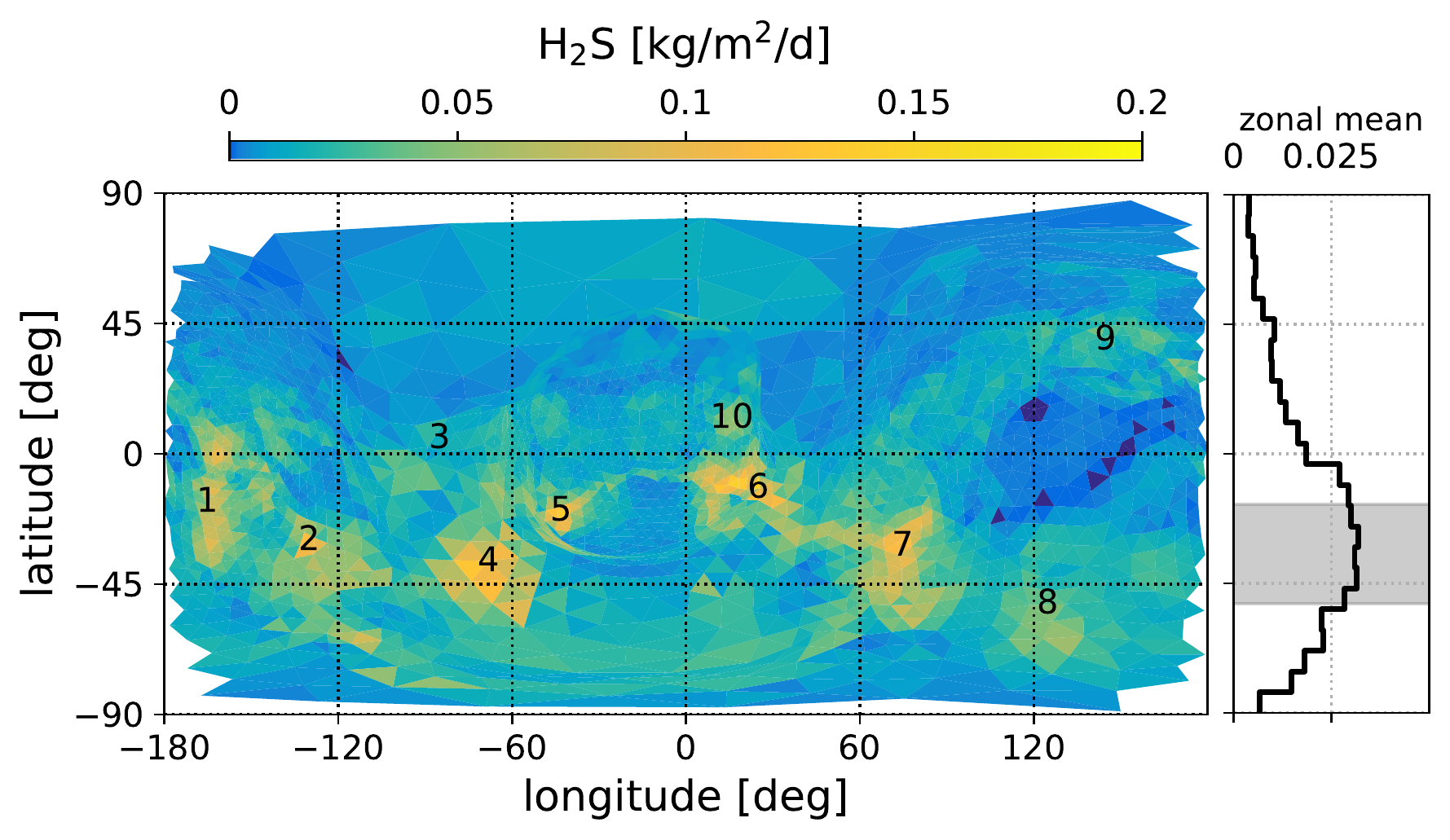}%
\hfill%
\includegraphics[width=0.5\textwidth]{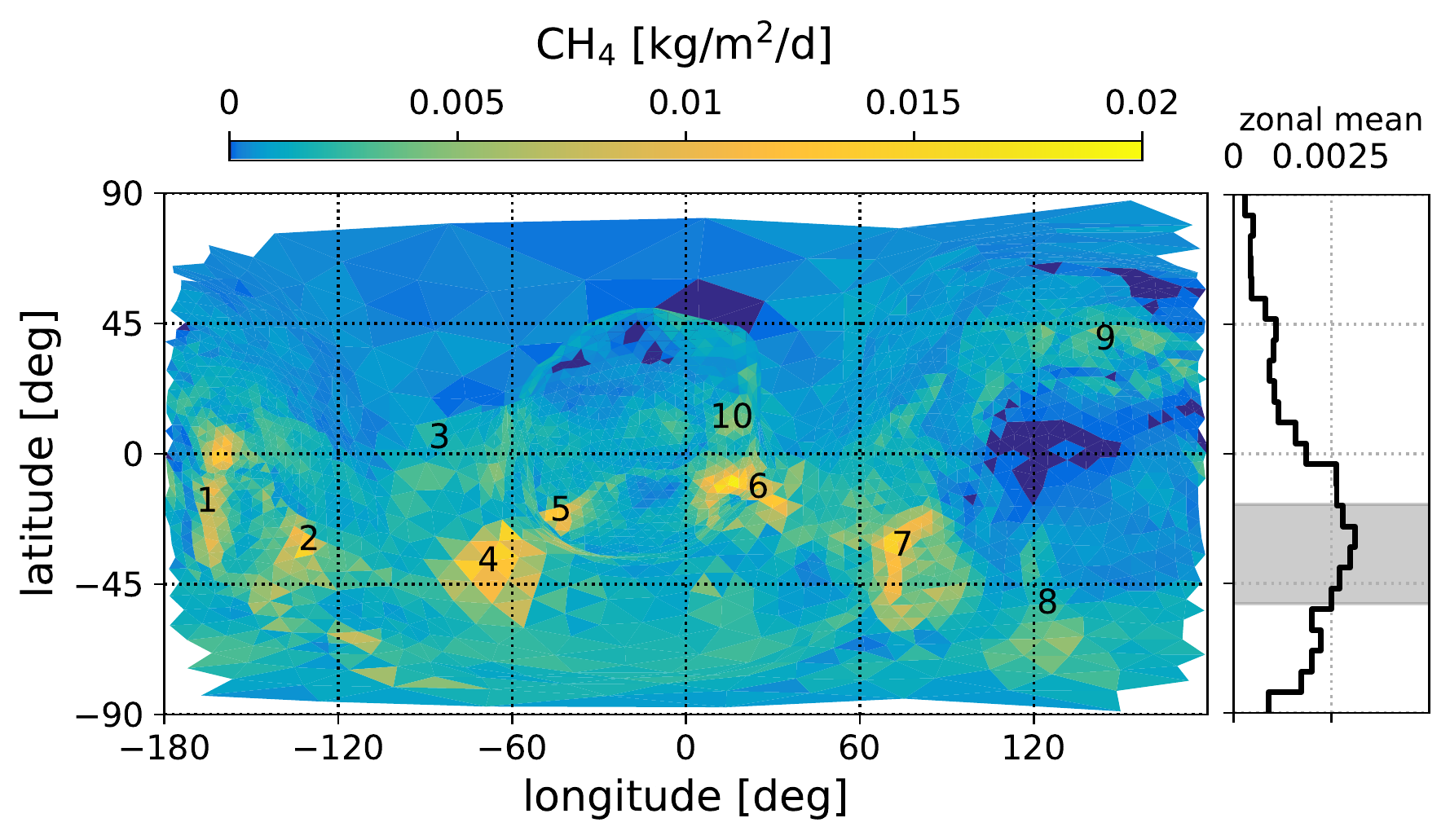}\\
\includegraphics[width=0.5\textwidth]{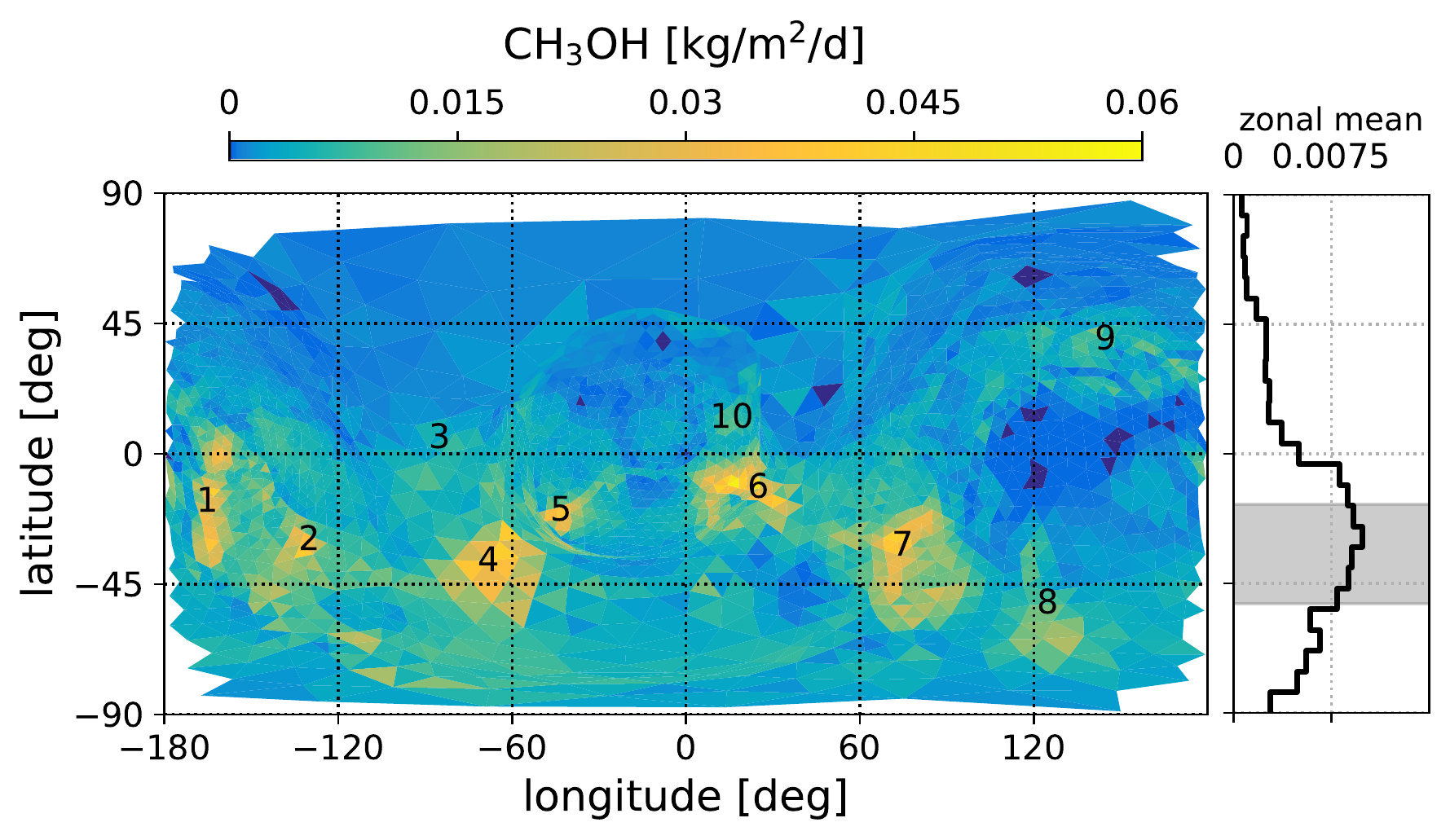}%
\hfill%
\includegraphics[width=0.5\textwidth]{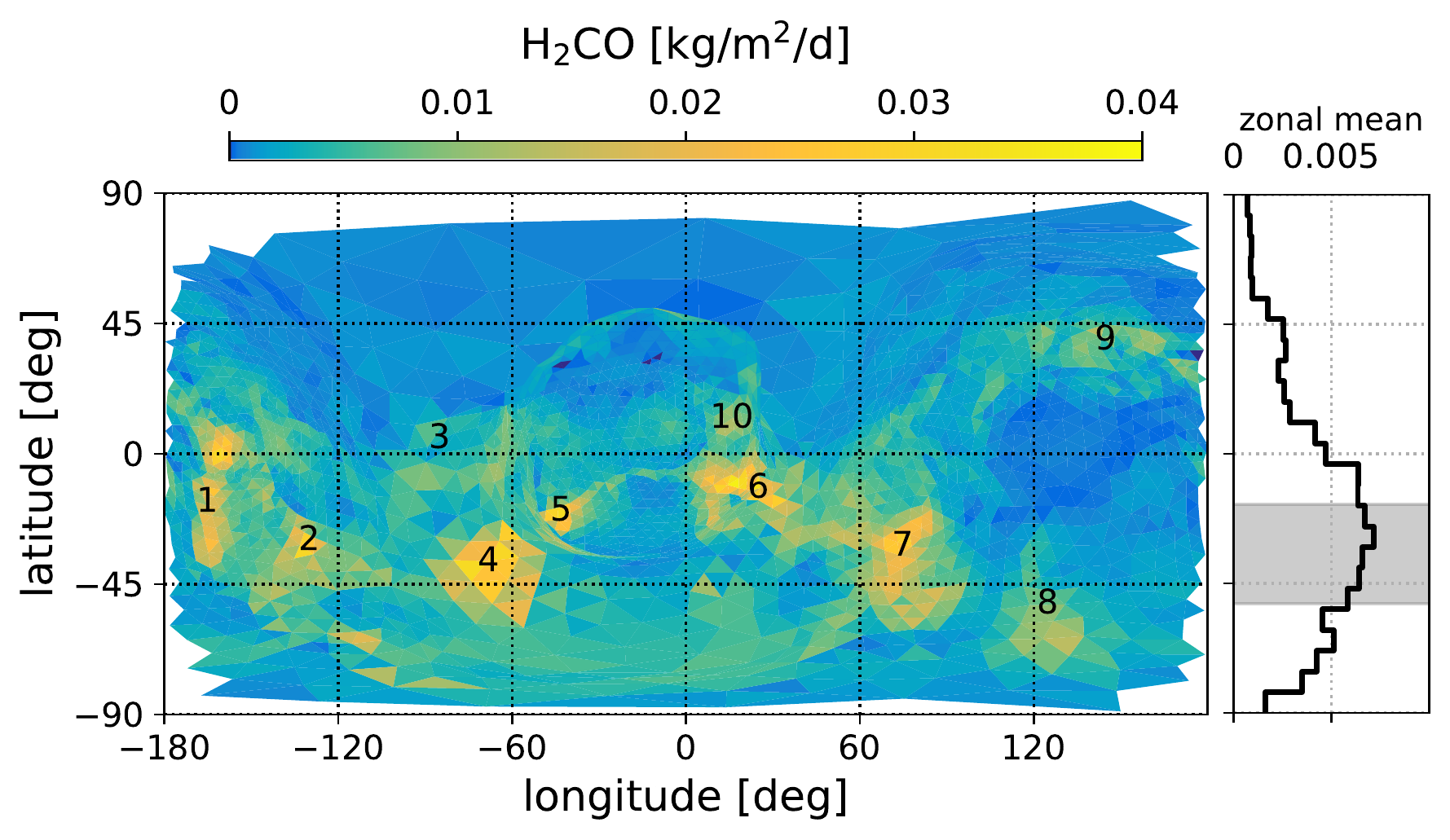}%
\caption{Surface emission rates for the minor gas species H$_2$S, CH$_4$, CH$_3$OH, and H$_2$CO, averaged from 50~days before perihelion to 50~days after perihelion (see $I_\mathrm{ph}$ in Table~\ref{tab:intervals}).%
}
\label{fig:minorperihel}
\end{figure}
In addition to an increased emission rate for water, the patches $A_1$--$A_{10}$ also correspond to regions of enhanced emission rate for the major species CO, and O$_2$ around perihelion in the interval $I_\mathrm{ph}$ \cite{Lauter2020}.
For CO$_2$, the enhanced patches are restricted to the southern hemisphere.
Around perihelion, Figure~\ref{fig:minorperihel} shows the spatial distribution of the emission rate for (H$_2$S, CH$_4$) in the CO$_2$ group and for (CH$_3$OH,  H$_2$CO) in the H$_2$O group.
For these gases, peak production rates \cite{Lauter2020} range from $4.4\times 10^{26}~\mathrm{molecules}/\mathrm{s}$ for H$_2$S down to $8.2\times 10^{25}~\mathrm{molecules}/\mathrm{s}$ for CH$_4$.
Similar to the major species, the minor species display their peak emission rate in all southern patches $A_1$, $A_2$, $A_4$--$A_8$. 
The activity ratio $a_p$ varies between 1.1 for CH$_4$ on patch $A_8$ and 3.2 for CH$_3$OH on $A_4$, as well as for CH$_4$ on $A_6$.
The activity within the most active patches A$_4$ and A$_6$ in the South and the peak of the zonally averaged emission rate in the southern mid latitudes correlates closely with the solar irradiation shown in Figure~\ref{fig:irradiation}, middle panel.

\begin{figure}
\includegraphics[width=0.5\textwidth]{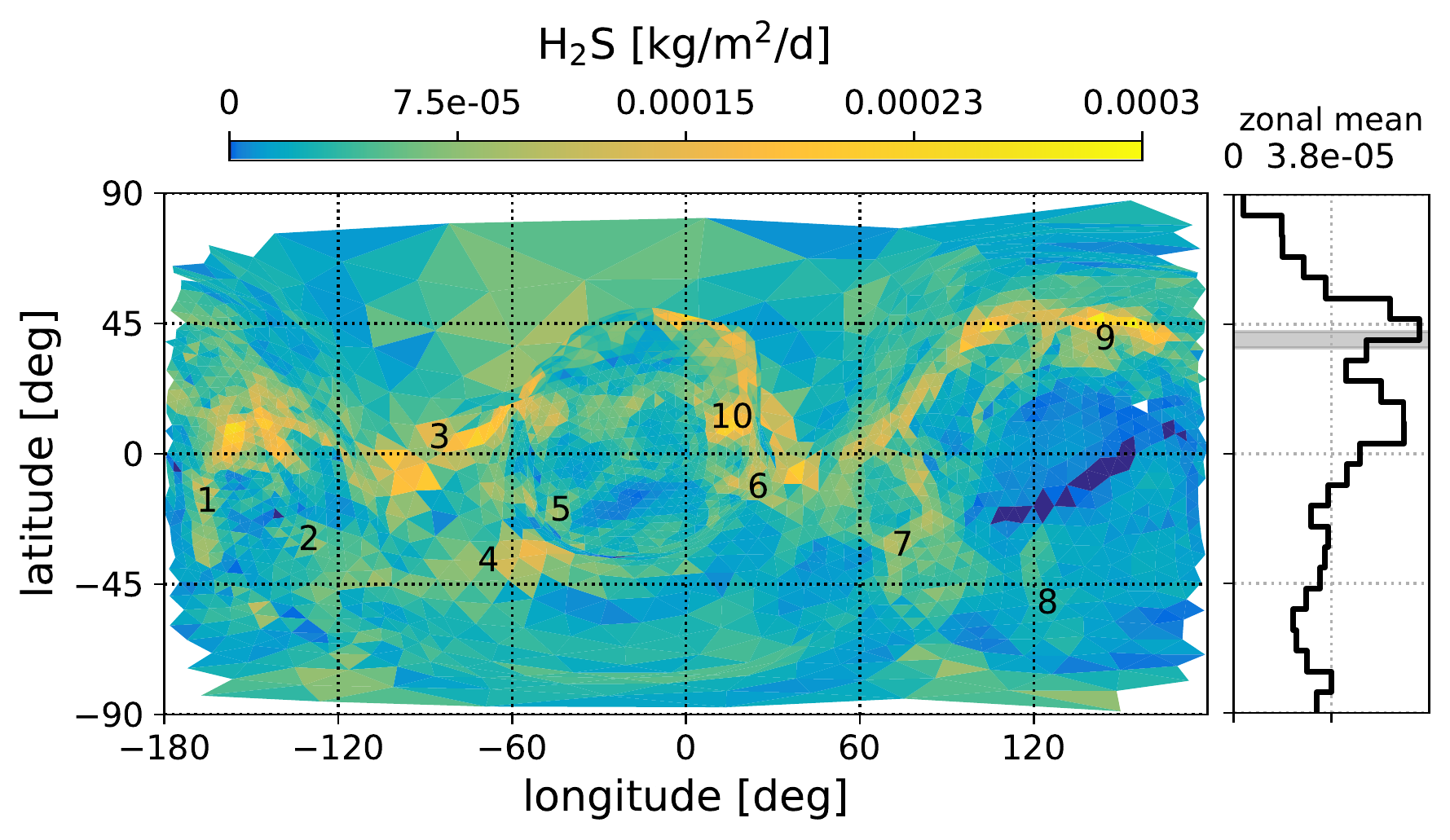}%
\hfill%
\includegraphics[width=0.5\textwidth]{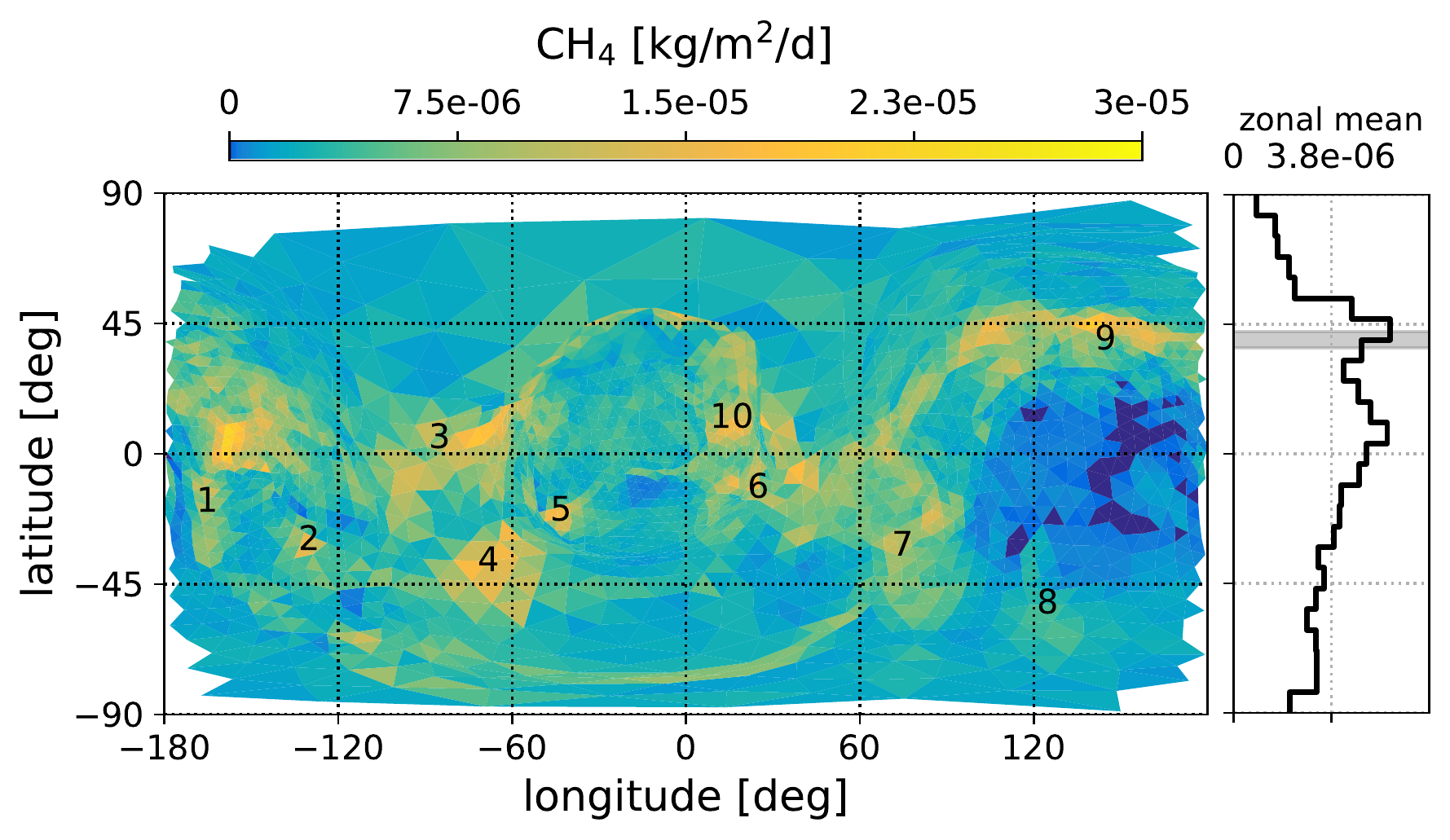}\\
\includegraphics[width=0.5\textwidth]{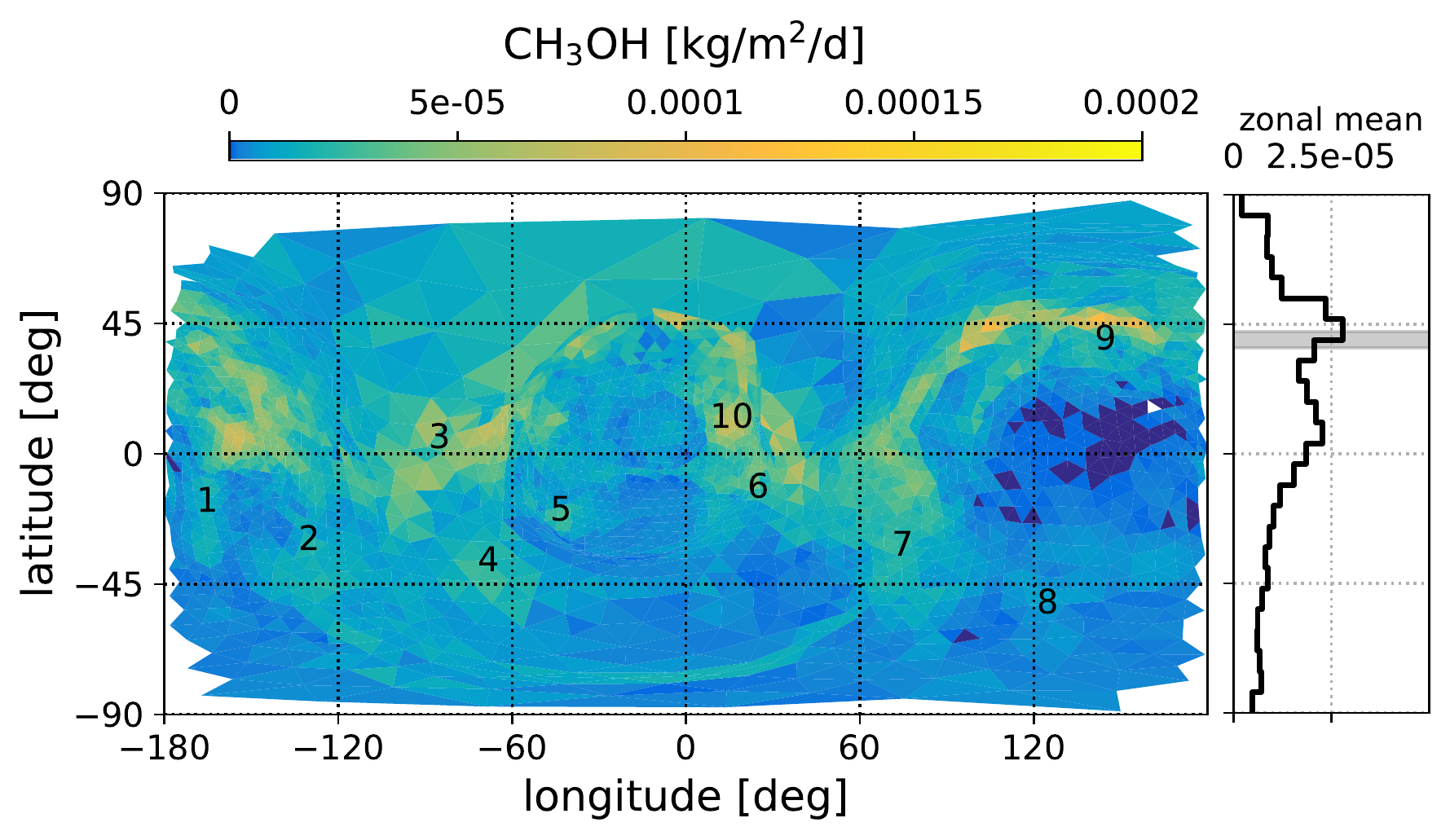}%
\hfill%
\includegraphics[width=0.5\textwidth]{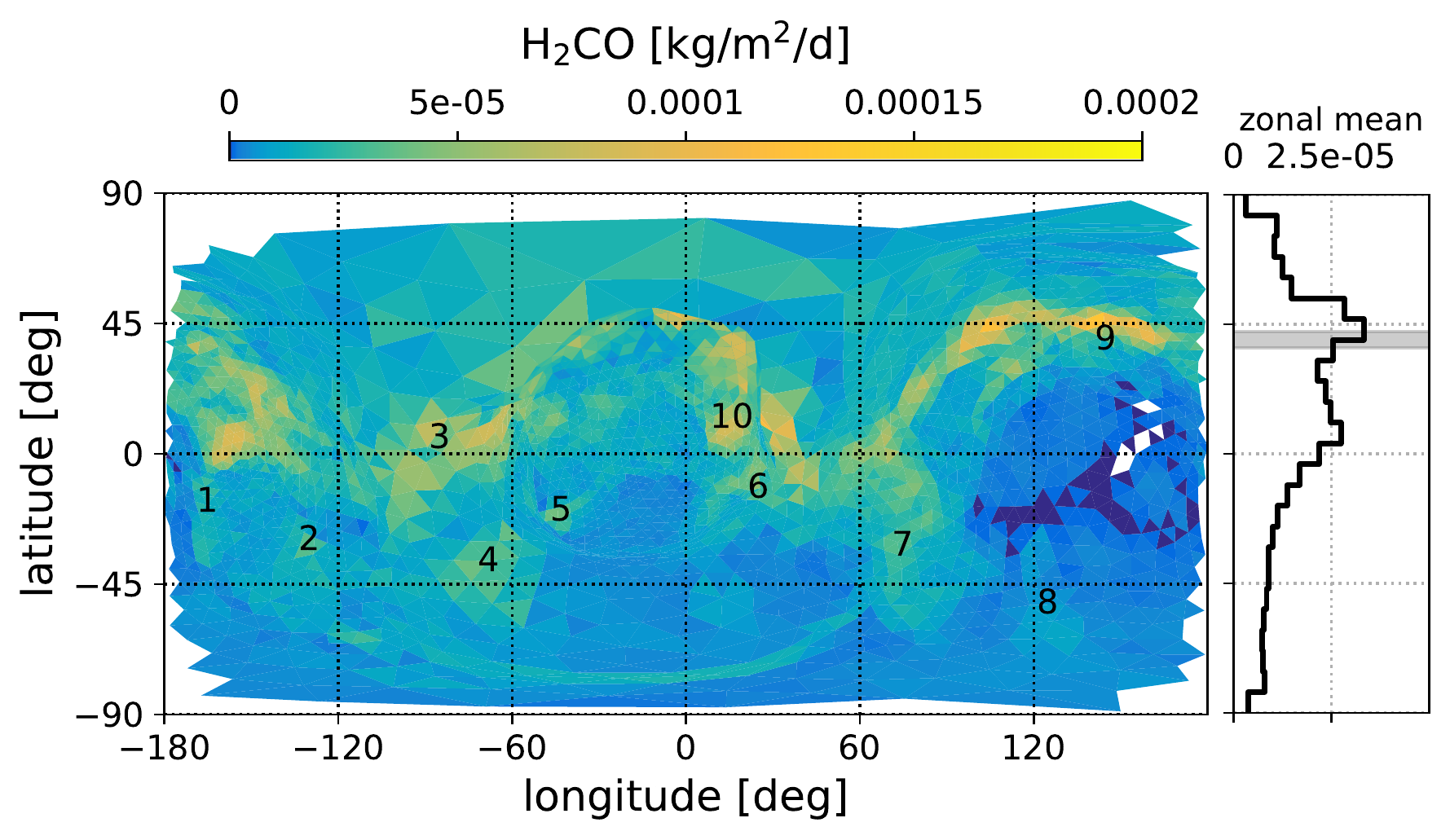}%
\caption{Surface emission rates for the minor gas species H$_2$S, CH$_4$, CH$_3$OH, and H$_2$CO, averaged from 330~days to 270~days before perihelion (see $I_\mathrm{inbound}$ in Table~\ref{tab:intervals}).%
}
\label{fig:minorinbound}
\end{figure}
Next we discuss the distribution of surface emission rates 
approximately 300 days before perihelion (interval $I_\mathrm{inbound}$ in Table~\ref{tab:intervals}), shown in Figure~\ref{fig:minorinbound}.
The peaks of the zonal mean emission rate for the minor gases H$_2$S, CH$_4$, CH$_3$OH and H$_2$CO are located close to northern subsolar latitudes between $+37^\circ$ and $+42^\circ$.
The bi-lobed shape of the nucleus, see Figure~\ref{fig:shape}, leads to zonally averaged irradiation conditions with significant illumination of the entire northern hemisphere and also partially of the southern hemisphere, see the left panel of Figure~\ref{fig:irradiation}.
This zonal mean of the irradiation closely correlates with the zonal mean of the emission rate across both hemispheres, see Figure~\ref{fig:minorinbound}.
For all four species we find increased activity ratios ($a_p>1.7$) on the northern patches $A_3$, $A_9$, and $A_{10}$.
The gases of the H$_2$O group show no significant active areas on the southern hemisphere, similar to H$_2$O and O$_2$ at that time \cite{Lauter2019}.
In contrast, CO$_2$ already displays active areas in the South \cite{Lauter2019}, although the solar irradiation does not peak at these latitudes.
This property is also shared by the other gases H$_2$S and CH$_4$ in the CO$_2$ group.
For both gases and CO$_2$ the activity ratios on the southern patches $A_1$, $A_4$, and $A_6$ are at least 1.3.
These three patches already show a presence of gas sources 300~days before perihelion, long before these areas become the most active emitters of the minor and major species.

\begin{figure}
\includegraphics[width=0.5\textwidth]{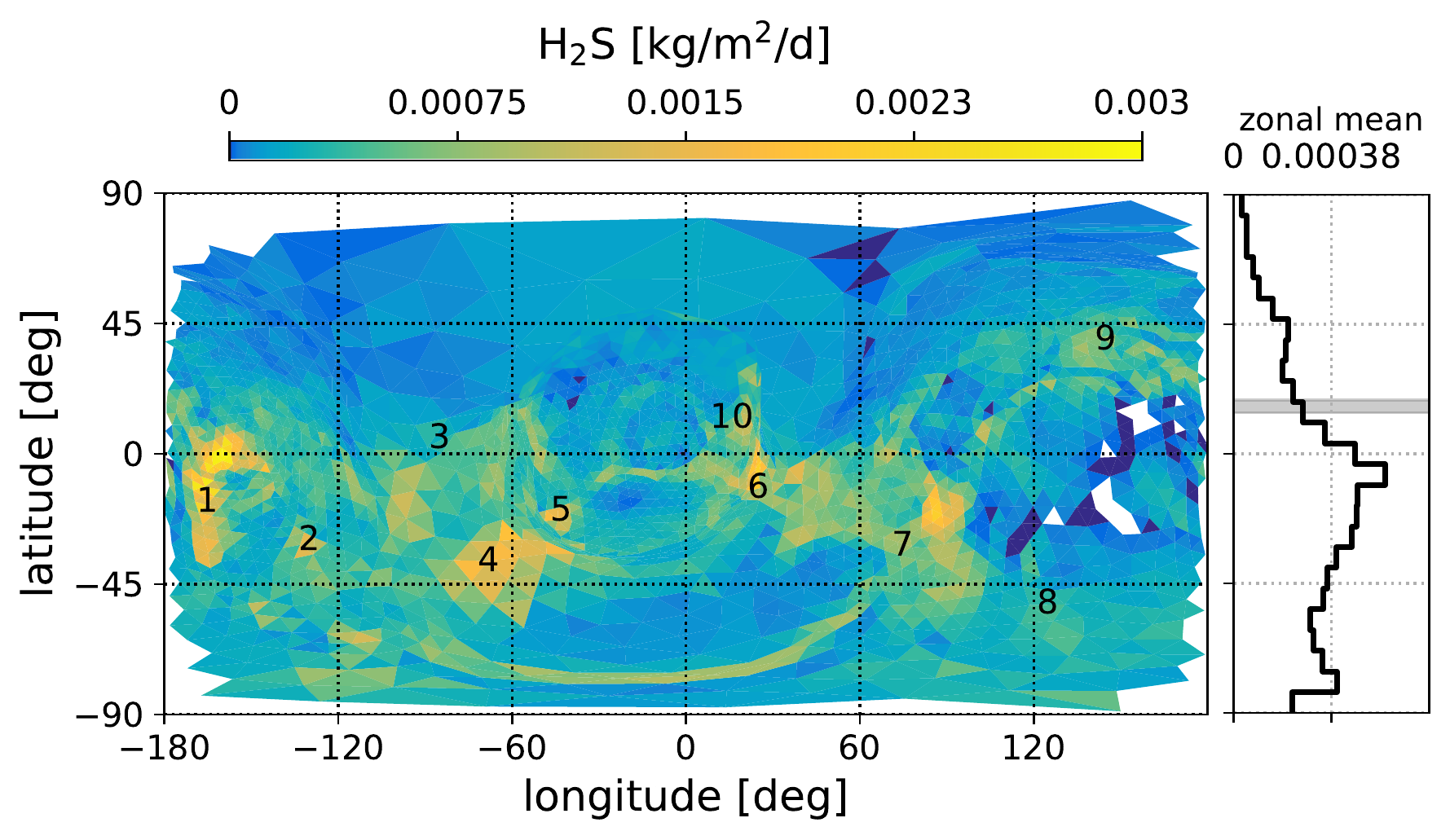}%
\hfill%
\includegraphics[width=0.5\textwidth]{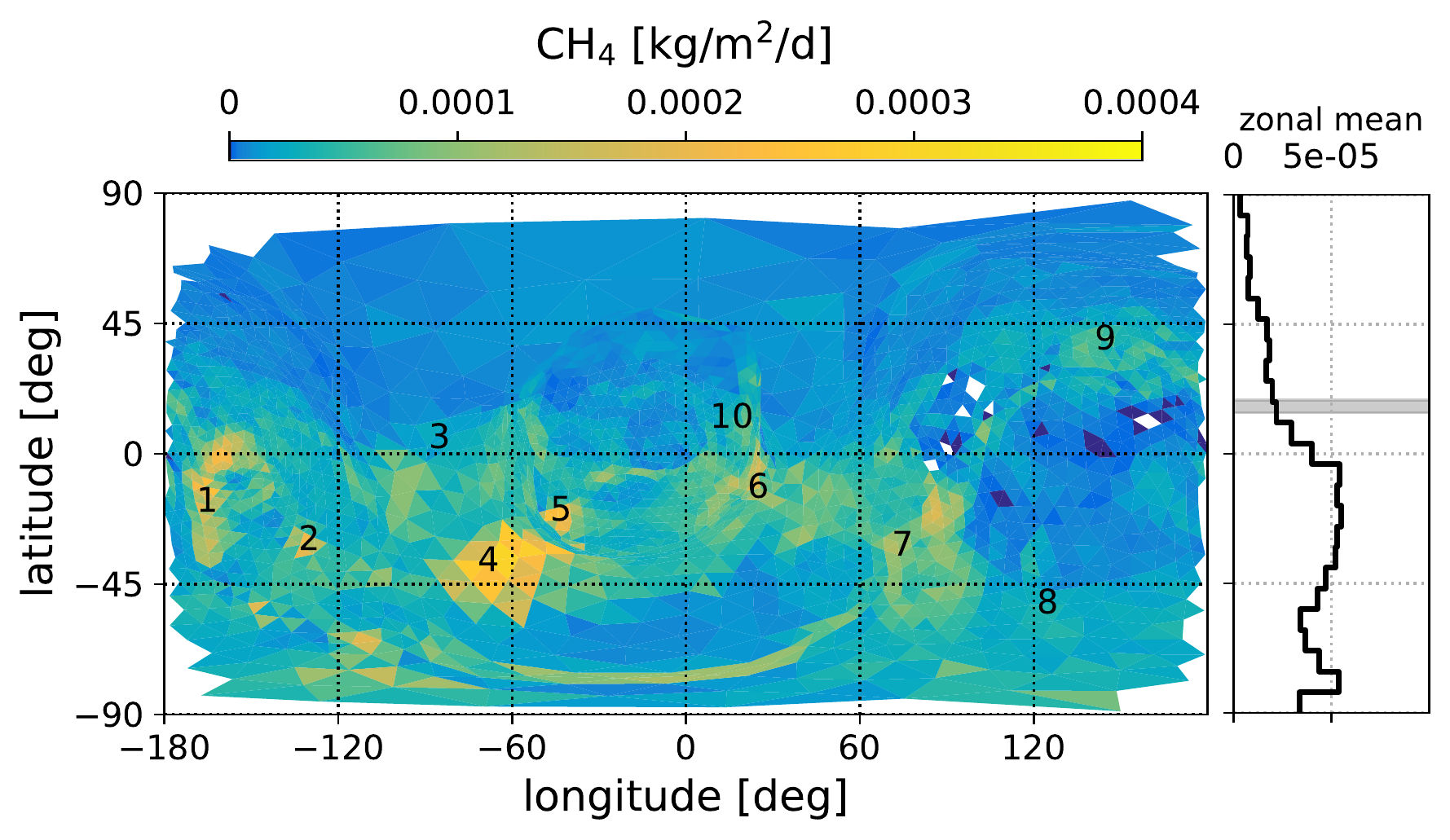}\\
\includegraphics[width=0.5\textwidth]{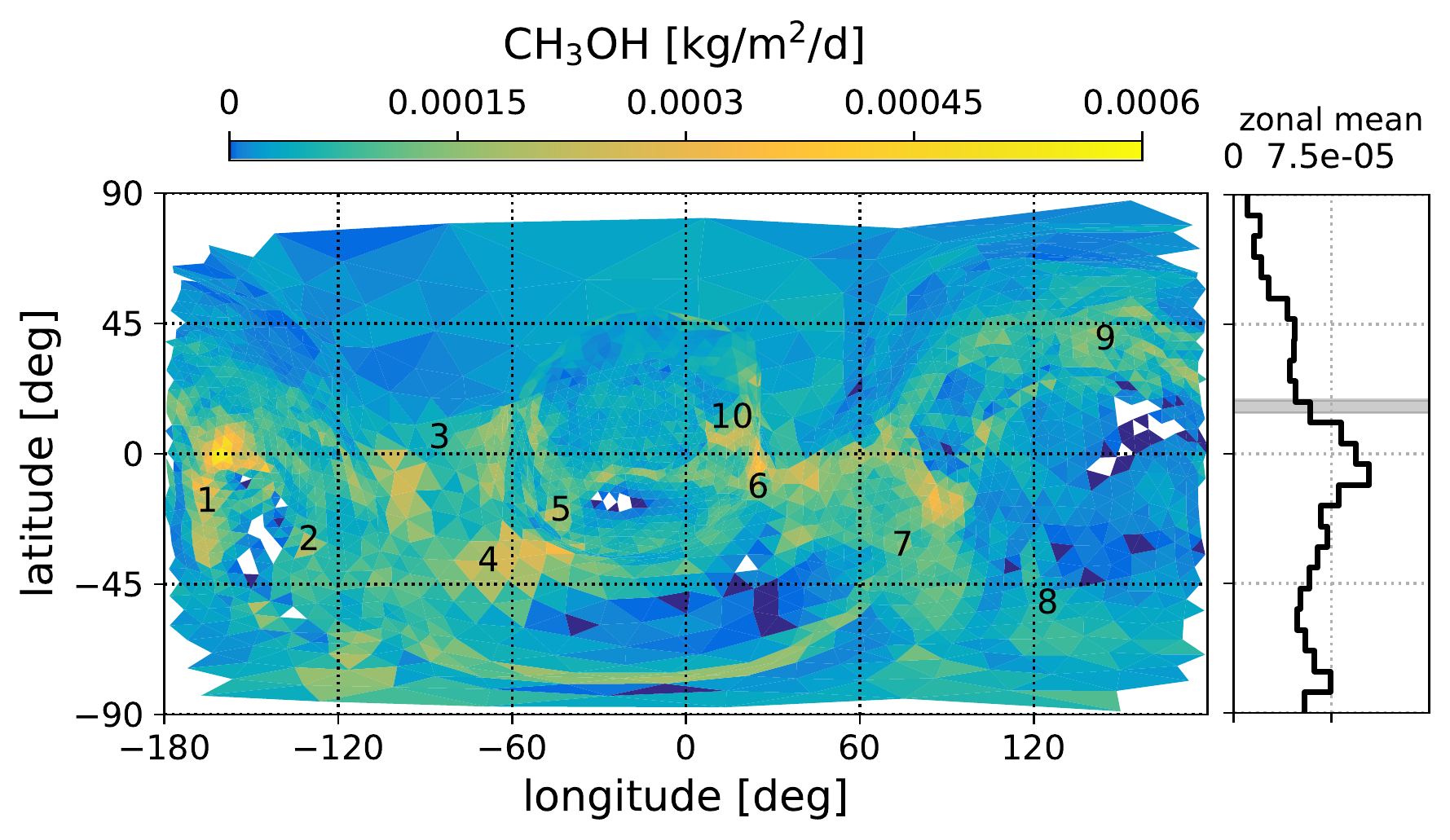}%
\hfill%
\includegraphics[width=0.5\textwidth]{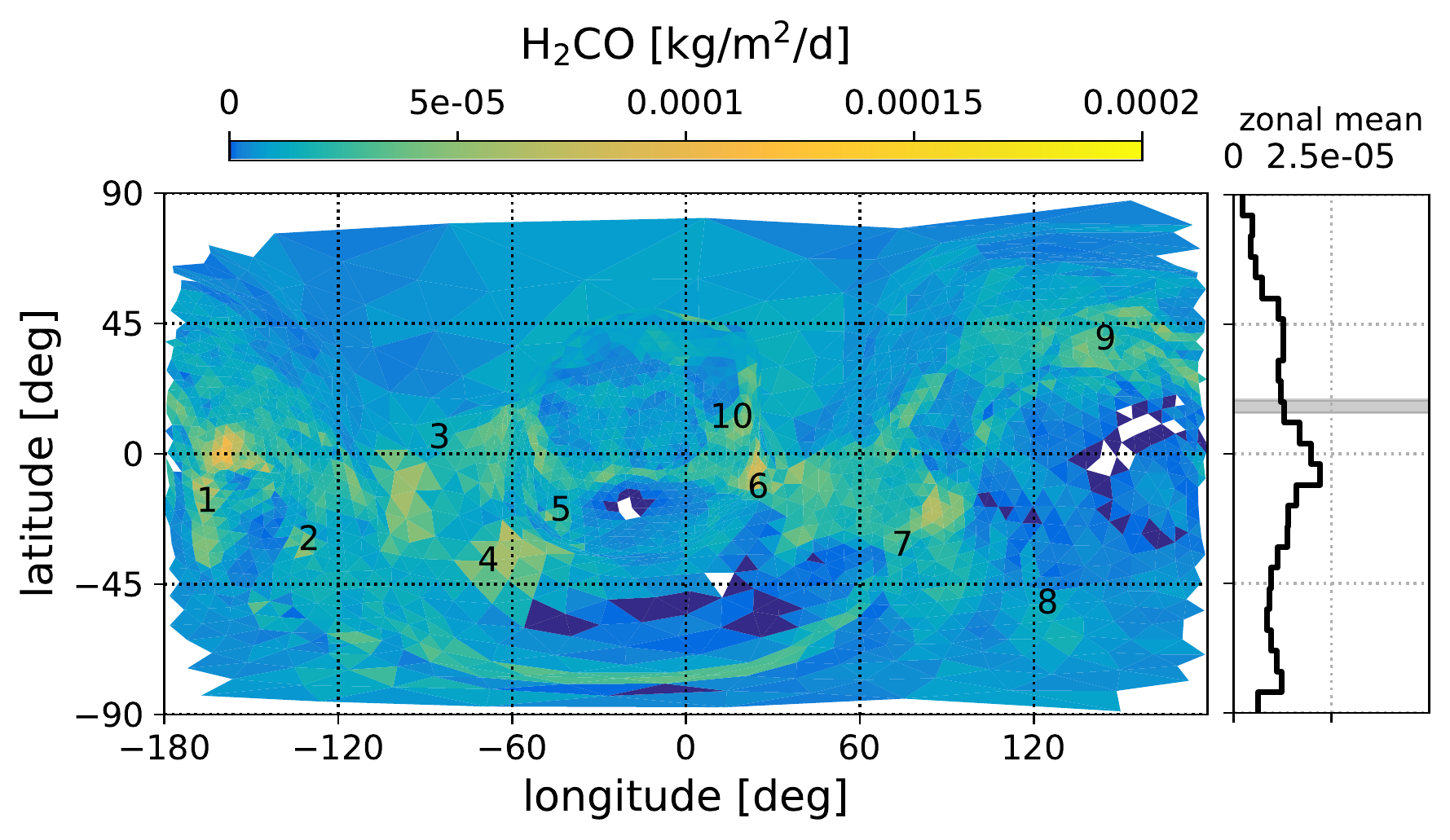}%
\caption{Surface emission rates for the minor gas species H$_2$S, CH$_4$, CH$_3$OH, and H$_2$CO, averaged from 340~days to 390~days after perihelion (see $I_\mathrm{outbound}$ in Table~\ref{tab:intervals}).%
}
\label{fig:minoroutbound}
\end{figure}
For the outbound time interval $I_{\mathrm{outbound}}$ (see Table~\ref{tab:intervals}) the discussion is complemented by the spatial distribution of emission rates in Figure~\ref{fig:minoroutbound} for the minor species H$_2$S, CH$_4$, CH$_3$OH and H$_2$CO.
300~days after perihelion, when the solar irradiation is diminished by more than a factor of 5 (see the right panel of Figure~\ref{fig:irradiation}).
The two northern patches $A_3$ and $A_9$ do not produce significantly more CO$_2$, H$_2$S and CH$_4$ than the mean with activity ratios between 1.0 and 1.2.
In the southern patches A$_1$, A$_4$, and A$_6$ at least twice as much of these three species is emitted with activity ratios between 2.2 and 5.5.
These areas with high emission rates during outbound time have already been reported \cite{Lauter2019} for CO$_2$, O$_2$, and CO.
For the species C$_2$H$_6$, NH$_3$, HCN, C$_2$H$_5$OH, OCS, and CS$_2$, the spatial distribution of emission rates is shown in the Figures of the Appendix.

\section{Sublimation and solar irradiation}
\label{sec:sublimation}

Solar irradiation provides the main energy influx to the icy components close to the cometary surface and drives the gas sublimation into the surrounding space.
In the last Section we examined the general correlation between localized emission rates on the surface and seasonal irradiation.
A quantitative analysis of this relation is the foundation for any thermophysical modeling \cite{Keller2020,Huebner2006}.
Thermophysical models predict the gas sublimation by evaluating all energy fluxes like incoming and outgoing radiation, heat conduction within the cometary material, and the energy going into the sublimation phase changes.
Thermal inertia due to energy exchange with subsurface layers results in retardation effects on the diurnal and seasonal scale.
This leads to a partial decoupling of instantaneous irradiation and gas release. 
In addition the presence of different volatiles, possibly refractory material and dust cover needs to be taken into account \cite{Blum2017,Attree2018a,Hu2019,Gundlach2020}.

Another consideration is the trapping and sealing off of ices consisting of different volatiles.
Amorphous water ice is able to trap minor species (like CO$_2$) and therefore it can affect the availability of volatile reservoirs at the sublimation front of water in the material column \cite{Davidsson2021b}.
Trapped species escape at higher temperature compared to their specific sublimation temperature \cite{Huebner1999,Filacchione2019}.
For comet 67P/C-G, we find that the temporal evolution of the minor gas species over the 2015 apparition suggests a grouping of minor species into those following CO$_2$ and H$_2$O, as defined in Section~\ref{sec:globalgas}.
The linked temporal evolution of the production rates of a respective minor species with a H$_2$O or CO$_2$ major species might indicate a trapping of the minor volatile \cite{Gasc2017}.

\begin{figure}
\includegraphics[width=0.5\textwidth]{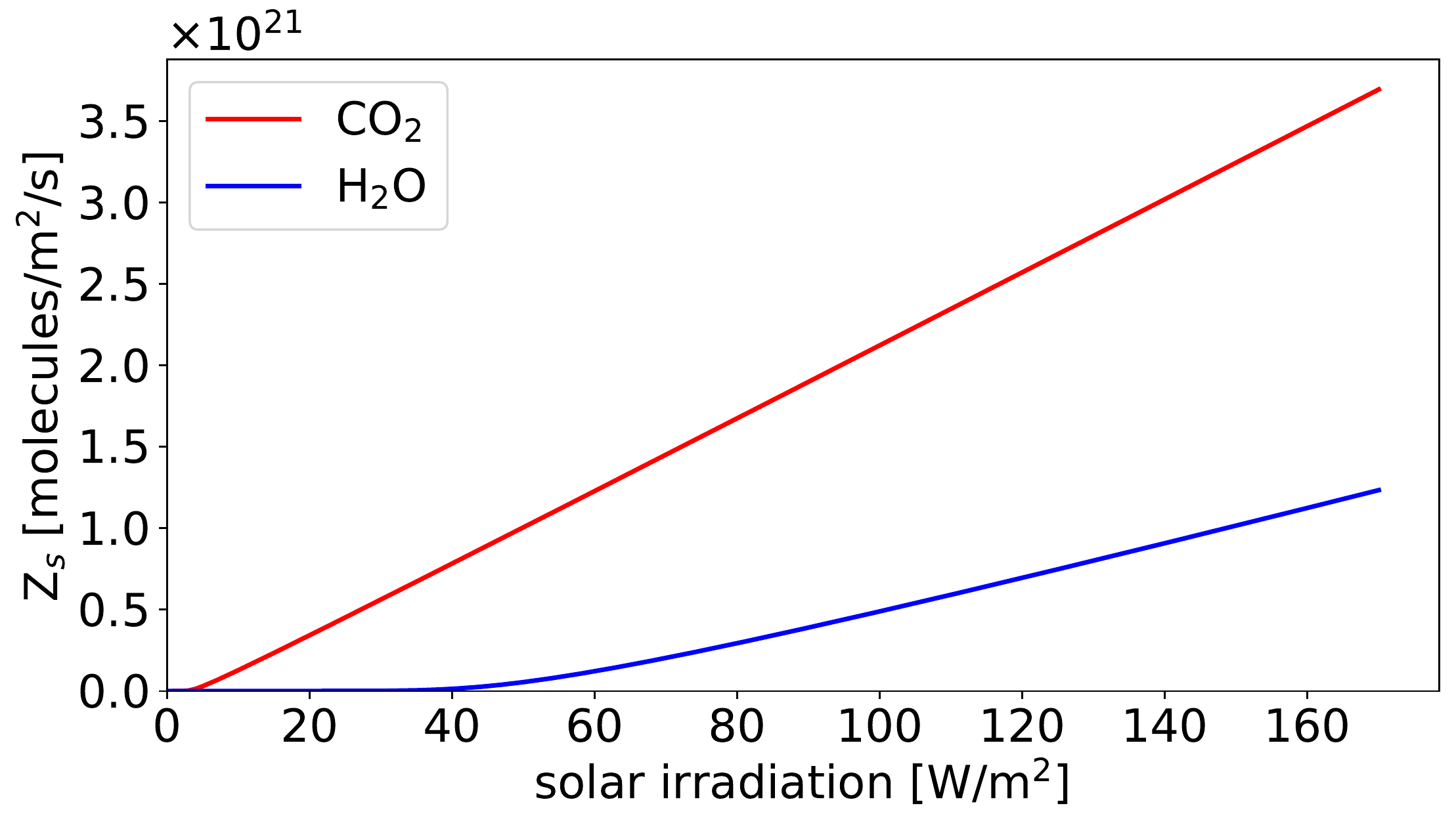}%
\caption{%
Sublimation rate for H$_2$O and CO$_2$ and for irradiation conditions on comet 67P/C-G based on the simplified sublimation model in Eq.~(\ref{eq:keller}).%
}
\label{fig:keller}
\end{figure}
The microscopic thermophysical models for cometary material composed of different ices and dust requires additional parameters, many of which are unknown or uncertain  \cite{Davidsson2021a}.
Therefore, we work in the following with a reduced parameters approach in a minimal thermophysical model for describing the sublimation of H$_2$O and CO$_2$ in response to the incoming radiation.
We consider each facet (average diameter 120~m) as representing a separate vertical column consisting of a porous mixture of ices together with dust.
We do not take amorphous water into account and assume that H$_2$O and CO$_2$ sublimate independently, and possibly at different depths. 
The non-trapping and non-blocking assumption between  H$_2$O and CO$_2$ is commonly applied in thermophysical models, see for instance \cite{Davidsson2021b}.
The independence of the H$_2$O and CO$_2$ sublimation leads to the coexistence of two vertically stacked sublimation fronts \cite{Davidsson2021a} at two distinct temperatures, with H$_2$O sublimating closer to the surface and with the CO$_2$ sublimation front at a larger depth (about 1.9\,m according to Davidsson et al. \cite{Davidsson2021a}).
The depth values of the sublimation front are seen as averaged values adjusted to match the observed CO$_2$ release.
To reduce the complexity we map the complex structure of the vertical columns into a direct relation between incoming irradiation and the gas production of a surface partially consisting of ice.
This approach allows one to determine the effective fraction of a fictitious icy surface which reproduces the observed emission rates \cite{Keller2015,Kramer2019}.
The effective surface fraction $f_p$ is determined independently for H$_2$O and CO$_2$ and with respect to a completely covered body by the respective ice species ($f_p=1$).
The sublimation rate at $f_p=1$ for both gas species $s=$ H$_2$O and CO$_2$ is determined from the energy balance between the solar irradiation $I$, the re-emitted radiation, and the sublimation rate $Z_s(T_s)$ together with the latent heat of sublimation ($L_s$)
\begin{equation}\label{eq:keller}
(1-A) I = \epsilon \sigma T_s^4 + Z_s(T_s) L_s.
\end{equation}
This relation provides a non-linear constraint for the temperature $T_s$.
Here we assume \cite{Keller2015} a bolometric Bond albedo of $A=0.01$, $\epsilon = 0.9$ for the emissivity, and $\sigma$ denotes the Stefan-Boltzmann constant.
Different values for the bond albedo between $0.01$ and $0.04$ have been considered in the literature  \cite{Keller2015,Statella2021,Davidsson2021b}, which do not change our analysis appreciably.
The latent heat of sublimation $L_s$ and Hertz-Knudsen relation $Z_s(T_s) = 2 p_s(T_s)/\pi/v_s(T_s)$ are specific for the gas species $s$ with the vapor pressure $p_s(T_s)$, the thermal velocity $v_s(T_s) = \sqrt{ 8 R T_s / \pi / \mu_s}$, the molar mass $\mu_s$, and the gas constant $R$.
With this model approach one obtains the temperature $T_s$ and the corresponding sublimation rate $Z_s(I)$ for any given irradiation $I$, see Figure~\ref{fig:keller}.
The temperature $T_s$ depends on the species and the gas emission is not affected by the occurrence of the other species.
We discuss the sublimation rates for the two major species H$_2$O and CO$_2$ based on
the volatile properties tabulated by Huebner et al. \cite{Huebner2006}.
For water we assume a temperature independent heat of sublimation $L_{\mathrm{H}_2\mathrm{O}} = 2.8 \times 10^6~ \mathrm{J}/\mathrm{kg}$, and take for the water vapor pressure $p_{\mathrm{H}_2\mathrm{O}}(T_{\mathrm{H}_2\mathrm{O}}) = 3.56 \times 10^{12} \exp^{-6141.67/T_{\mathrm{H}_2\mathrm{O}}}~\mathrm{Pa}$.
For $\mathrm{CO}_2$ we take as heat of sublimation $L_{\mathrm{CO}_2} = 5.94 \times 10^5~ \mathrm{J}/\mathrm{kg}$ and set for the vapor pressure $p_{\mathrm{CO}_2}(T_{\mathrm{CO}_2}) = 1.079\times 10^{12} \exp^{-3148/T_{\mathrm{CO}_2}}~\mathrm{Pa}$.
For both gases, $Z_s(I)$ follows a flat plateau for low irradiation conditions, see Figure~\ref{fig:keller}.
After a threshold of approximately $5~\mathrm{W}/\mathrm{m}^2$ for CO$_2$ and $50~\mathrm{W}/\mathrm{m}^2$ for H$_2$O, the sublimation curve becomes steeper and attains an almost constant slope with increasing irradiation.
The globally received solar irradiation follows the relation $I\sim r_{\mathrm{h}}^{-2}$ with respect to the heliocentric distance $r_{\mathrm{h}}$, while the local illumination conditions also depend on the rotation axis orientation.

\begin{figure}
\includegraphics[width=0.5\textwidth]{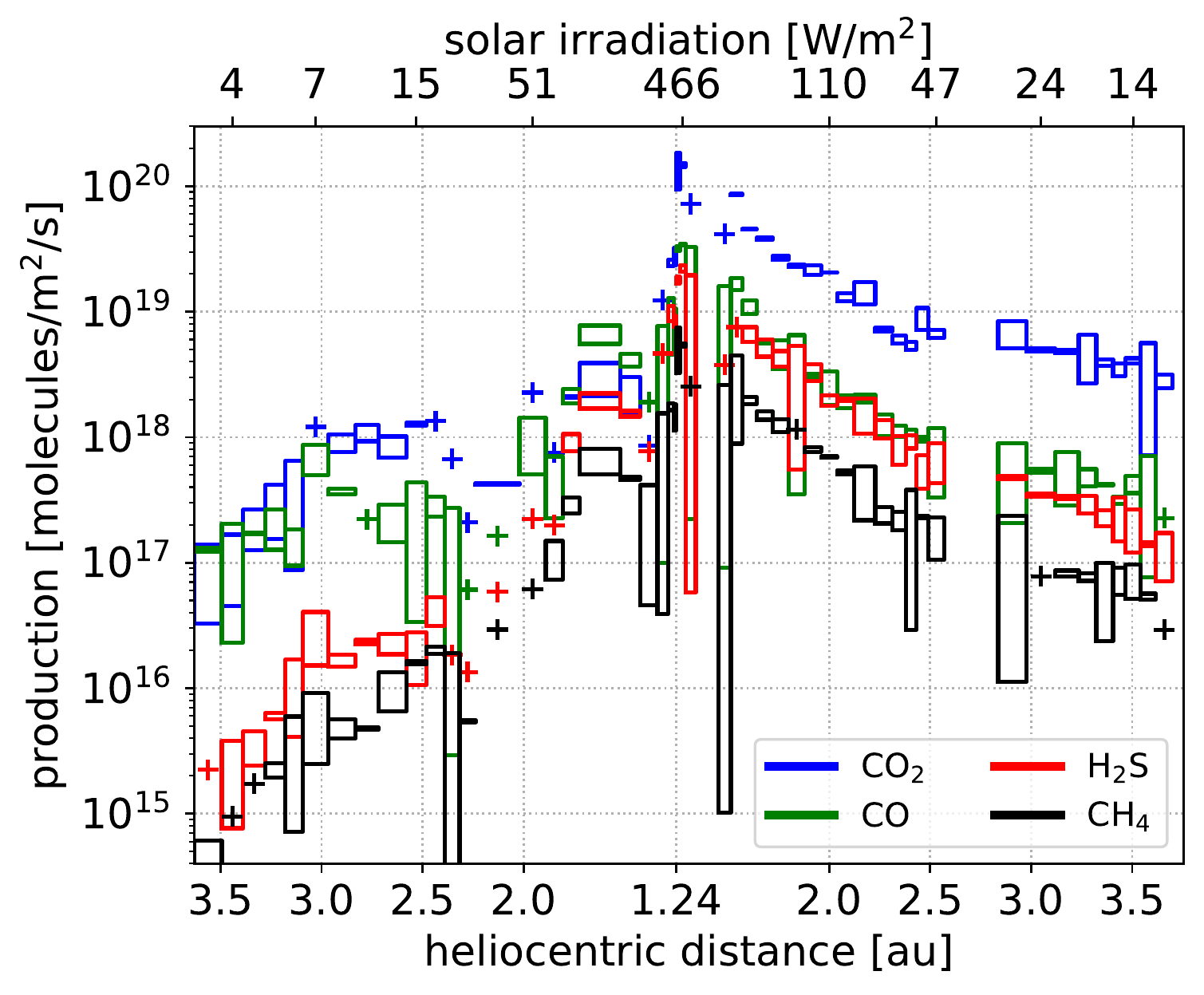}%
\includegraphics[width=0.5\textwidth]{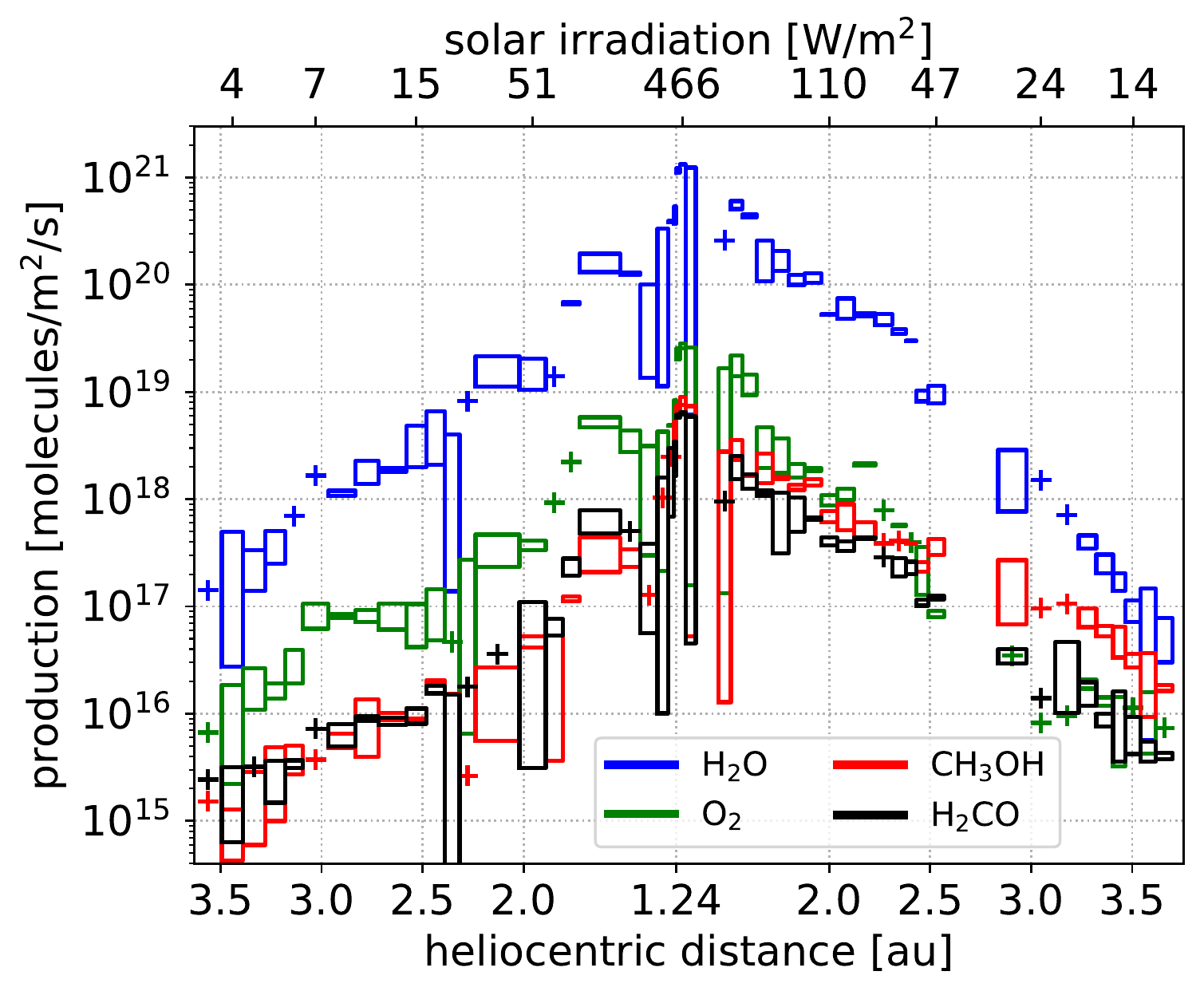}
\caption{%
Surface emission rate $\dot\rho_{s,i,j}$ averaged over the patch $A_4$.
The top axis denotes the diurnally and area-averaged solar irradiation on patch $A_4$.
Left panel: CO$_2$, CO, H$_2$S, CH$_4$, right panel: H$_2$O, O$_2$, CH$_3$OH, H$_2$CO.%
}
\label{fig:patchtree}
\end{figure}
The analysis for the global production rates $Q_s$, e.g. as shown in Figure 2 of Läuter et al. \cite{Lauter2020}, is complemented by the analysis of the surface emission rates $\dot\rho_{s,i,j}$ (see Section~\ref{sec:model}) averaged over the most active patch $A_4$ in Figure~\ref{fig:patchtree}.
On the same patch, the production of H$_2$O (as well as other gases like O$_2$, CH$_3$OH, H$_2$CO) diminishes faster than for CO$_2$, especially for heliocentric distances exceeding 2.5~au.
This finding is related to the increasing heliocentric distance and the retreat of the subsolar latitude to the northern hemisphere (see Figure~\ref{fig:irradiation}), which reduce the average radiation below $50~\mathrm{W}/\mathrm{m}^2$ and shut down the water sublimation (Figure~\ref{fig:keller}).
\begin{figure}
\includegraphics[height=0.275\textwidth]{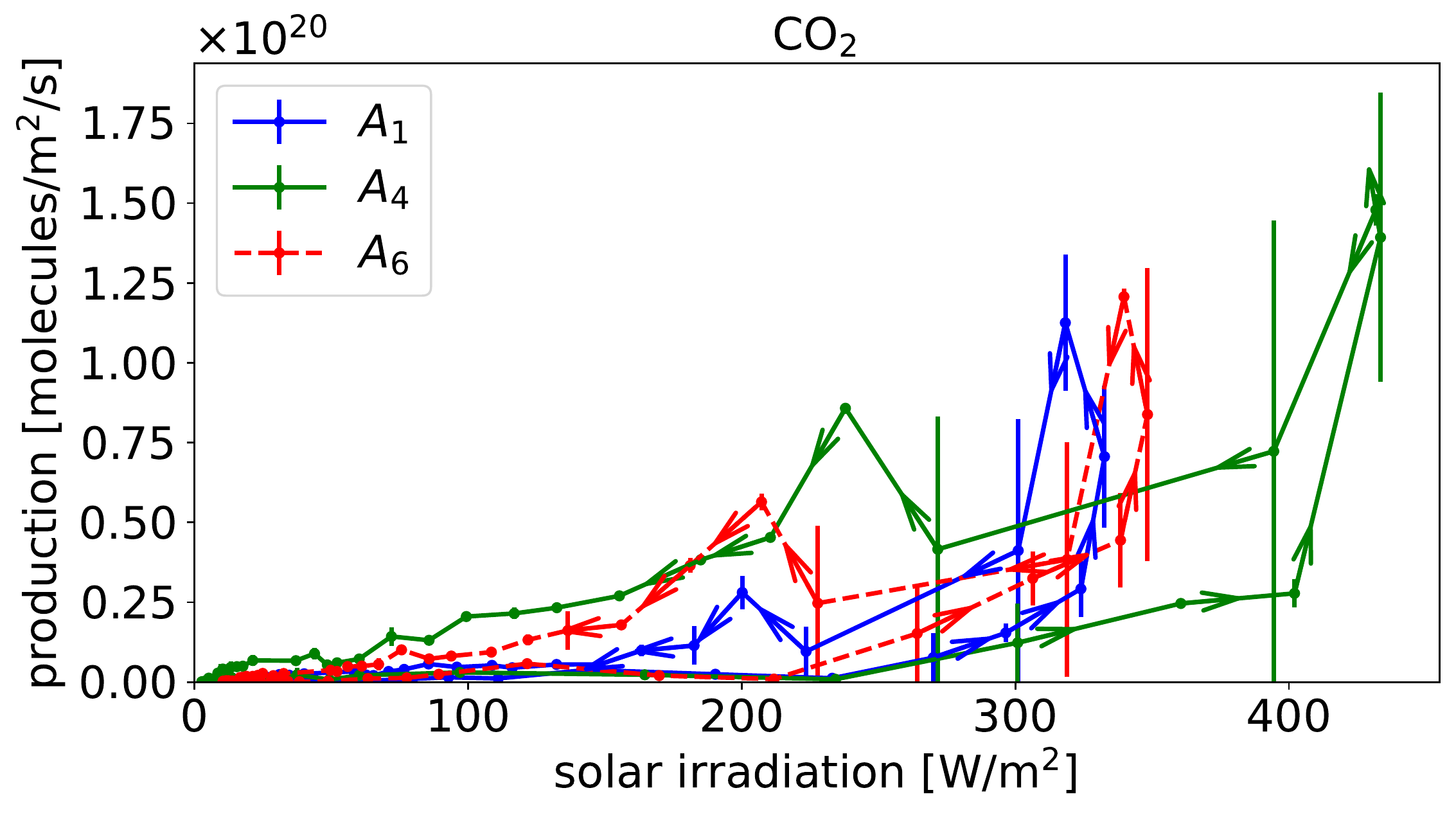}%
\hfill%
\includegraphics[height=0.275\textwidth]{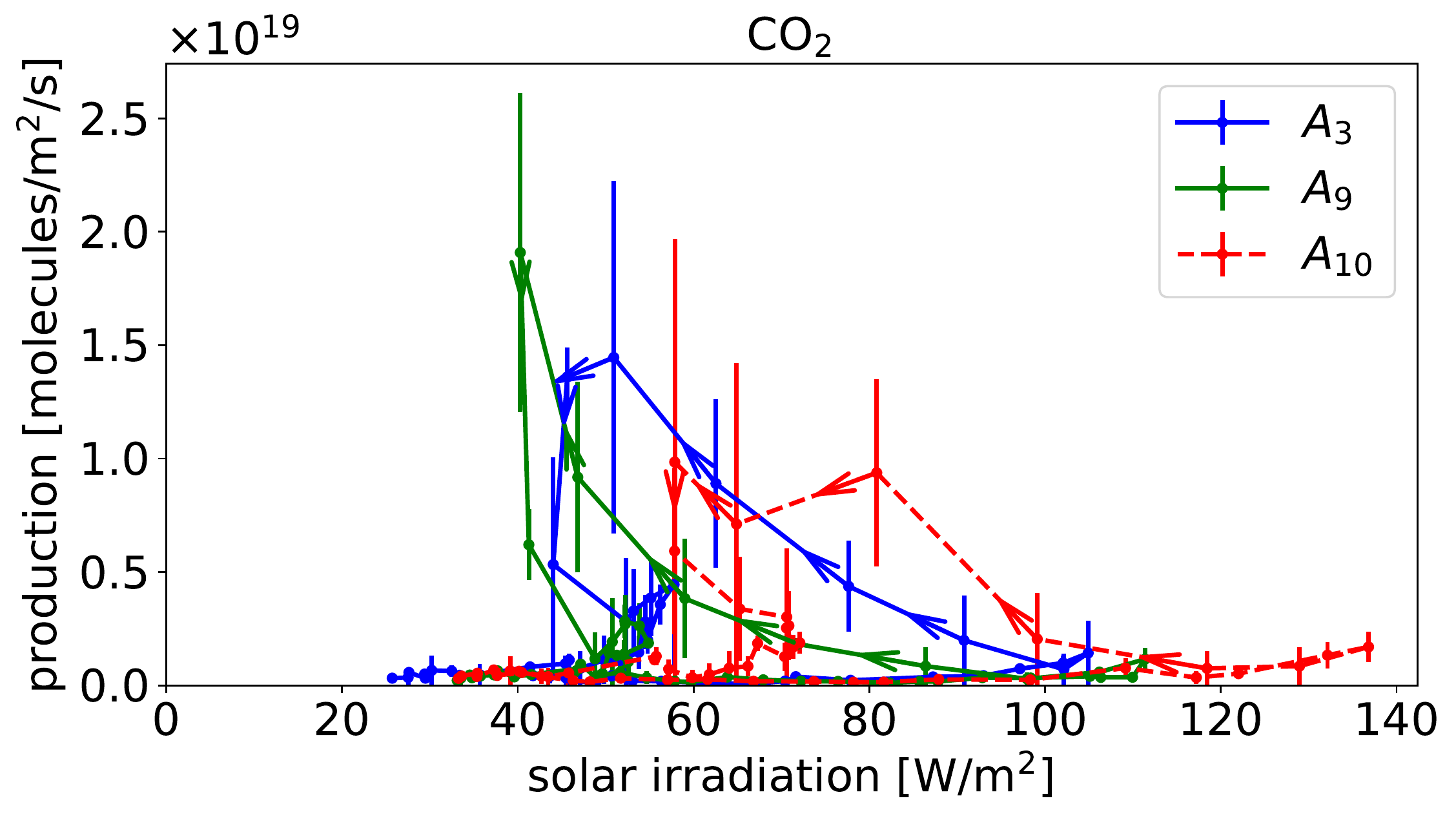}%
\caption{%
Surface emission rate $\dot\rho_{\mathrm{CO}_2,i,j}$ averaged over patches
as a function of diurnally averaged solar irradiation,
inbound orbital arc, around perihelion and outbound orbital arc.
The arrows denote the chronological sequence of the data points.
The data points are connected to guide the eye.
Left panel: on the southern hemisphere patches A$_1$, A$_4$, and A$_6$,
right panel: on the northern hemisphere patches A$_3$, A$_9$, and A$_{10}$.%
}
\label{fig:co2full}
\end{figure}
Similar to the evolution on patch A$_4$, each patch undergoes a specific temporal evolution of the solar irradiation on the one hand and of the gas production on the other hand.
The relation between irradiation and observed sublimation rate can be viewed as an 
\textit{effective sublimation curve}.
Figure~\ref{fig:co2full}, left panel, shows the effective sublimation of CO$_2$ for the three most active patches A$_1$, A$_4$, and A$_6$ in the southern hemisphere.
Around perihelion the patches receive a maximum irradiation between $300\,\mathrm{W}/\mathrm{m}^2$ and $450\,\mathrm{W}/\mathrm{m}^2$, which coincides with the maximum CO$_2$ production.
The relationship between the diurnally averaged radiation and the gas production is non-linear before and after perihelion and shows a hysteresis
effect when one traces the production chronologically. 
The arrows in Figure~\ref{fig:co2full} connect consecutive observation times and show that the same amount of solar irradiation for the inbound and outbound orbital arcs lead to a higher gas production at the outbound arc.
The right panel of Figure~\ref{fig:co2full} focuses on the northern patches A$_3$, A$_9$, and A$_{10}$.
Toward perihelion passage the irradiation decreases and the CO$_2$ production increases, see also Figure 4 in Läuter et al. \cite{Lauter2020}.
This indicates a failure for thermophysical models driven only by the instantaneous irradiation for the gas sublimation and requires a more comprehensive model.
Additional parameters possibly include diurnal and seasonal retardation effects and the consideration of back falling material.
Most COPS/DFMS measurements originate from terminator orbits, and therefore the measured densities do not sample the local time equally over one rotation, but preferentially from local morning and evening times. 
Only for a linear relationship between instantaneous solar irradiation and sublimation, the morning and evening conditions correspond to a diurnal mean of the gas production.

\begin{figure}
\includegraphics[height=0.275\textwidth]{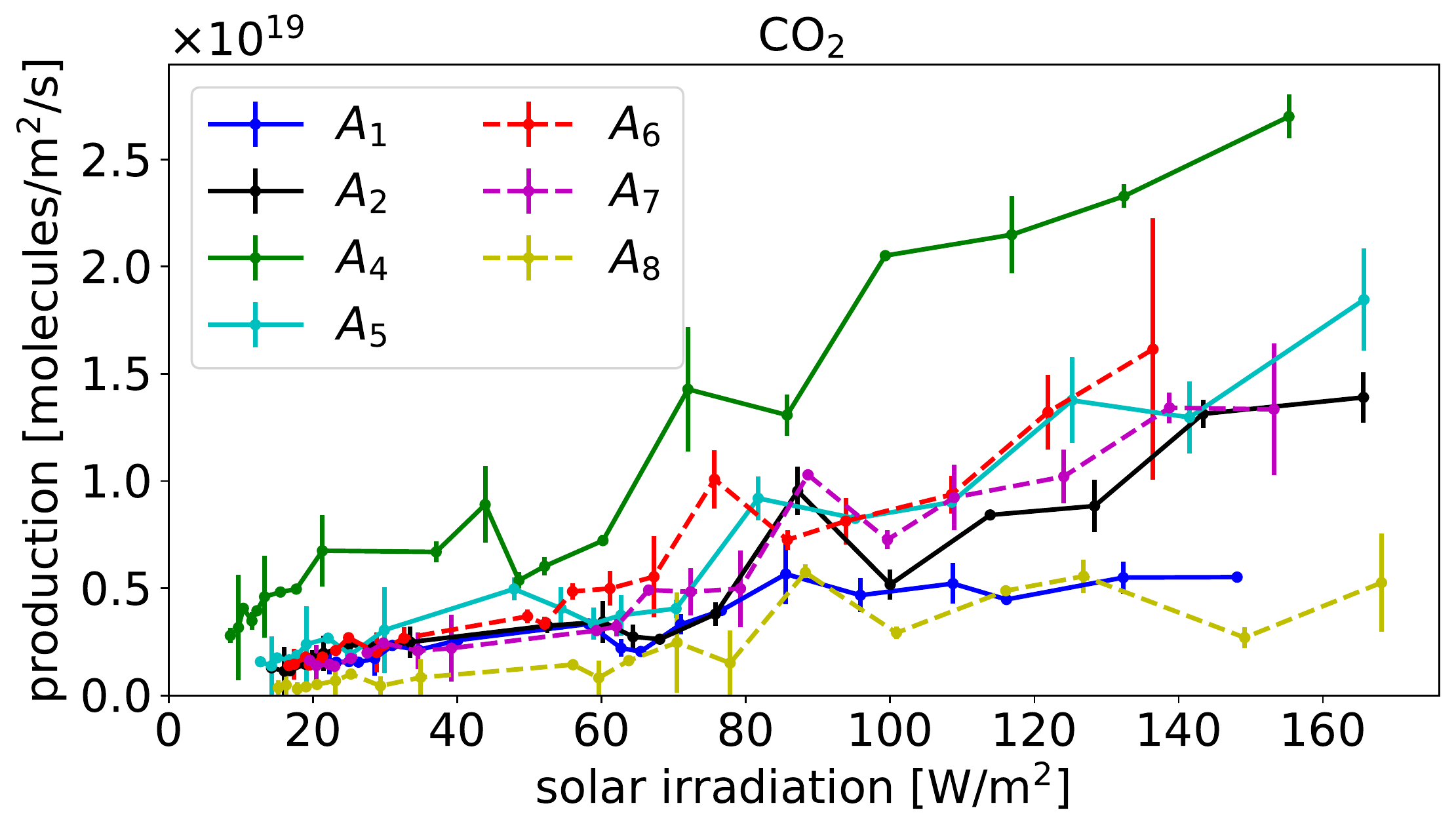}%
\hfill%
\includegraphics[height=0.275\textwidth]{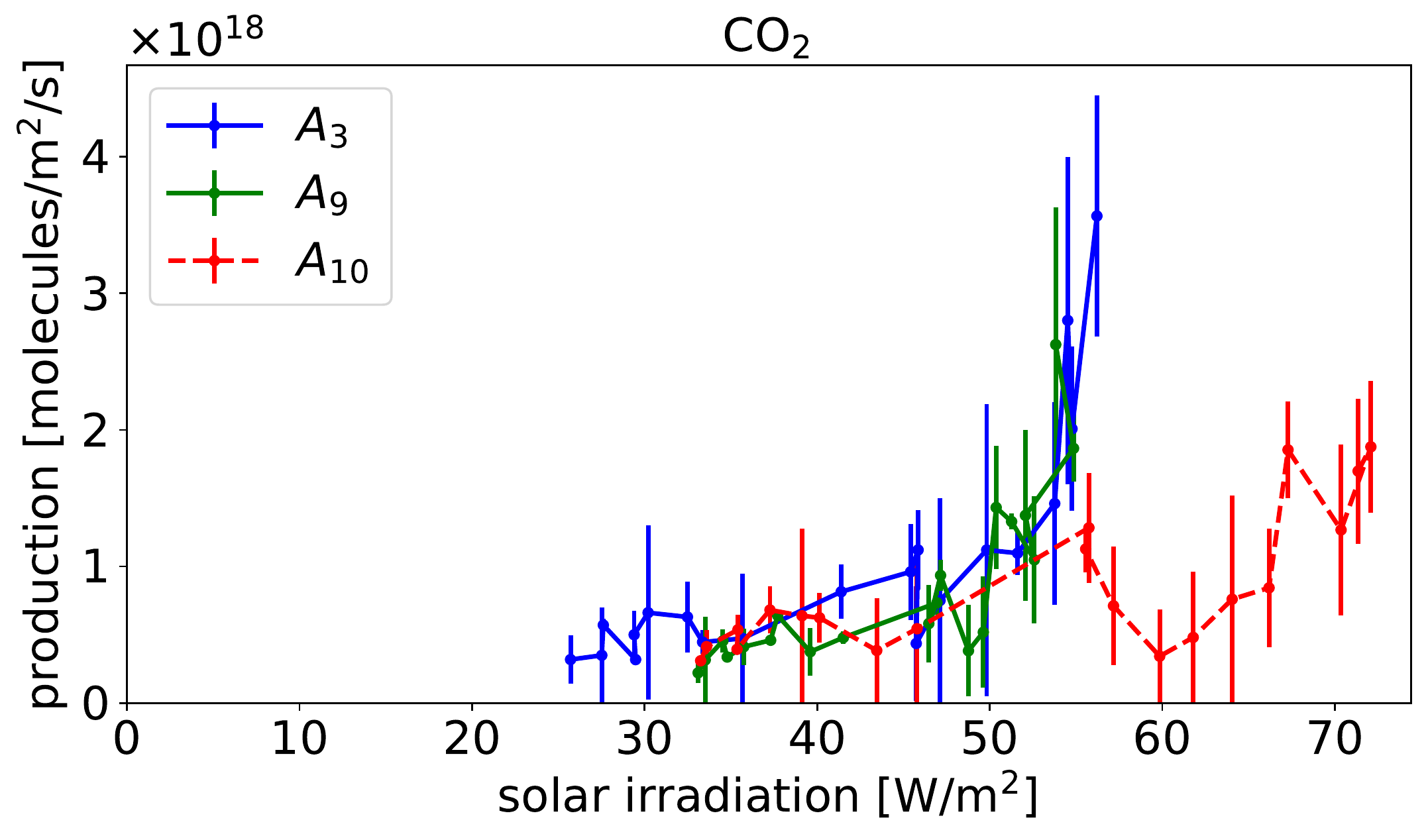}%
\caption{%
Surface emission rate $\dot\rho_{\mathrm{CO}_2,i,j}$ averaged over patches
as a function of diurnally averaged solar irradiation, between 100~days after and 390~days after perihelion.
The data points are connected to guide the eye.
Left panel: patches A$_1$, A$_2$, A$_4$--A$_8$
right panel: patches A$_3$, A$_9$, A$_{10}$.%
}
\label{fig:co2outbound}
\end{figure}
The minimal model used in this Section cannot explain the non-linear increase of the gas production around perihelion and the observed hysteresis.
To circumvent this problem we restrict ourselves to the outbound orbital arc 100~days to 390~days after perihelion, where the minimal model yields a better agreement with the measurements.
For all southern patches we observe an almost linear relationship between CO$_2$ production and irradiation, see the left panel of Figure~\ref{fig:co2outbound}.
We quantify the effective surface active fraction $f_p$ by comparing the observed relation of irradiation and emission rate to the theoretical sublimation function $Z_{\mathrm{CO}_2}(I)$ at $f_p=1$ shown in Figure~\ref{fig:keller}.
To determine $f_p$ for each southern patch $p$, the sublimation function $Z_{\mathrm{CO}_2}(I)$ is multiplied with $f_p$ to reproduce the observed data $Z_{\mathrm{CO}_2,p}(I)$ in Figure~\ref{fig:co2outbound}, namely to fit best the relation $f_{p} \cdot Z_{\mathrm{CO}_2}(I)=Z_{\mathrm{CO}_2,p}(I)$.
In our minimal parameters model $f_{p}$ signifies the  sublimation activity from a patch (which includes all exhumed gases in the column beneath the surface) to the incoming solar irradiation with respect to a fully ice-covered surface used to compute Figure~\ref{fig:keller}.
On the patches $A_1$, $A_2$, $A_{4}$--$A_8$ this value ranges between 0.2\% on $A_8$ and 0.8\% on $A_4$.
Filacchione et al. \cite{Filacchione2016} reported a fraction of 0.1\% of the area to show spectral signatures of CO$_2$ ice on one region (red dot in Figure~\ref{fig:h2operihel}) located in the patch $A_7$ (Anhur).

\begin{figure}
\includegraphics[height=0.275\textwidth]{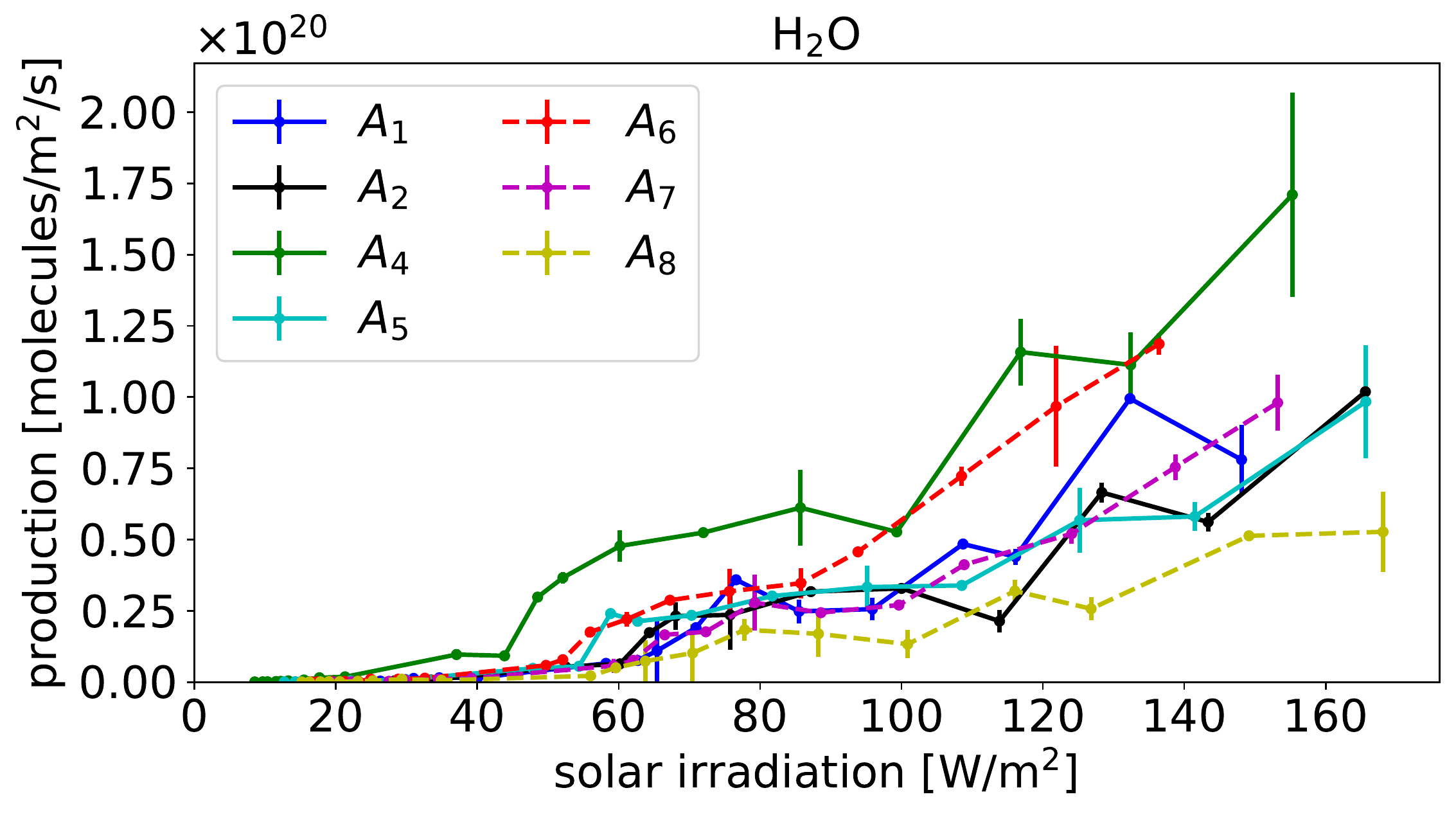}%
\hfill%
\includegraphics[height=0.275\textwidth]{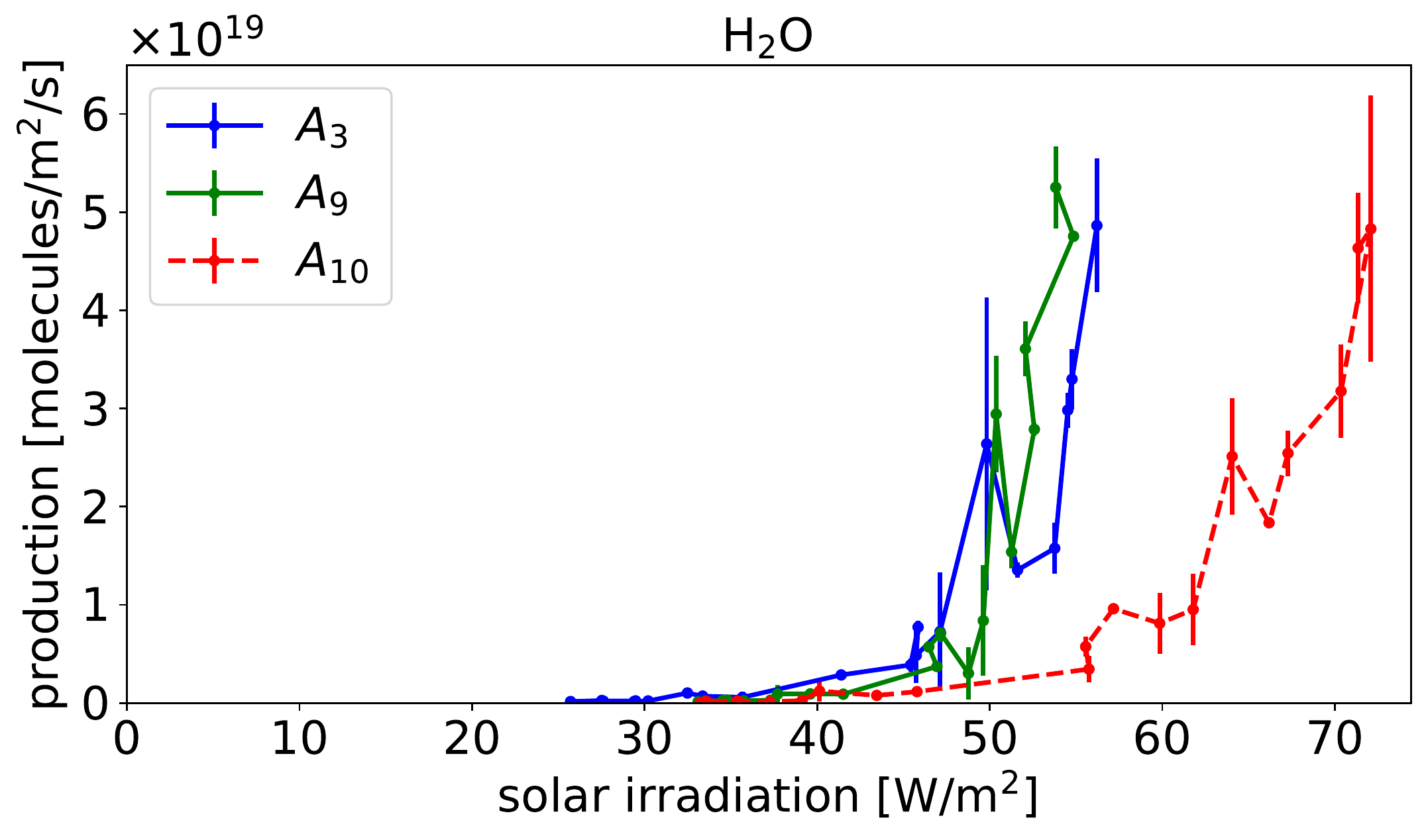}%
\caption{%
Surface emission rate $\dot\rho_{\mathrm{H}_2\mathrm{O},i,j}$ averaged over patches
as a function of diurnally averaged solar irradiation, between 100~days after and 390~days after perihelion.
The data points are connected to guide the eye.
Left panel: patches A$_1$, A$_2$, A$_4$--A$_8$
right panel: patches A$_3$, A$_9$, A$_{10}$.%
}
\label{fig:h2ooutbound}
\end{figure}
Figure~\ref{fig:h2ooutbound} shows the H$_2$O productions for the period 100 days after perihelion for all patches.
The analysis for H$_2$O shows similar hysteresis features as for CO$_2$ in Figure~\ref{fig:co2full} and also shows an increasing emission for a decreasing irradiation on the northern patches.
Quantitatively for H$_2$O both effects are smaller compared to CO$_2$.
In contrast to CO$_2$, the water sublimation is sharply reduced to values below $10^{19}\,\mathrm{molecules}/\mathrm{m}^2/\mathrm{s}$ for irradiation smaller than $50\,\mathrm{W}/\mathrm{m}^2$.
This is in agreement with the minimal thermophysical model for H$_2$O in Figure~\ref{fig:keller}.
The decline in water production occurs at heliocentric distances of 2.5\,au--3\,au \cite{Biver2019,Combi2020,Lauter2020}.
The surface active fraction $f_p$ for H$_2$O is evaluated in the same way as for CO$_2$,
by comparing the slopes in the Figure~\ref{fig:h2ooutbound} to the reference with $f_p=1$ shown in Figure~\ref{fig:keller}.
On the patches $A_1$, $A_2$, $A_{4}$--$A_8$ this value ranges between 4\% on $A_8$ and 15\% on $A_4$.

The surface active fraction $f_p$ obtained from our thermophysical model can be put in context with independent estimates of the active area.
Keller et al. \cite{Keller2015} include a fixed dust layer in the same model approach.
They obtain fractions of the active surface between 6\% and 25\%, for the northern and southern hemisphere.
Attree et al. \cite{Attree2019} consider a time varying surface active fraction in the range between 4\% and 35\% in different regions on the surface.
Kramer et al. \cite{Kramer2019} estimate surface active fractions between 14\% and 23\% constrained by changes of the rotation axis.
Filacchione et al. \cite{Filacchione2019} report fractions of the active area between 1\% and 30\% for water ice.
Herny et al. \cite{Herny2021} require 20\%  active area to fit the water production within a complete orbit.

The evaluation of the surface active fraction for the species H$_2$O and CO$_2$ on the patches $A_1$, $A_2$, $A_{4}$--$A_8$ only works at times later than 100 days after perihelion.
Then the monotonous relation between solar irradiation and emission rate allows us to apply the minimal thermophysical model.
The northern patches $A_3$, $A_9$, and $A_{10}$ display a more complex behaviour during this time period and the emission rates cannot be explained within the minimal model.
Additional factors might include the accumulation of dust (fallback).

\section{Ice composition on hemispheres and on both lobes}
\label{sec:patches}

The evolution of sublimation rates as a function of the received solar irradiation reflects a surface property which we have quantified as the surface active fraction with respect to one of the two major gases H$_2$O and CO$_2$.
Another property we can derive from the observed ROSINA data is the ice composition close to the surface.
The observed low water production for an average irradiation below $50\,\mathrm{W}/\mathrm{m}^2$ is linked to a change of the coma composition from a water dominated coma to a CO$_2$ dominated one 250~days after perihelion and later \cite{Gasc2017,Lauter2020}.
The redeposition of dust onto the nucleus surface and the resulting layered structure of the nucleus material with different sublimation fronts for water and CO$_2$ causes relative abundances in the released gas which can differ from the ice composition within the nucleus \cite{Huebner2006}.
Thus relative abundances of volatiles in the coma at a specific time may not reflect the ice composition of the nucleus directly \cite{Filacchione2019}.

Independent of the temporal evolution of the gas composition, the ``activity paradox'' \cite{Blum2014,Vincent2019,Keller2020,Ip2021} concerns the well known prevalence of a consistent and repetitive activity pattern over separated apparitions of a comet.
While the sustained activity is an observed fact, the theoretical modeling of comets does not provide a simple mechanism to keep a comet active across multiple apparitions \cite{Blum2014}.
The main challenge is the continuous removal of the dry dust mantle, which is required to keep ices sublimating. 
Some models assume a time-independent thickness of a dust-layer \cite{Blum2017} or adapt an empirically chosen dust erosion rate to match the observation \cite{Davidsson2021a,Davidsson2021b}.
In particular, the activity paradox refers to the limited understanding of the mechanisms for dust lift-off and fallback, and the dust composition.
E.g. if the fallback was dry (without ice available for sublimation) it would cover the surface and the activity would be choked off.
Consequently, a repeating emission rate across apparitions requires periodic and repetitive sublimation processes with similar molecular compositions and a largely unchanged ice composition.
To quantify the ice composition in the nucleus from the ROSINA data in the coma requires additional assumptions.

One approach is to take the coma composition at a specific time close to perihelion passage \cite{Rubin2020} as representative for the nucleus ice composition.
Around perihelion the highest erosion rates are expected and deeper layers of material might be continuously exposed on the cometary surface.
However, the sublimation rate is strongly affected by the local illumination conditions and the decoupling of CO$_2$ and H$_2$O sublimation leads to a variation of the relative abundances of the major and minor species in the coma over time.
A second approach, and this is our assumption for this section, is to link the ice composition to the integrated gas emissions over an entire orbital period which we refer to as the orbital production of the gas.
Even if the comet accumulates back falling material, the hypothesis of a repetitive activity across apparitions implies that fallback from a previous apparition will be shed off during the next one, and thus all sublimating materials are accounted for if one considers a complete orbital period.
The pronounced maxima for the gas production around perihelion \cite{Lauter2020} lead to a significant representation of this time period as well.
Thus the ice compositions of both approaches are comparable and within the range of the uncertainties of the gas production.

Our approach takes the relative mass-fractions of the orbital volatiles release to be representative of the ice composition close to the surface.
The entire orbital production has not been directly measured and can only be approximated from the observed production \cite{Lauter2020} $P_s$ integrated over the mission time between 2014 and 2016.
For all species we have estimated the uncertainty corresponding to the section of the orbital arc which was not covered by the Rosetta mission.
The assumption of constant global outgassing based on the latest observed production rates in September 2016 yielded uncertainties in the range of the preexisting uncertainties for $P_s$ (see Table 1 in Läuter et al. \cite{Lauter2020}).
That is why we consider the observed production $P_s$ to be a good approximation for the orbital production.

\begin{table}
\begin{tabular}{lcc}
$s$ & $P_s(A_{\mathrm{SH}})/|A_{\mathrm{SH}}|~(\mathrm{kg}/\mathrm{m}^2)$ &
$P_s(A_{\mathrm{NH}})/|A_{\mathrm{NH}}|~(\mathrm{kg}/\mathrm{m}^2)$ \\ \hline
H$_2$O     & $118\pm 22$ & $65\pm 6$ \\
CO$_2$     & $29\pm 7$   & $4.9\pm 1.6$ \\
CO         & $5.3\pm 1.2$    & $3.2\pm 0.8$ \\
H$_2$S     & $4.4\pm 1.5$    & $1.7\pm 0.6$ \\
O$_2$      & $4.7\pm 1.0$    & $2.7\pm 0.3$ \\
C$_2$H$_6$ & $2.1\pm 0.5$ & $[4.8\pm 1.4]\times 10^{-1}$ \\
CH$_3$OH   & $1.3\pm 0.3$ & $[4.6\pm 0.9]\times 10^{-1}$ \\
H$_2$CO    & $[9.5\pm 2.1]\times 10^{-1}$ & $[4.6\pm 1.2]\times 10^{-1}$ \\
CH$_4$     & $[4.8\pm 1.1]\times 10^{-1}$ & $[2.1\pm 0.6]\times 10^{-1}$ \\
NH$_3$     & $[4.0\pm 1.1]\times 10^{-1}$ & $[2.8\pm 0.8]\times 10^{-1}$ \\
HCN        & $[3.4\pm 0.8]\times 10^{-1}$ & $[1.6\pm 0.3]\times 10^{-1}$ \\
C$_2$H$_5$OH & $[4.2\pm 1.1]\times 10^{-1}$ & $[1.7\pm 0.3 ]\times 10^{-1}$ \\
OCS        & $[3.3\pm 0.8]\times 10^{-1}$ & $[1.2\pm 0.3]\times 10^{-1}$ \\
CS$_2$     & $[1.1\pm 0.2]\times 10^{-1}$ & $[4.2\pm 1.1]\times 10^{-2}$
\end{tabular}
\caption{%
Observed gas production $P_s$ per area for all species for the complete Rosetta mission time ranging from -377~days before to 390~days after perihelion.
The gas production is given separately for the southern hemisphere $A_{\mathrm{SH}}$ (left column with the area $|A_{\mathrm{SH}}| =20.71\times 10^6\,\mathrm{m}^2$) and the northern hemisphere $A_{\mathrm{NH}}$ (right column with the area $|A_{\mathrm{NH}}| = 23.60\times 10^6\,\mathrm{m}^2$).
}
\label{tab:hemidata}
\end{table}
Based on the values in Table 1 of Läuter et al. \cite{Lauter2020} and the entire surface area $|A_{\mathrm{67P}}|$ from Section~\ref{sec:globalgas} the accumulated production (the sum $\sum P_r$ over all species $r$) is $121\pm 13\,\mathrm{kg}/\mathrm{m}^2$.
The same summations of the gas production (separately for both hemispheres in Table~\ref{tab:hemidata}) leads to the observed production of $168\pm 23\,\mathrm{kg}/\mathrm{m}^2$ and $80\pm 6\,\mathrm{kg}/\mathrm{m}^2$ on the southern and on the northern hemisphere, $A_\mathrm{SH}$ and $A_\mathrm{NH}$, respectively.
These significant differences are related to the higher irradiation of the southern hemisphere, since southern solstice occurs just 23~days after perihelion (see  Figure~\ref{fig:irradiation}).

\begin{figure}
\includegraphics[width=0.49\textwidth]{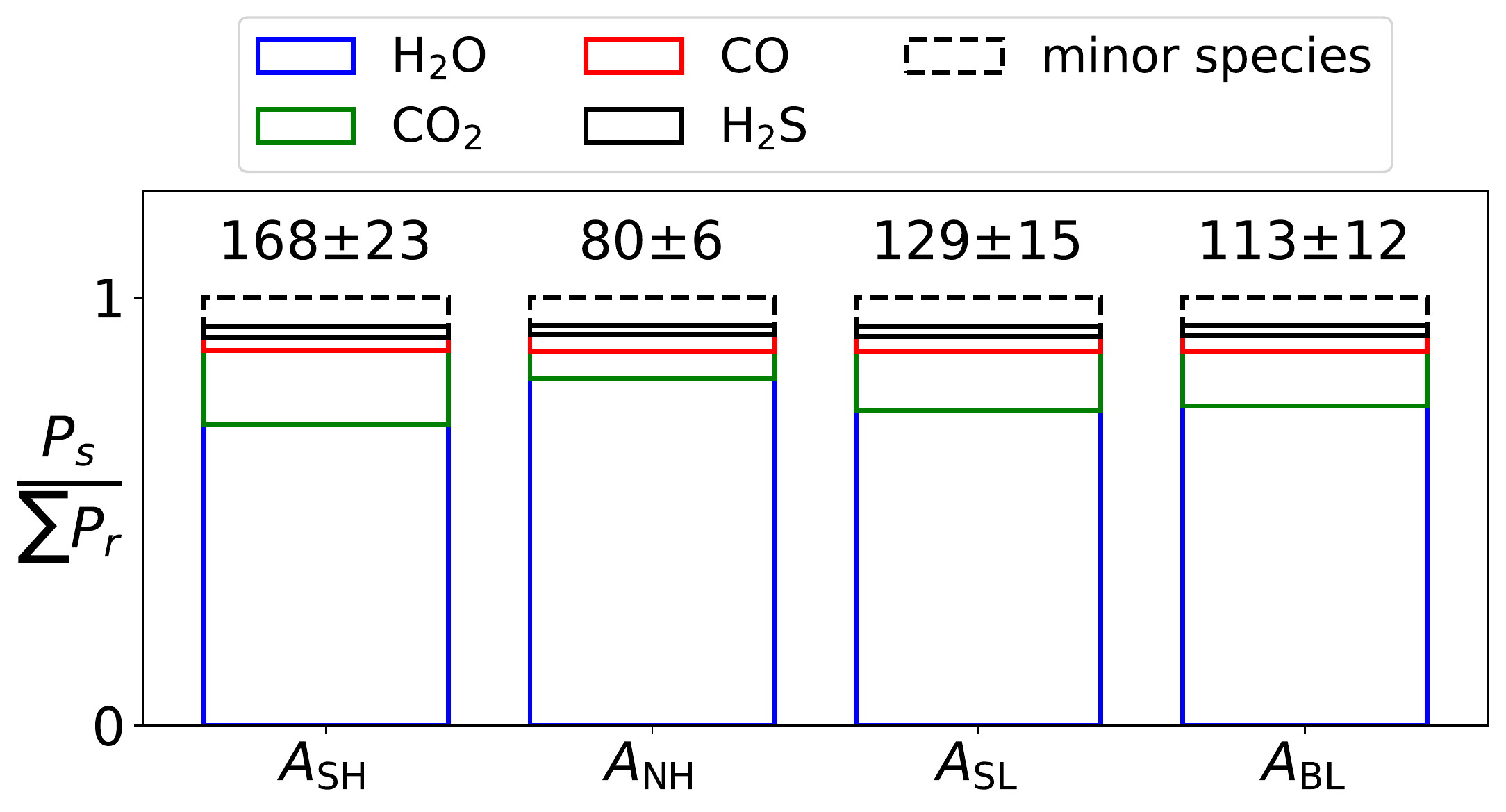}
\caption{%
Observed gas production $P_s$ (integrated emission rates) over the entire 2015 apparition for the southern ($A_\mathrm{SH}$) and northern hemisphere $A_\mathrm{NH}$,
and the small ($A_\mathrm{SL}$) and big lobe $A_\mathrm{BL}$, see Table~\ref{tab:geomorph}.
Mass-fraction of a species $P_s/\sum P_r$ is reflected by the segment size in the bar.
On top of each bar, the sum (over the species) per area $\sum P_r / |A|~[\mathrm{kg}/\mathrm{m}^2]$ is given. The minor species cover O$_2$, C$_2$H$_6$, CH$_3$OH, H$_2$CO, CH$_4$, NH$_3$, HCN, C$_2$H$_5$OH, OCS, and CS$_2$ and are expanded in Figure~\ref{fig:region_minor}.%
}
\label{fig:region_major}
\end{figure}

\begin{figure}
\includegraphics[height=0.26\textwidth]{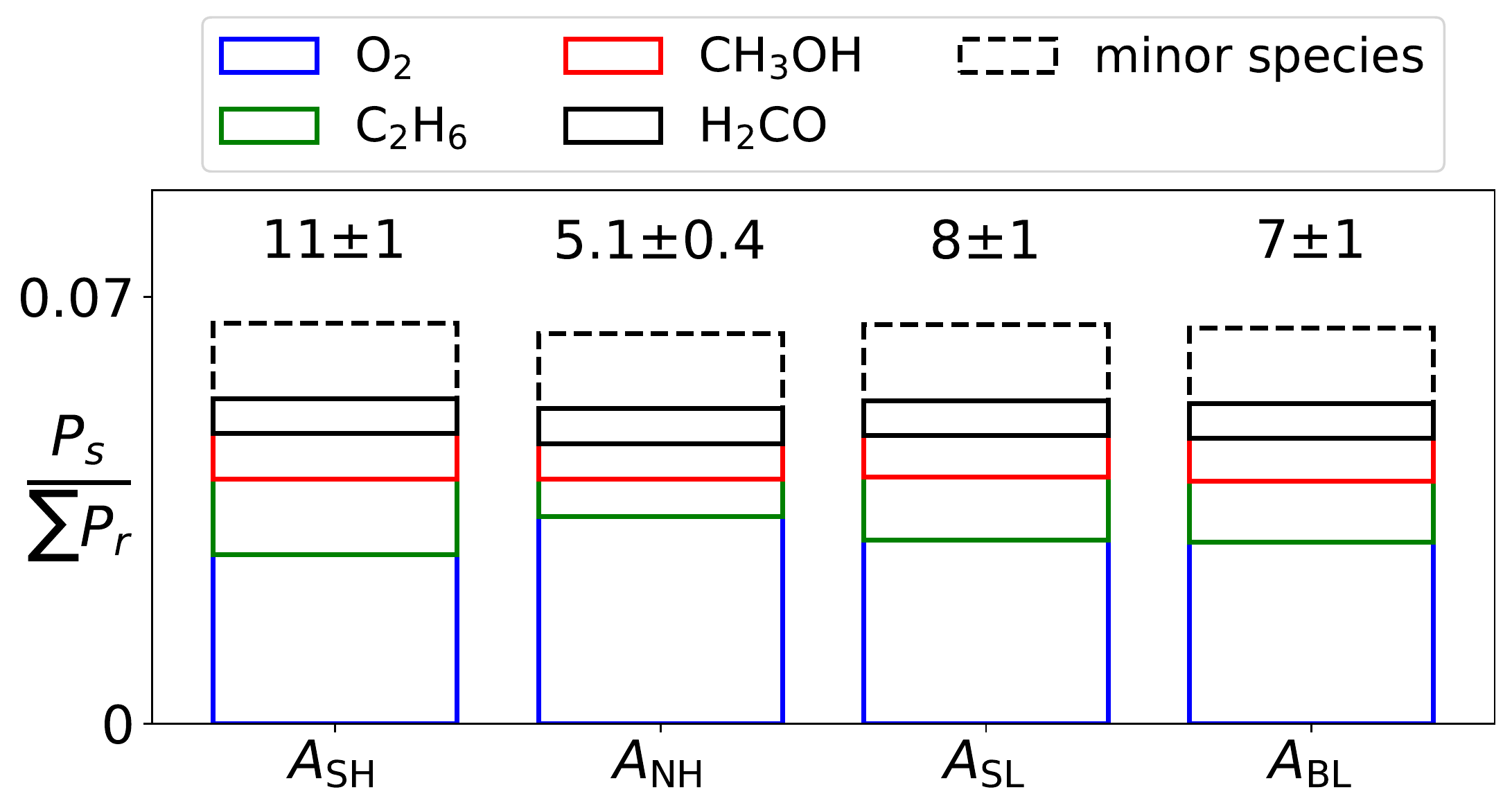}%
\hfill%
\includegraphics[height=0.26\textwidth]{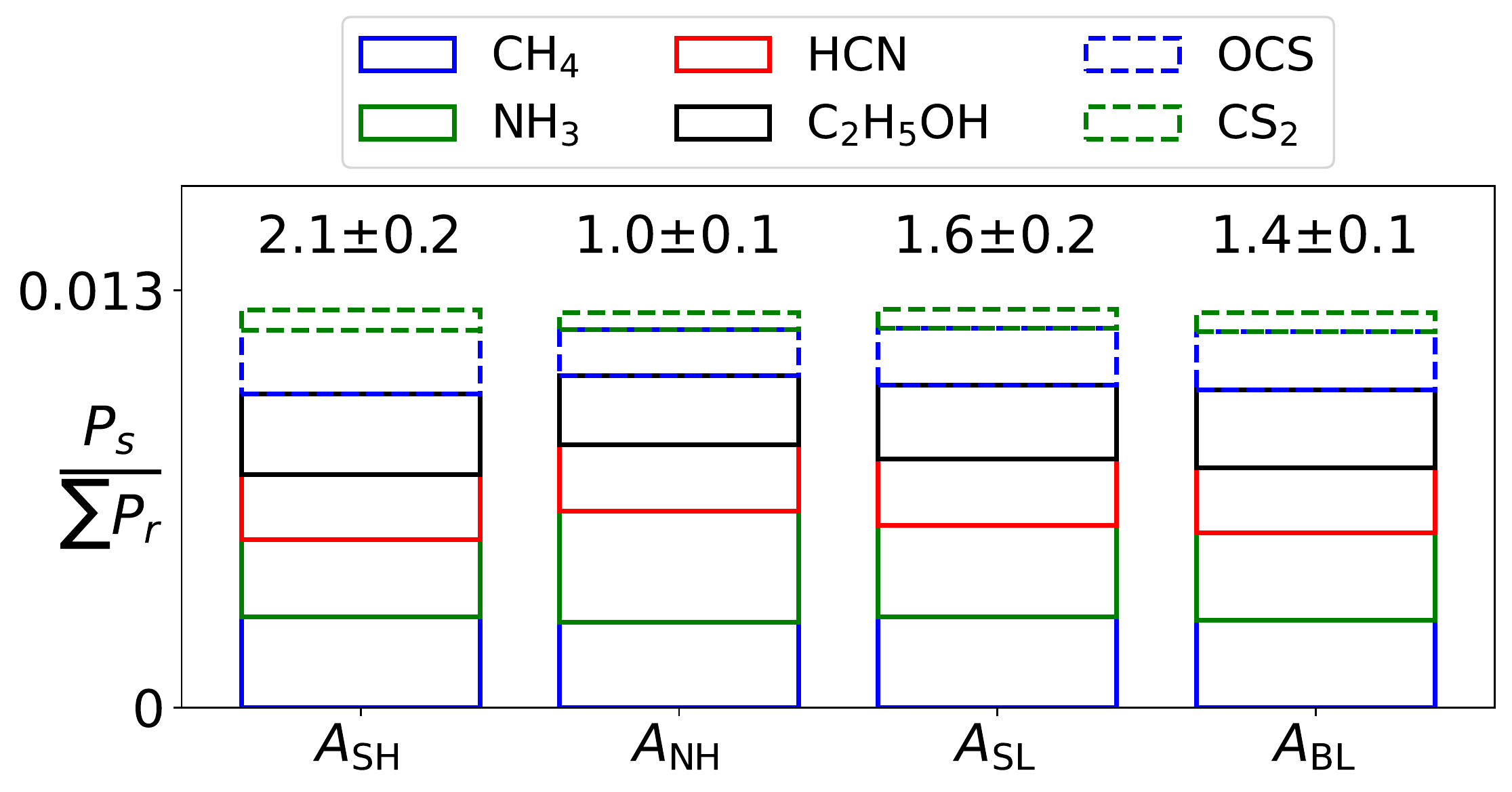}%
\caption{%
Observed production for minor species $P_s$ (integrated emission rates) over the entire 2015 apparition for the southern ($A_\mathrm{SH}$) and northern hemisphere $A_\mathrm{NH}$, and the small ($A_\mathrm{SL}$) and big lobe $A_\mathrm{BL}$, see Table~\ref{tab:geomorph}.
Mass-fraction of a species $P_s/\sum P_r$ is reflected by the segment size in the bar.
On top of each bar, the sum (over the species) per area $\sum P_r / |A|~[\mathrm{kg}/\mathrm{m}^2]$ is given.
Left panel: The minor species cover CH$_4$, NH$_3$, HCN, C$_2$H$_5$OH, OCS, and CS$_2$ which are given in the right panel.%
}
\label{fig:region_minor}
\end{figure}

Figures~\ref{fig:region_major} and \ref{fig:region_minor} visualize the production data for both hemispheres given in Table~\ref{tab:hemidata} and additionally they display the productions from the small and big lobes, $A_\mathrm{SL}$ and $A_\mathrm{BL}$.
For both lobes, we find similar values for the observed productions, namely $129\pm 15\,\mathrm{kg}/\mathrm{m}^2$ on $A_\mathrm{SL}$ and $113\pm 12\,\mathrm{kg}/\mathrm{m}^2$ on $A_\mathrm{BL}$.
We associate the relative mass-fractions of $P_s$ in the Figures~\ref{fig:region_major} and \ref{fig:region_minor} with the ice composition in the region of interest and deduce a similar composition of  both lobes.

Averaged over the complete surface the mass-fractions in the surface ice for H$_2$O and CO$_2$ are $74\pm 13\%$ and $13\pm 4\%$, respectively.
These mass-fractions correspond to a ratio of $7\pm 2\%$ for CO$_2$/H$_2$O (taken as mole-fraction, see Table 1 in Läuter et al. \cite{Lauter2020}).
Based on the same DFMS data Combi et al. \cite{Combi2020} reported a CO$_2$/H$_2$O ratio of 7.3\%.
The CO$_2$/H$_2$O ratio changes with time (see Figure 5 in Läuter et al. \cite{Lauter2020}) and can be compared with reported relative abundances at specific time periods.
For April 2015 Migliorini et al. \cite{Migliorini2016} give a value for the CO$_2$/H$_2$O ratio of 2.4\%--3.9\%,
Fink et al. \cite{Fink2016} report 3.7\%--7.4\% between February and April 2015, 
Rubin et al. \cite{Rubin2019} report $4.7\pm 1.4\%$ for May 2015, and
Bockel{\'e}e-Morvan et al. \cite{Bockelee-morvan2016} obtain 14\%--32\% between July and September 2015.

In addition to the reported difference of observed production on the southern and on the northern hemisphere, also the mass-fractions differ between both hemispheres.
Based on the CO$_2$ values in Table~\ref{tab:hemidata} and Figure~\ref{fig:region_major}, the mass-fraction $6\pm 2\%$ in the ice on $A_\mathrm{NH}$ is decreased compared to $17\pm 5\%$  on $A_\mathrm{SH}$.
This is related to CO$_2$/H$_2$O ratios (mole-fraction) of $3.1\pm 1.0\%$ on $A_\mathrm{NH}$ and $10\pm 3\%$ on $A_\mathrm{SH}$. 
The distinctive feature of a decreased CO$_2$ fraction on $A_\mathrm{NH}$ is also discussed by other authors \cite{Fink2016,Bockelee-morvan2016}.
Without a conclusive explanation they claim that the surface ice on $A_\mathrm{NH}$ does not reflect the original composition of the nucleus.
The illumination history could induce a stratified ice composition and affects a devolatilization process for CO$_2$ in the top layers \cite{Marboeuf2014}.
A complementing explanation points to lifted ice and dust particles due to surface gas activity (from the South) which later fall back in another region (in the North) due to their gravitational binding \cite{Fulle2019a,Combi2020,Davidsson2021}.
Similarly to water and CO$_2$, the mass-fractions (in relation to all ices) of the minor species O$_2$, C$_2$H$_6$, and NH$_3$ differ on $A_\mathrm{SH}$, with the values $2.8\pm 0.7\%$, $1.2\pm 0.3\%$ and $0.24\pm 0.07\%$, and on $A_\mathrm{NH}$, with the values $3.4\pm 0.5\%$, $0.6\pm 0.2\%$ and $0.35\pm 0.10\%$, respectively.
On both lobes the mass-fractions for H$_2$O and CO$_2$ are almost identical, namely $74\pm 14\%$ and $14\pm 4\%$ on $A_\mathrm{SL}$ and $75\pm 13\%$ and $13\pm 3\%$ $A_\mathrm{BL}$, respectively.
Similarly, the mass-fractions for all minor species shown in Figure~\ref{fig:region_minor} are comparable between both lobes.

\begin{figure}
\includegraphics[width=0.49\textwidth]{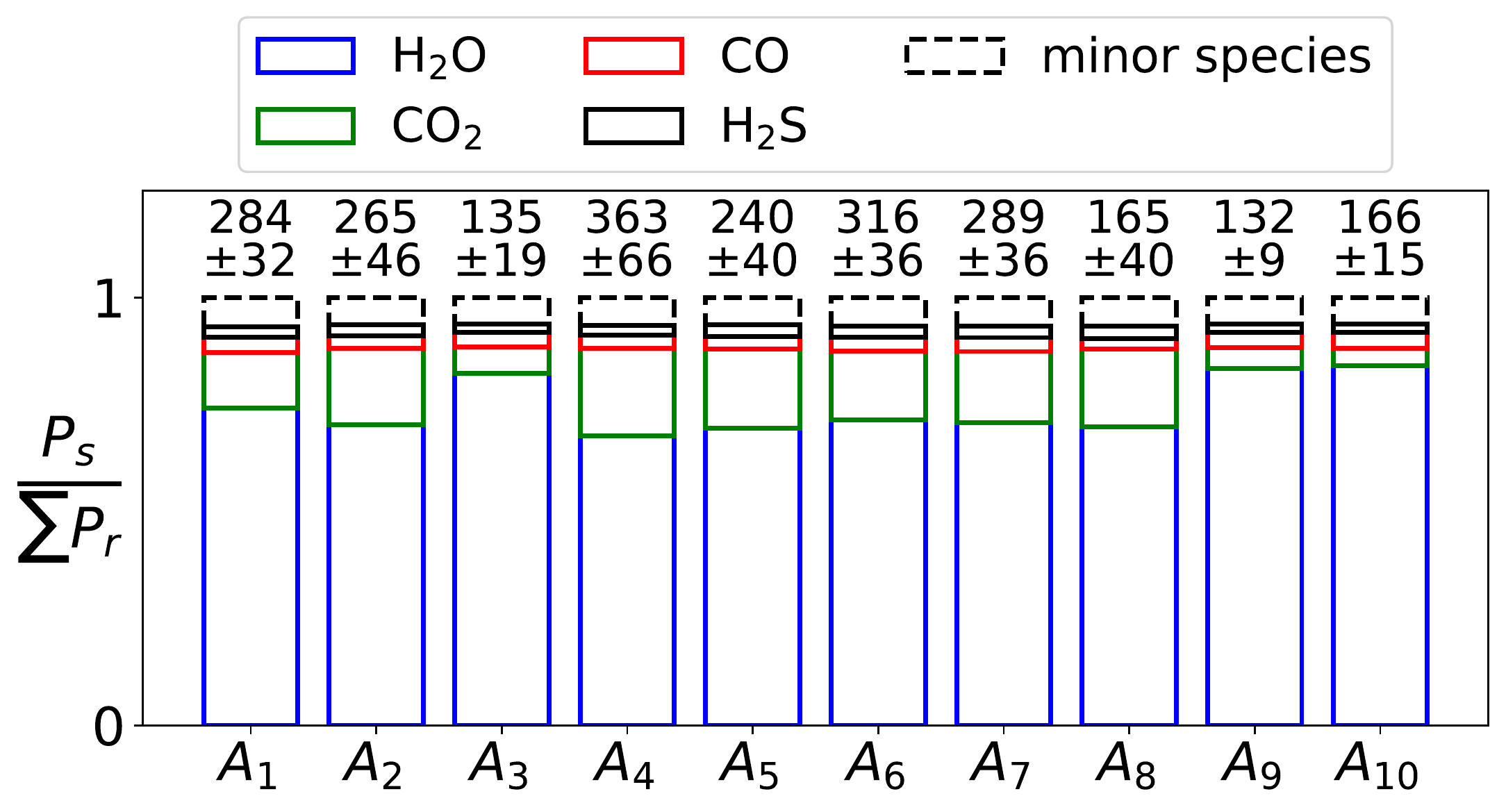}%
\caption{
Observed gas production $P_s$ (integrated emission rates) over the entire 2015 apparition for the 10 patches in Table~\ref{tab:geomorph}.
Mass-fraction of a species $P_s/\sum P_r$ is reflected by the segment size in the bar.
On top of each bar, the sum (over the species) per area $\sum P_r / |A|~[\mathrm{kg}/\mathrm{m}^2]$ is given.
The minor species cover O$_2$, C$_2$H$_6$, CH$_3$OH, H$_2$CO, CH$_4$, NH$_3$, HCN, C$_2$H$_5$OH, OCS, and CS$_2$.%
}
\label{fig:patch_major}
\end{figure}
Averages over large regions (hemispheres and lobes) in the Figures~\ref{fig:region_major} and \ref{fig:region_minor} provide only the mean value of the gas productions in active areas.
High resolution OSIRIS images and the VIRTIS spectroscopic analysis show small-scale areas holding ice-containing material (icy and ice-rich patches) which can be localized on the spatial scale of metres up to several tens of metres \cite{Barucci2016,Filacchione2016,Oklay2017,Filacchione2019}.
With areas between $4\times 10^5 ~\mathrm{m}^2$ and $1.6\times 10^6 ~\mathrm{m}^2$, the consideration of the patches $A_1$--$A_{10}$ from Figure~\ref{fig:h2operihel} allows us to increase the lower bounds for the observed production on the large regions in Figure~\ref{fig:region_major}.
Figure~\ref{fig:patch_major} denotes the most active patches (with the highest observed production), namely $A_4$, $A_6$, $A_7$, $A_1$, and $A_2$, with production values of up to $363\pm 66\,\mathrm{kg}/\mathrm{m}^2$.
Assuming a bulk density \cite{Preusker2017} of $538~\mathrm{kg}/\mathrm{m}^3$ this corresponds to an erosion due to ice sublimation of up to $0.7~\mathrm{m}$.
The smallest observed productions within the set of patches occurs on the northern patches $A_3$ and $A_9$ with $135\pm 19\,\mathrm{kg}/\mathrm{m}^2$ and $132\pm 9\,\mathrm{kg}/\mathrm{m}^2$ which corresponds to a $0.2~\mathrm{m}$ height loss.
Additional dust components increase the height loss depending on the dust-to-ice ratio in the nucleus.
The averaged loss of 0.55~m height for cometary material \cite{Keller2017} (ice and dust) shows the difference between very local changes and the gas release averaged over larger areas.
The higher productions across patches located on the southern hemisphere (compared to the ones located on the northern hemisphere) corresponds to the higher overall gas production on the entire southern hemisphere $A_\mathrm{SH}$, shown in Figure~\ref{fig:region_major}.

In Figure~\ref{fig:patch_major} within the group of all southern patches $A_1$, $A_2$, $A_4$--$A_8$ the mass-fractions for H$_2$O and CO$_2$ vary between $68\pm 21\%$ and $20\pm 7\%$ on A$_4$ and $74\pm 13\%$ and $13\pm 5\%$ on A$_1$, respectively.
The corresponding CO$_2$/H$_2$O ratios (mole-fraction) range between $7\pm 3\%$ on A$_1$ and $12\pm 5\%$ on A$_4$ close to the averaged value on $A_\mathrm{SH}$.
This agreement across patches also prevails on the northern hemisphere which is expressed in low
mass-fractions of CO$_2$ between $4.1\pm 2.9\%$ on $A_{10}$ and $6\pm 3\%$ on $A_{3}$ (related to CO$_2$/H$_2$O mole-fractions between $2.0\pm 1.4\%$ and $3.0\pm 1.7\%$).

\section{Conclusion}

We have derived spatial maps of the emission rates on the surface of comet 67P/C-G, augmenting the globally integrated gas production given in Läuter et al. \cite{Lauter2020}.
Due to the uncertainties of the forward model and the partially limited surface coverage by measurement data, the spatial resolution of the grid ($\sim 120~\mathrm{m}$) provides only a lower bound for the resolution of the physical signal on the surface.
The extracted spatial distribution of water emissions around perihelion leads us to identify 10 active patches on the surface with a typical size of $700~\mathrm{m}$ (the smallest patch is $4\times 10^5 ~\mathrm{m}^2$).
7 out of 10 patches are located on the southern hemisphere and all 10 patches (taken together) represent almost half of the global water emission.
Both lobes of the concave surface harbor enhanced emission patches.
The increased outgassing activity on the southern patches $A_1$, $A_2$, $A_4$--$A_8$ is not limited to H$_2$O emissions only, but goes hand in hand with enhanced emission of CO$_2$, CO and O$_2$ \cite{Lauter2019} and also of all minor species analyzed here.

The analysis of gas emissions for different times and for different patches confirms the qualitative relationship between solar irradiation on the one hand and gas emission on the surface on the other hand.
Our quantitative study of the emissions on the cometary surface is suitable for the validation of thermophysical models on comet 67P/C-G.
The non-linear characteristic of such a model holds the property, that diurnally averaged irradiation might not be sufficient to describe observed emission rates.
Periodic condensation of water in the surface layer can change insolation properties and thus delimit sublimation during night time when the interior is warmer \cite{DeSanctis2015}.
Our analysis relates solar irradiation and emission rates and reveals complex phenomena depending on material properties (ice composition and dust fraction) and possibly on a global redistribution of dust from the South to the North.
A hysteresis effect on the seasonal scale yields increased CO$_2$ emissions during the outbound orbit compared to the inbound phase with corresponding irradiation.
On the northern patches, the maximum CO$_2$ emission does not coincide with the local maximum irradiation condition.

We established a minimal thermophysical model for the outbound orbital arc starting 100~days after perihelion, which allows one to determine the surface active fractions of H$_2$O and CO$_2$ on the southern patches.
The CO$_2$ emission follows an almost linear relation with irradiation, which can be used to obtain the fraction of the active surface in the range of 0.2\% and 0.8\%.
For H$_2$O, the sublimation requires higher irradiation exceeding $50~\mathrm{W}/\mathrm{m}^2$, which is seen in the data.
We determine a fraction of the active surface for water between 4\% and 15\%.

Due to the complexity of the thermophysical processes on the surface and within the layers below, the ice composition is probably not reflected by the relative abundances in the instantaneous gas emissions.
The observation of the ``activity paradox'', a repetitive activity over several apparitions, relies on the assumption that the emission rates, integrated in time (gas production), serve as a good quantification for the ice composition.
Our large scale analysis confirms the known phenomenon that gas production on comet 67P/C-G is predominantly located in the southern hemisphere,
with twice as much gas production as in the northern hemisphere during the two-year mission time.
During the same time the southern hemisphere received 1.4 as much solar irradiation as the northern one.
Our method based on a localization on patches yields a lower estimate of $363\pm 66~\mathrm{kg}/\mathrm{m}^2$ for localized peak production during one apparition.
This corresponds to a height loss of $0.7~\mathrm{m}$ due to ice sublimation only.
On the northern hemisphere the mass-fraction of CO$_2$ in the ice close to the surface is $6\pm 2\%$ which goes down to $4.1\pm 2.9\%$ on the northern patch A$_{10}$.
This is in contrast to the higher CO$_2$ mass-fractions, $13\pm 4\%$ globally averaged (corresponds to a mole-fraction of $7\pm 2\%$ for CO$_2$/H$_2$O) and $17\pm 5\%$ on the southern hemisphere.
A maximum mass-fraction of $20\pm 7\%$ for CO$_2$ in the ice material is found on the southern patch A$_4$. 
In addition the minor species O$_2$, C$_2$H$_6$, and NH$_3$ change their mass-fractions in the ice from the northern hemisphere with the values $3.4\pm 0.5\%$, $0.6\pm 0.2\%$, and $0.35\pm 0.10\%$ to the southern hemisphere with the values $2.8\pm 0.7\%$, $1.2\pm 0.3\%$, and $0.24\pm 0.07\%$, respectively.
The asymmetric irradiation conditions for both hemispheres, with southern solstice occurring 23 days after perihelion, lead to a redistribution of surface material, including dust and ice.
Outgassing during the transfer of material from the southern to the northern hemisphere  suggests that CO$_2$ is not a significant component in the deposited fallback \cite{Davidsson2021}.
On the northern hemisphere, CO$_2$ is subjected to longer lasting sublimation processes during the orbital arc around aphelion, which might partially deplete northern reservoirs of CO$_2$ ices.
For the small and the big lobe, the comparisons of total gas production and of mass-fraction in the ice close to the surface do not show significant differences.

\section*{Acknowledgements}

We acknowledge NHR@ZIB for computing time and support.
Rosetta is an ESA mission with contributions from its member states and NASA.
We acknowledge the work of the whole ESA Rosetta team.
ROSINA would not have produced such outstanding results without the work of the many engineers, technicians, and scientists involved in the mission, in the Rosetta spacecraft team and in the ROSINA instrument team over the past 20~years, whose contributions are gratefully acknowledged.
Work on ROSINA at the University of Bern was funded by the State of Bern and by the European Space Agency PRODEX program.
The author M.R. was funded by the Swiss National Science Foundation (SNSF grant 200020\_182418).

\section*{Appendix}

\begin{figure}
\includegraphics[width=0.5\textwidth]{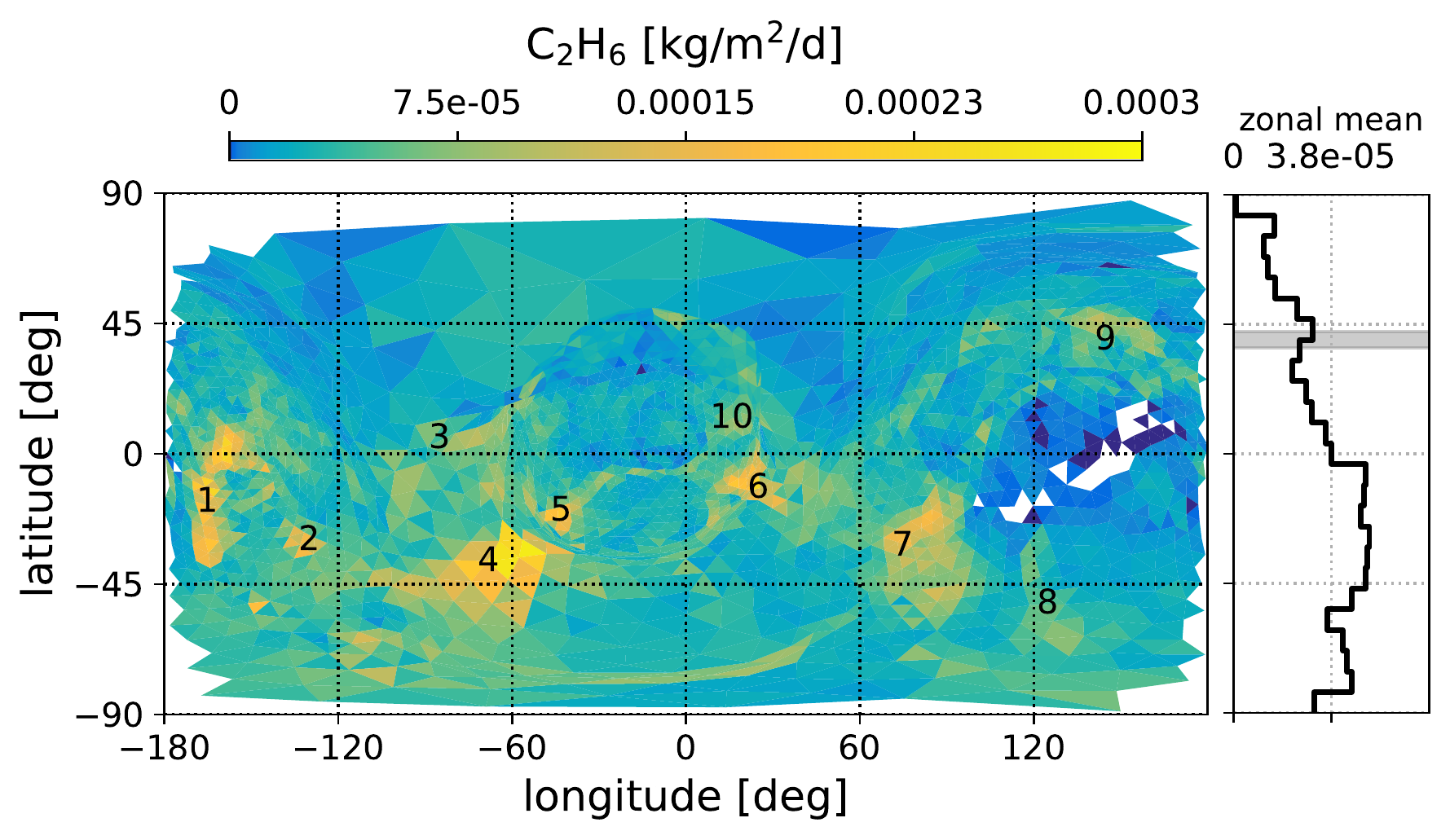}%
\hfill%
\includegraphics[width=0.5\textwidth]{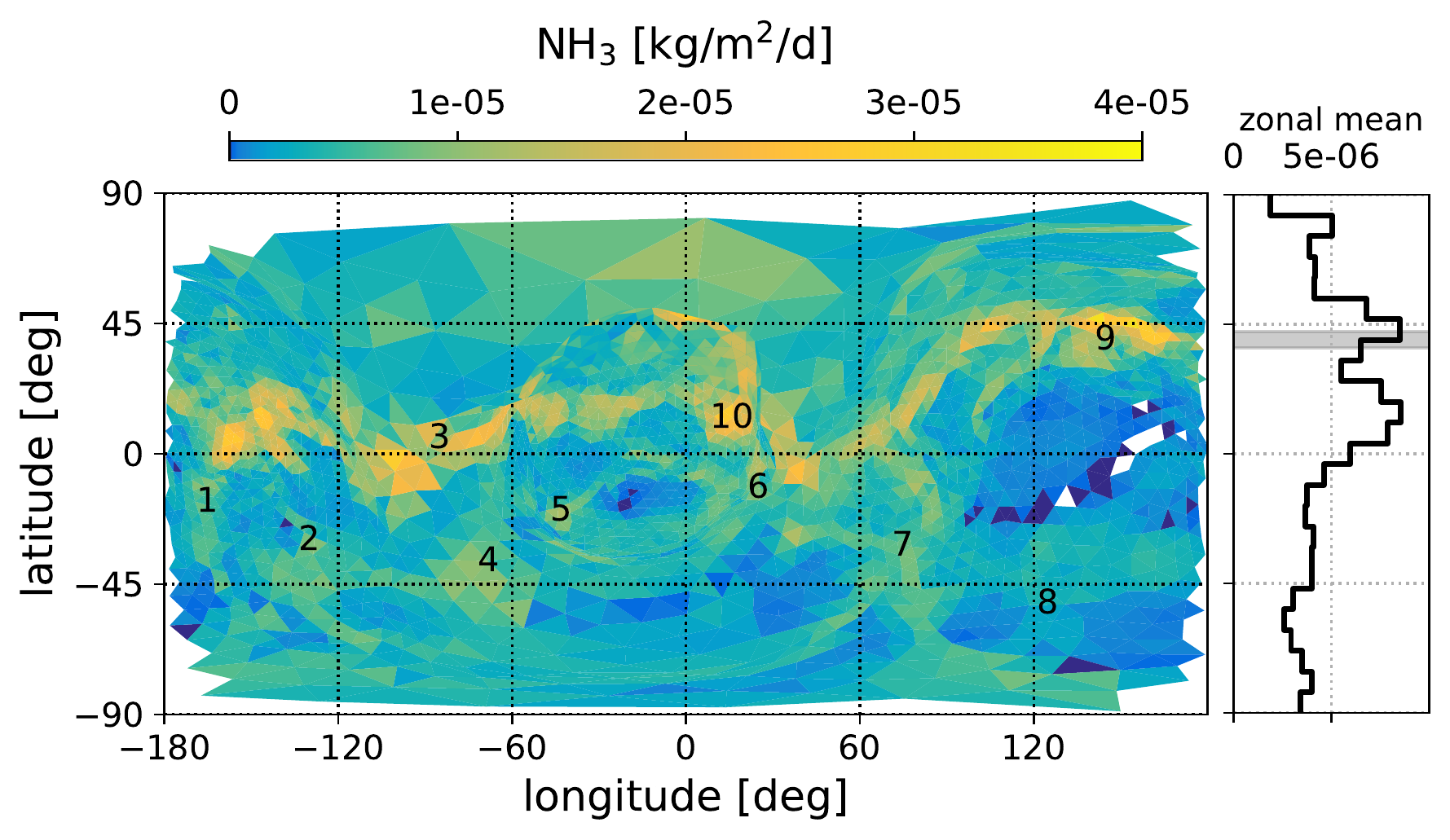}%
\\
\includegraphics[width=0.5\textwidth]{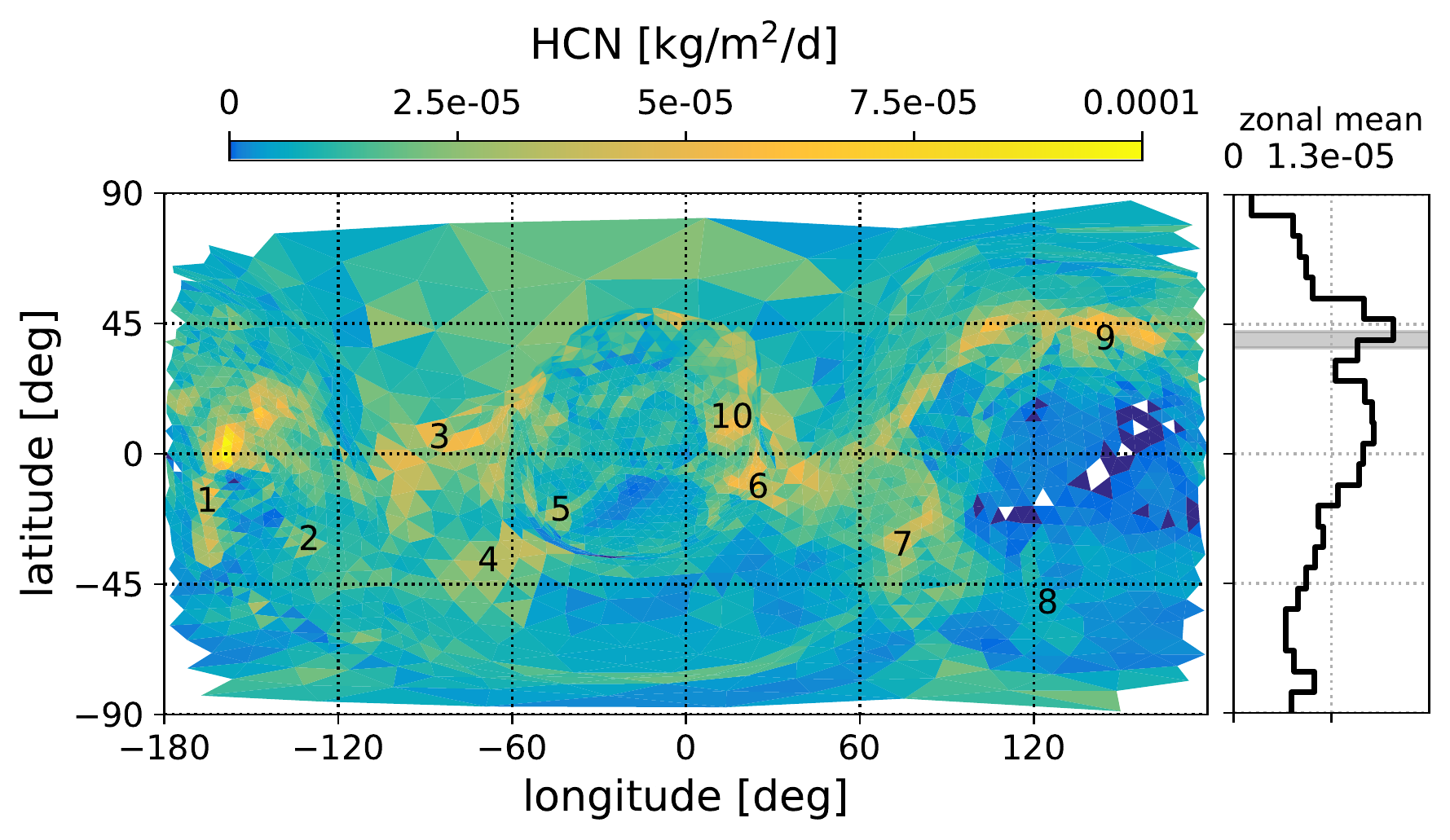}%
\hfill%
\includegraphics[width=0.5\textwidth]{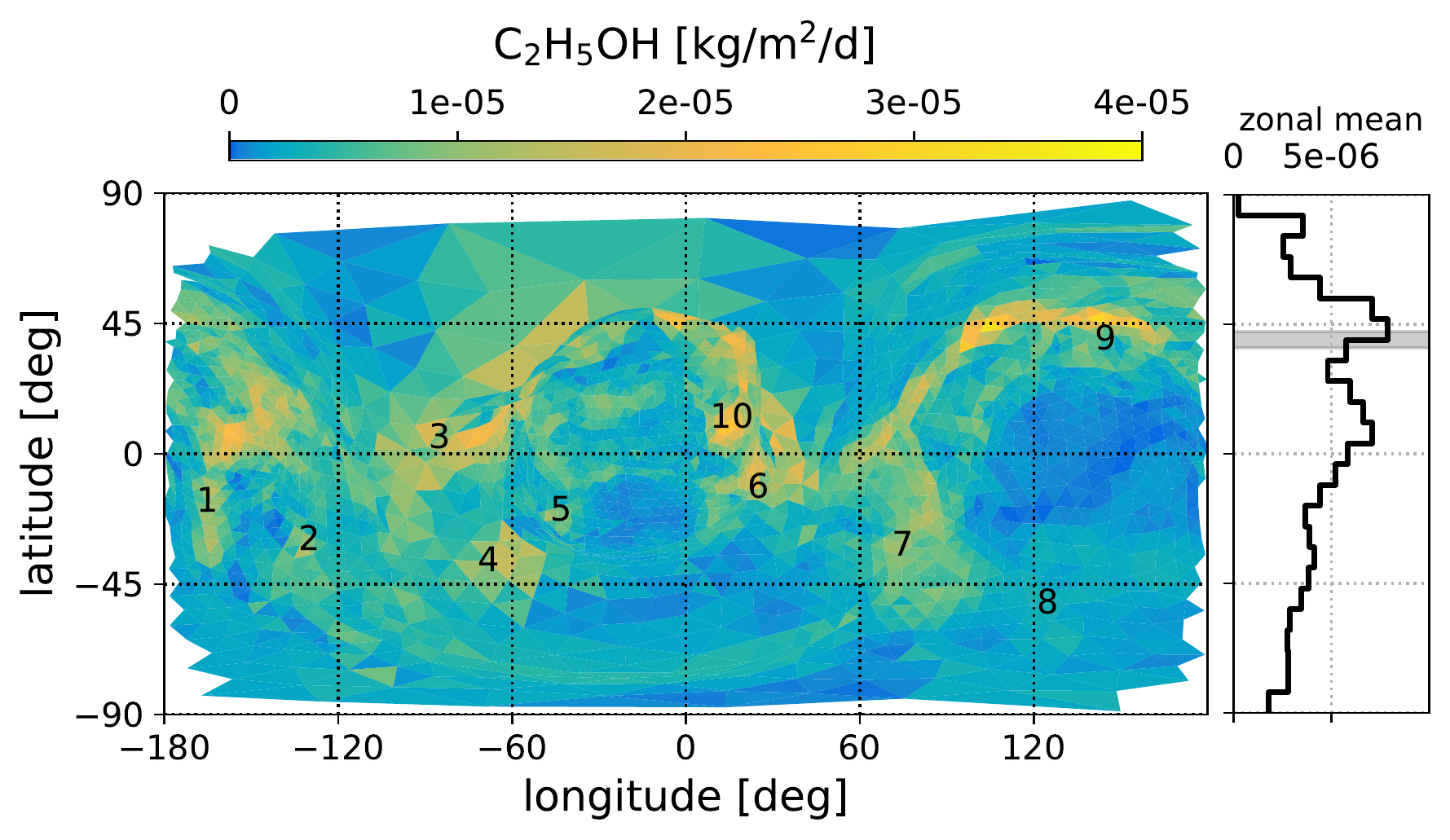}%
\\
\includegraphics[width=0.5\textwidth]{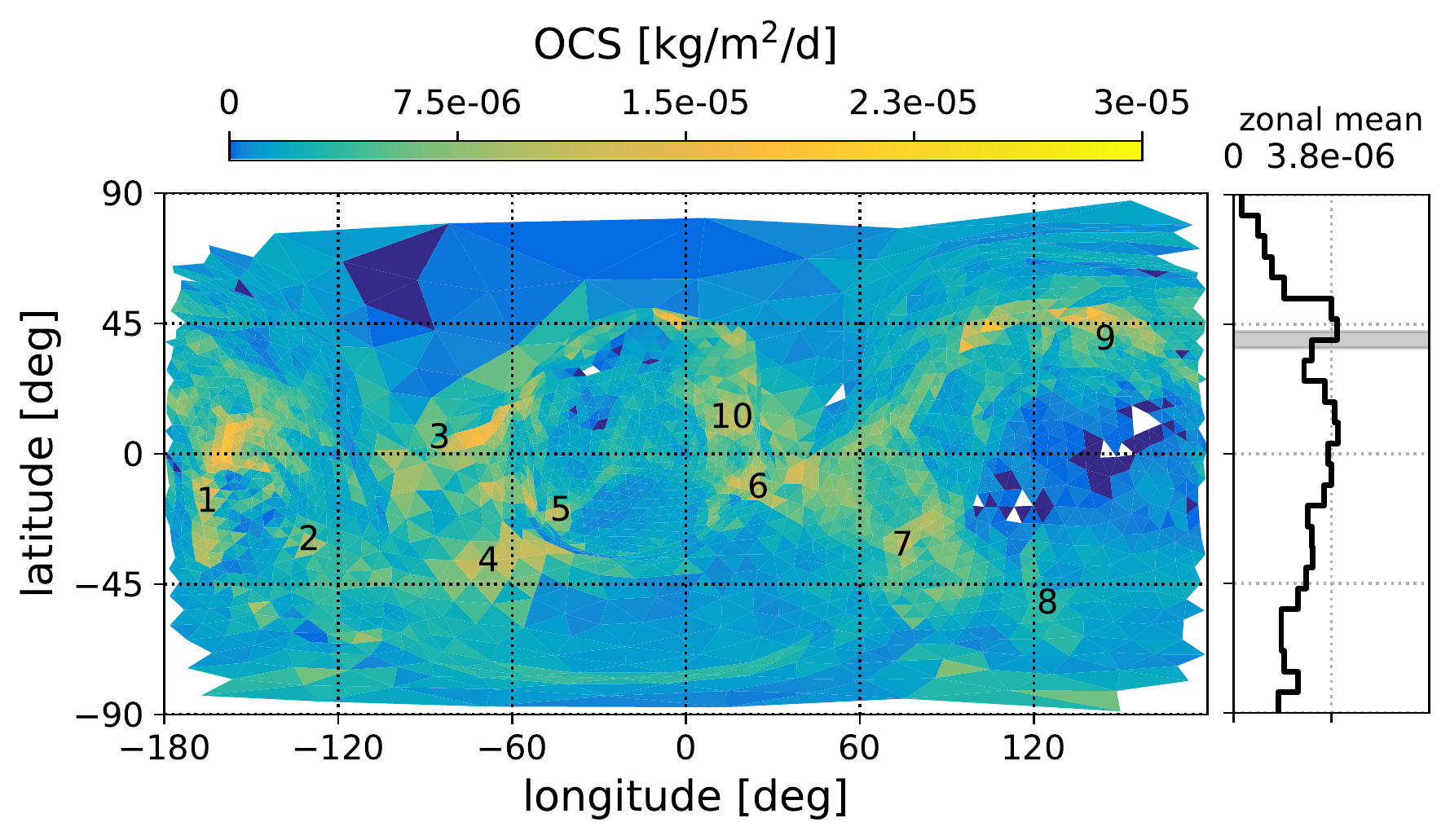}%
\hfill%
\includegraphics[width=0.5\textwidth]{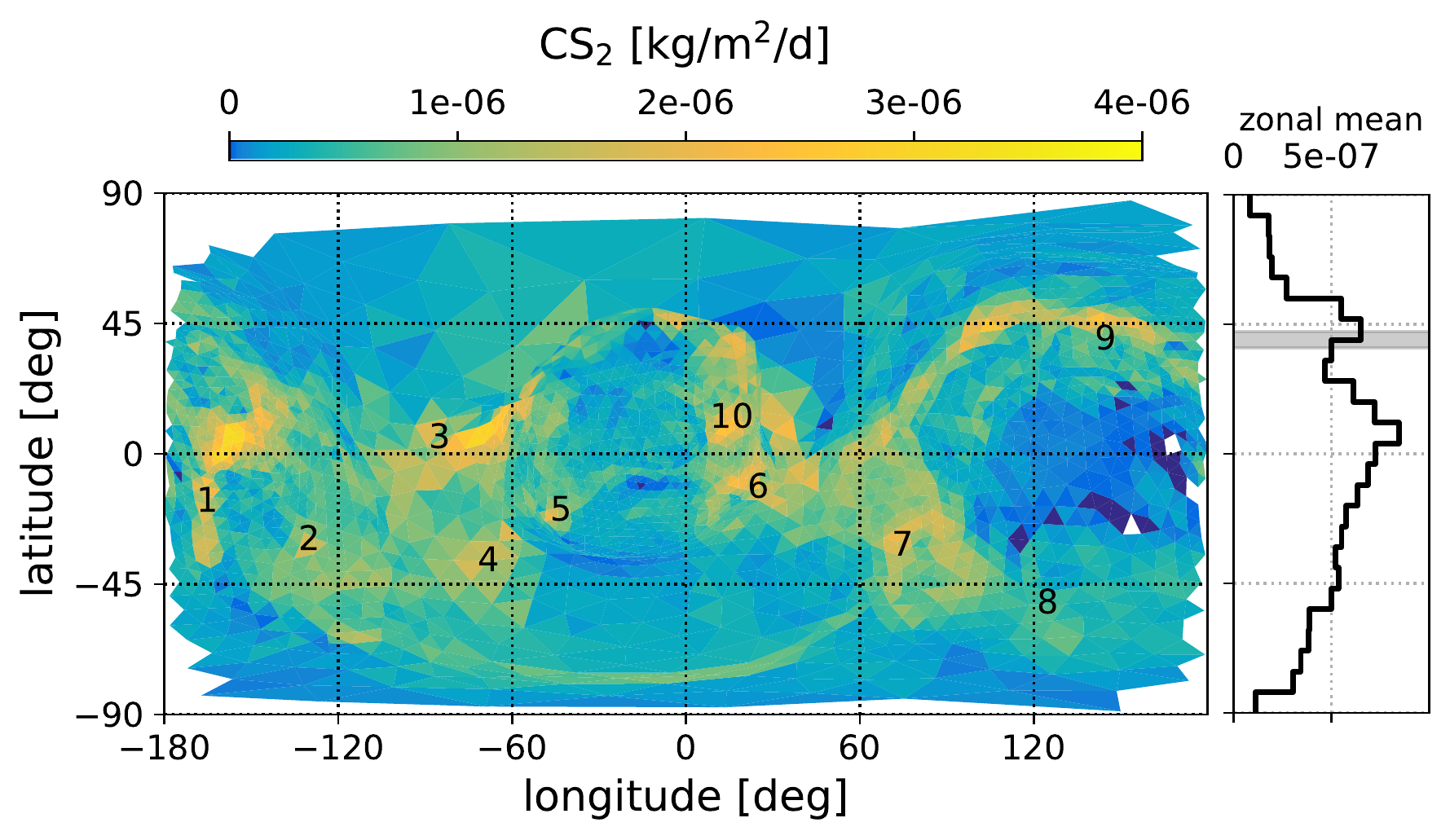}%
\caption{Surface emission rates for the minor gas species C$_2$H$_6$, NH$_3$, HCN, C$_2$H$_5$OH, OCS, and CS$_2$, averaged from 330~days to 270~days before perihelion, see $I_\mathrm{inbound}$ in Table~\ref{tab:intervals}.%
}
\label{fig:restinbound}
\end{figure}

\begin{figure}
\includegraphics[width=0.5\textwidth]{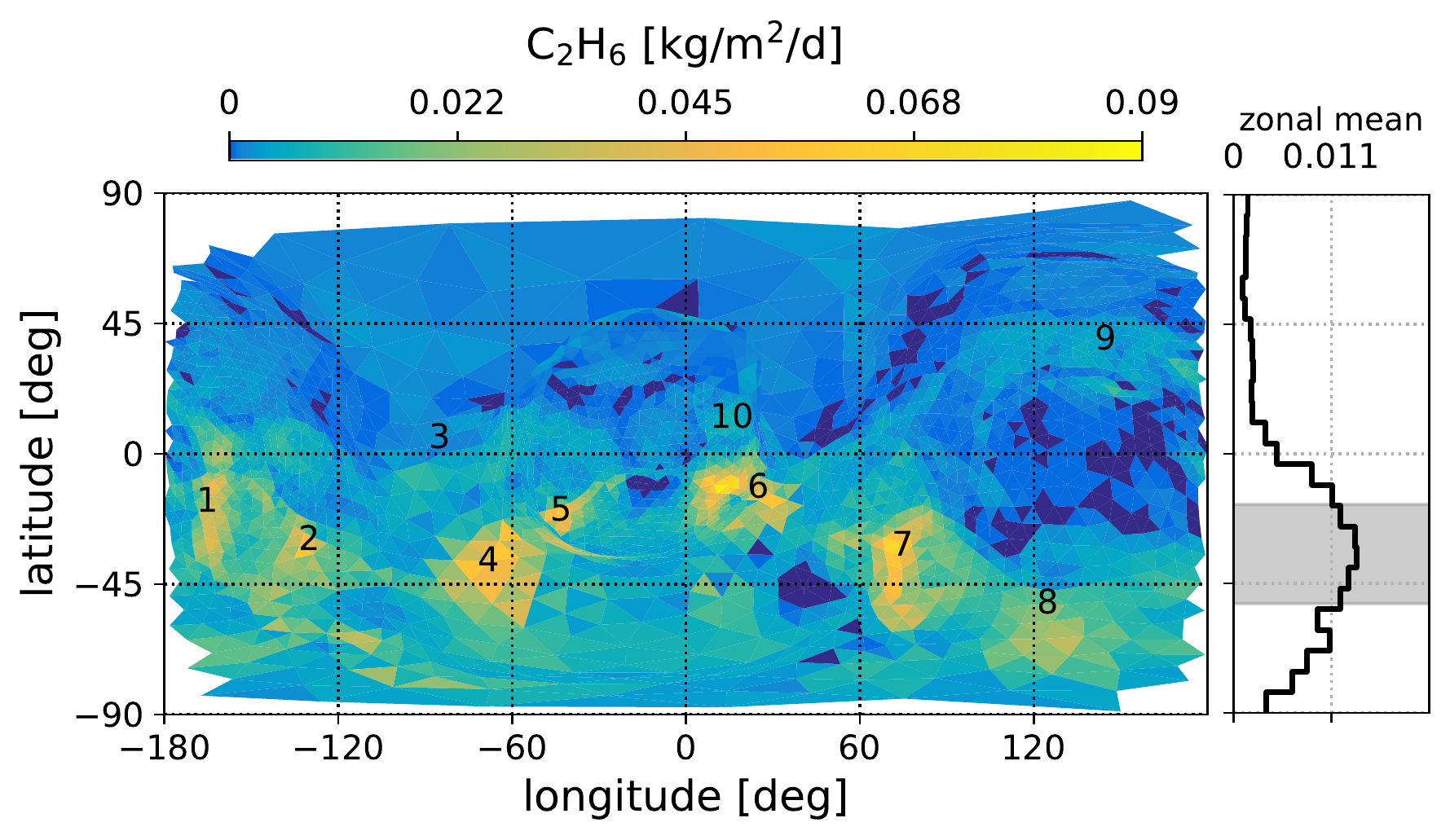}%
\hfill%
\includegraphics[width=0.5\textwidth]{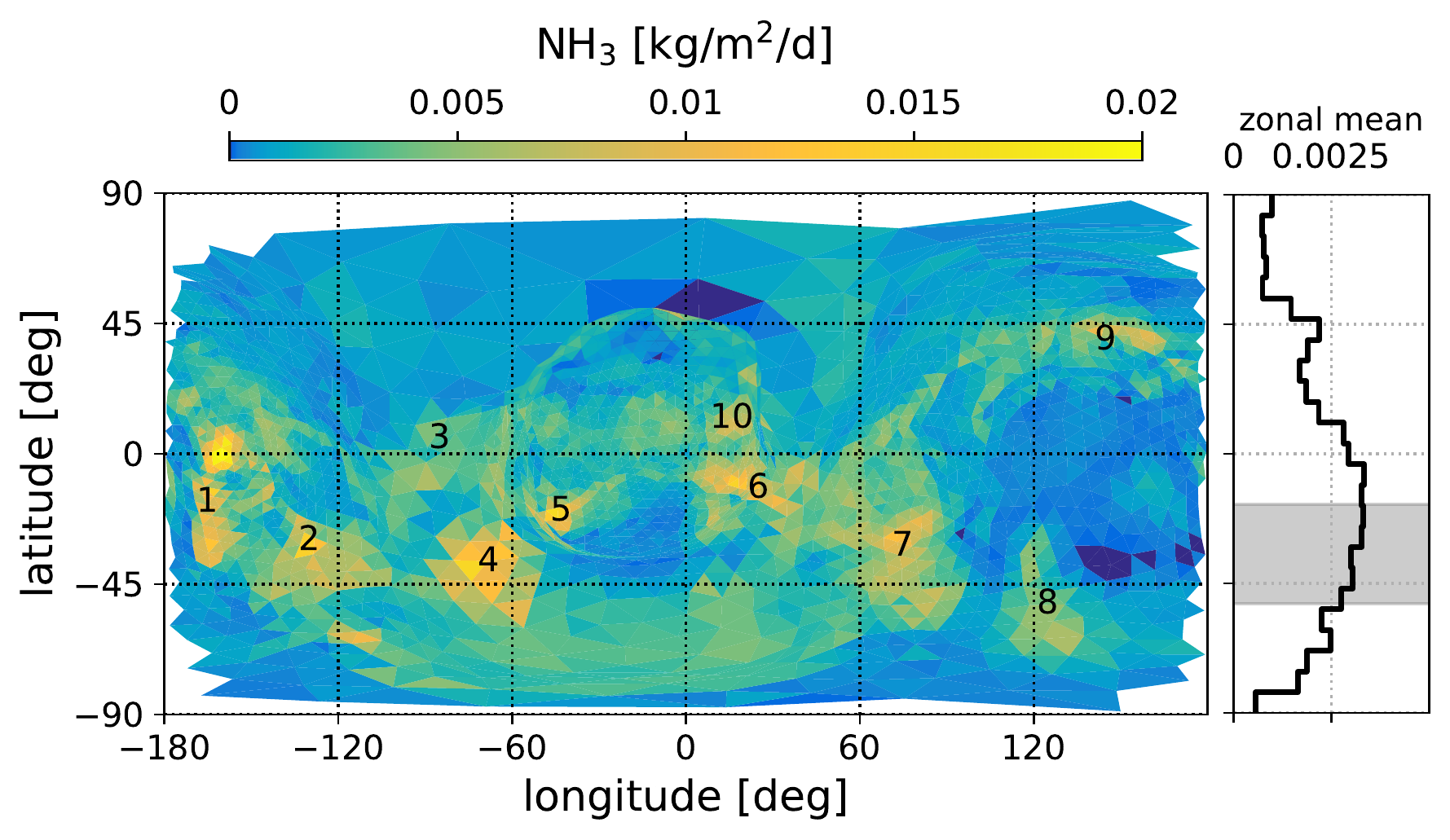}%
\\
\includegraphics[width=0.5\textwidth]{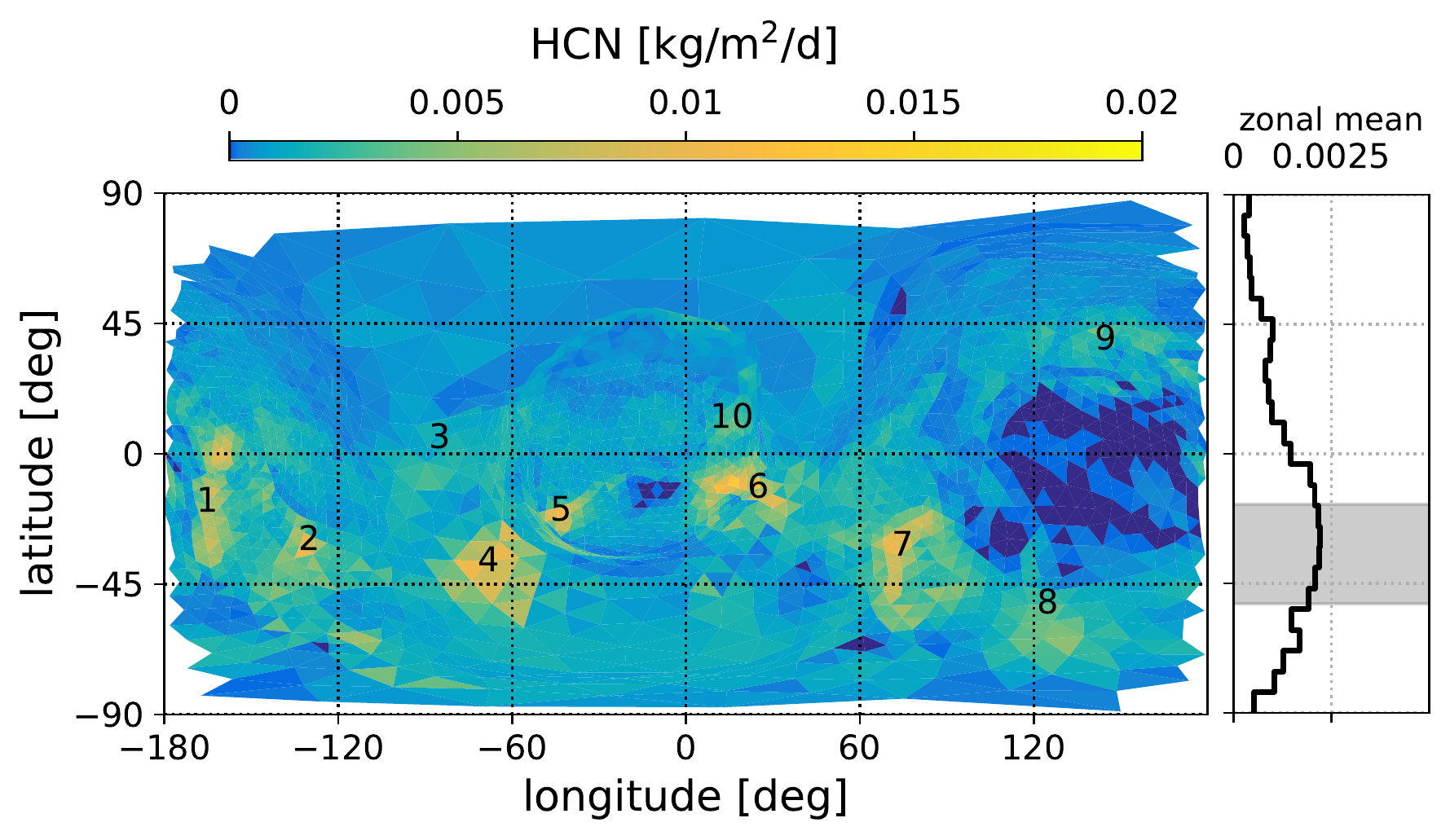}%
\hfill%
\includegraphics[width=0.5\textwidth]{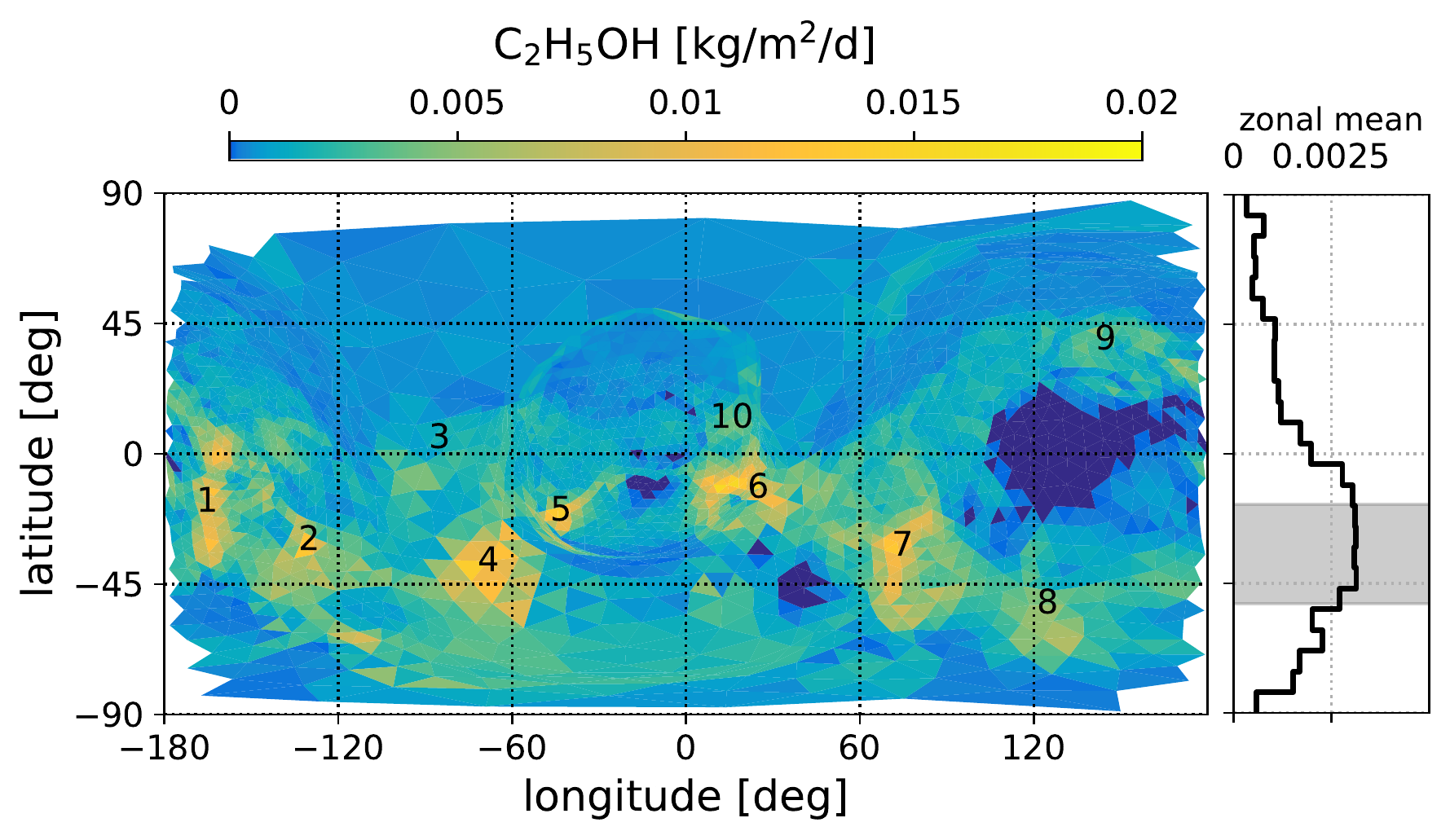}%
\\
\includegraphics[width=0.5\textwidth]{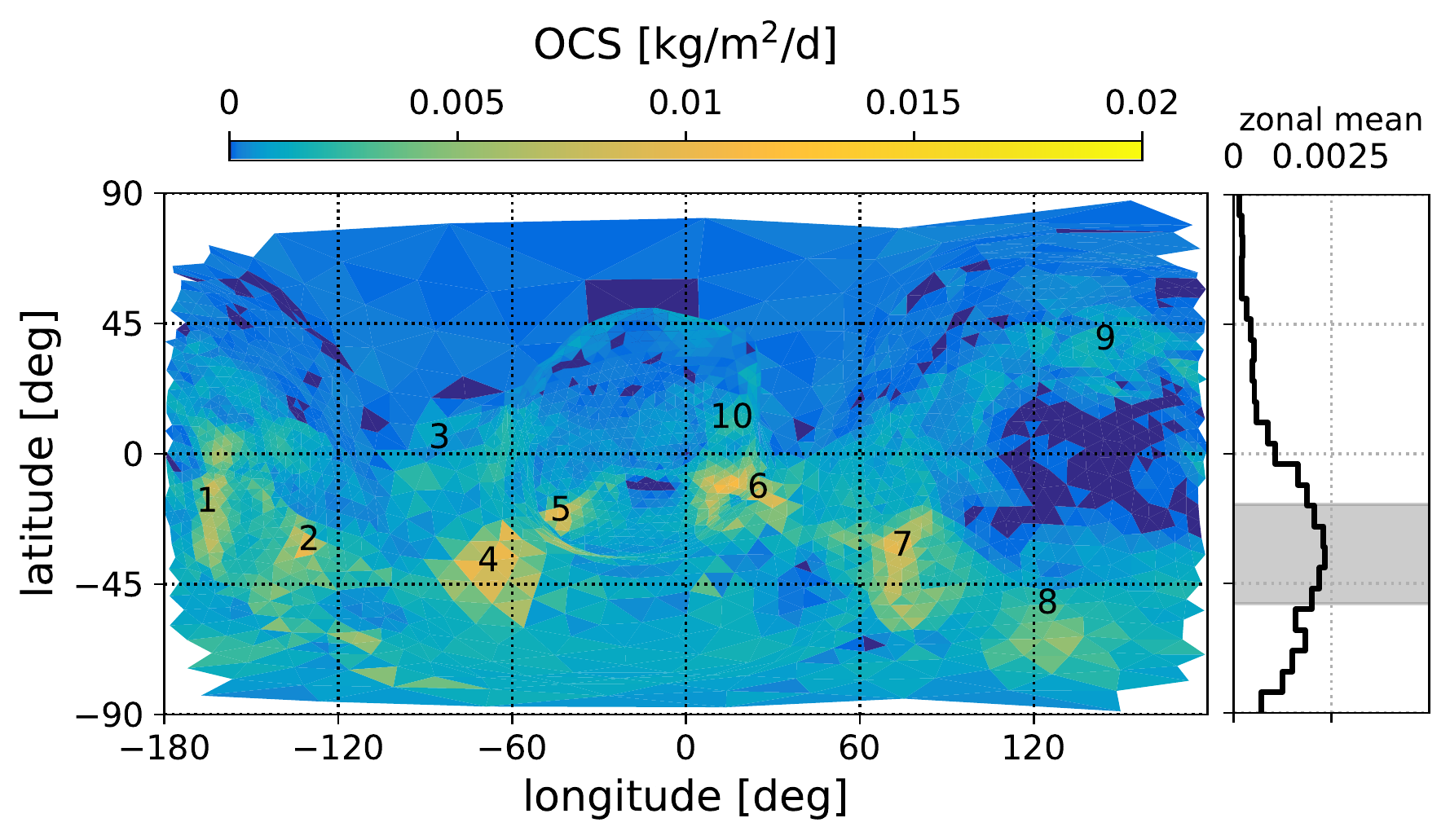}%
\hfill%
\includegraphics[width=0.5\textwidth]{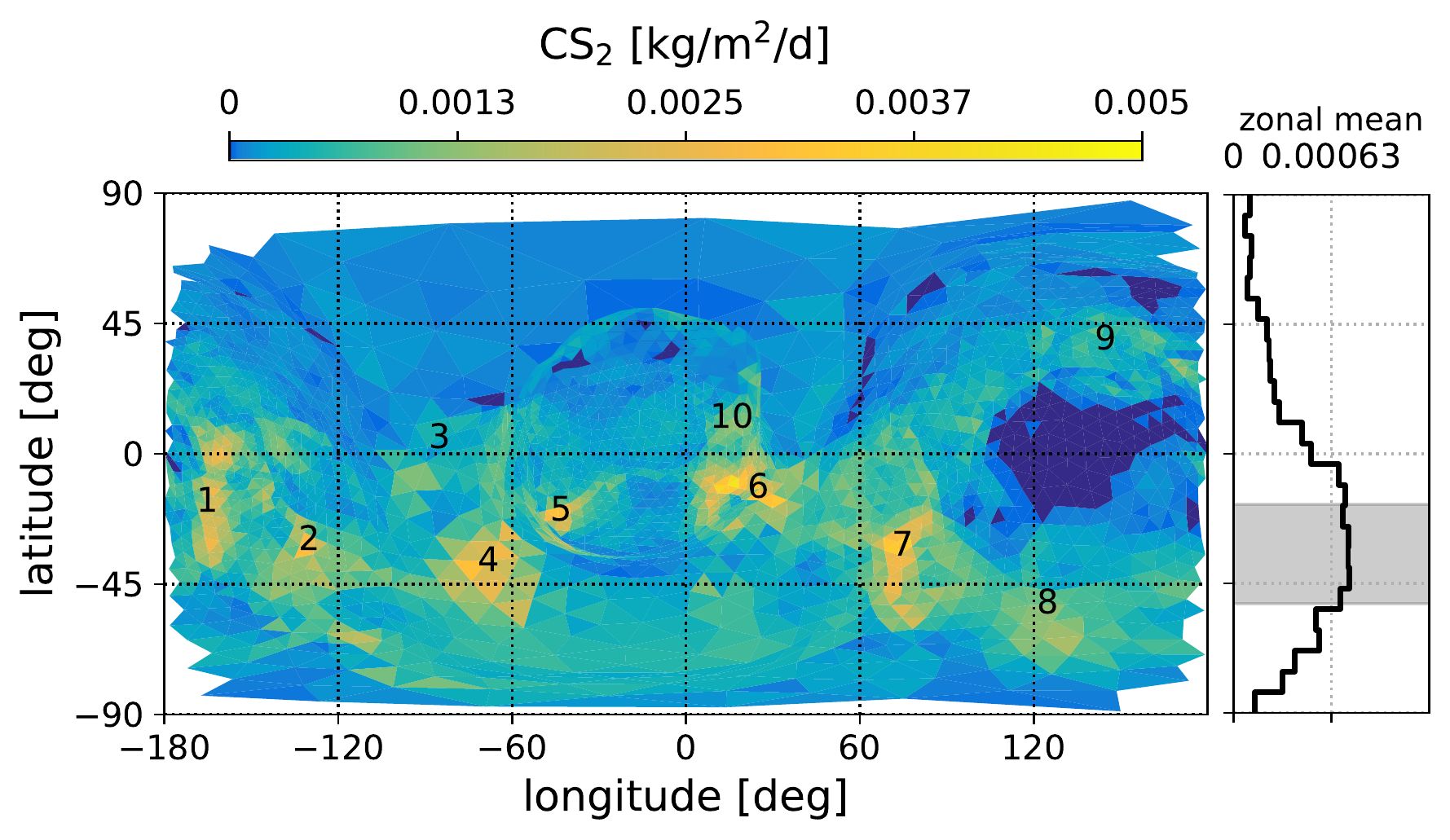}%
\caption{Surface emission rates for the minor gas species C$_2$H$_6$, NH$_3$, HCN, C$_2$H$_5$OH, OCS, and CS$_2$, averaged from 50~days before to 50~days after perihelion, see $I_\mathrm{ph}$ in Table~\ref{tab:intervals}.%
}
\label{fig:restperi}
\end{figure}

\begin{figure}
\includegraphics[width=0.5\textwidth]{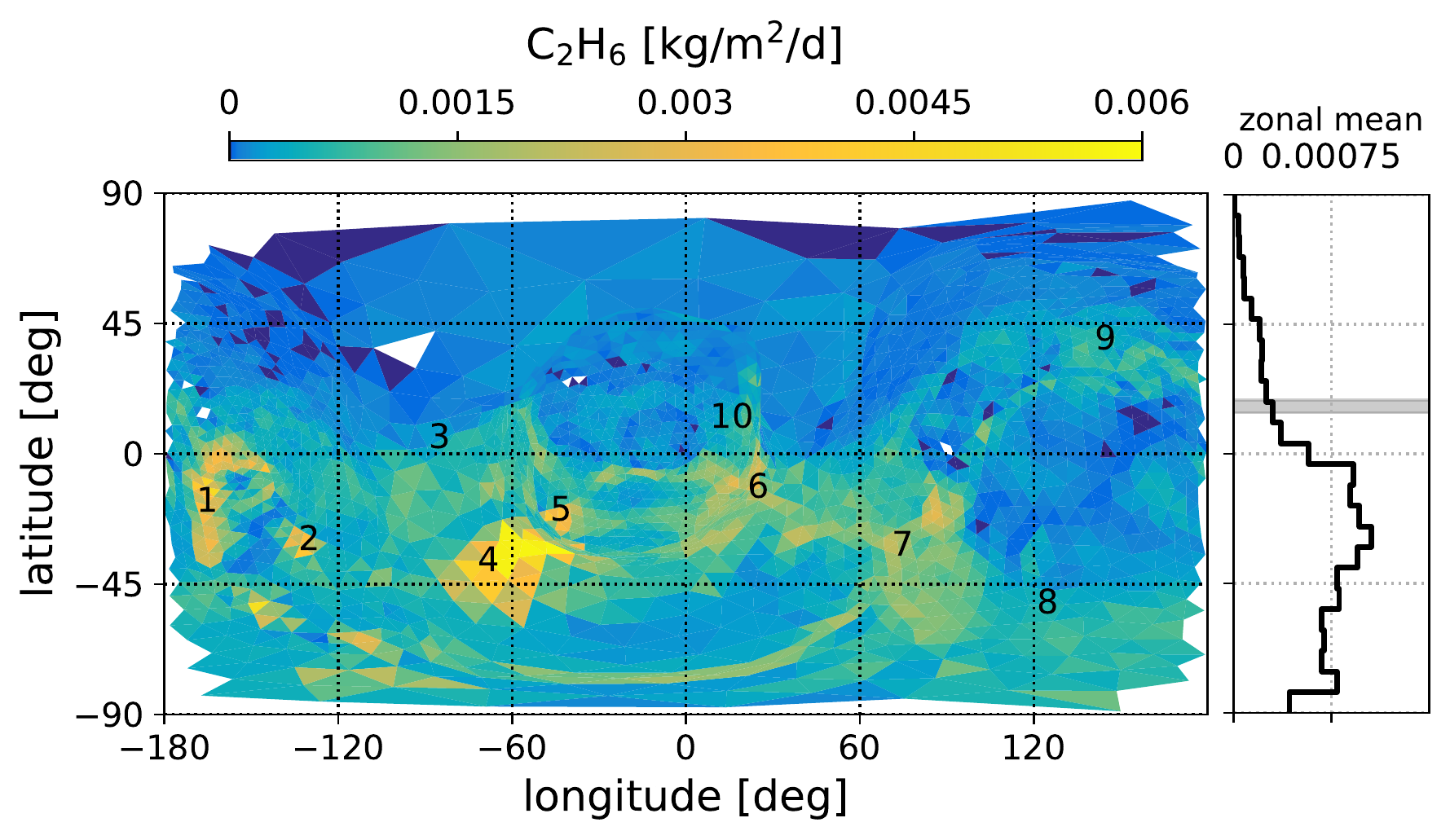}%
\hfill%
\includegraphics[width=0.5\textwidth]{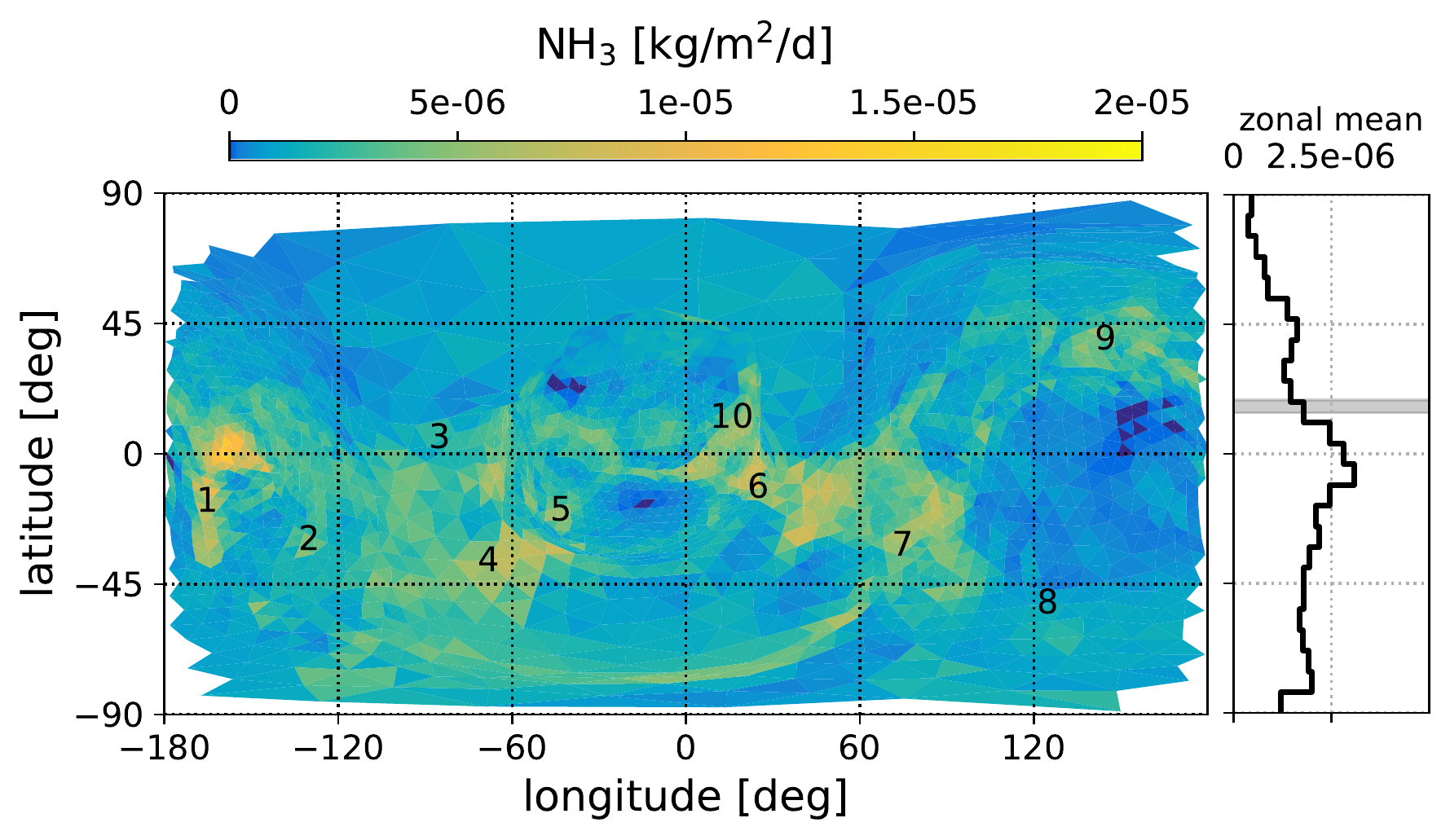}%
\\
\includegraphics[width=0.5\textwidth]{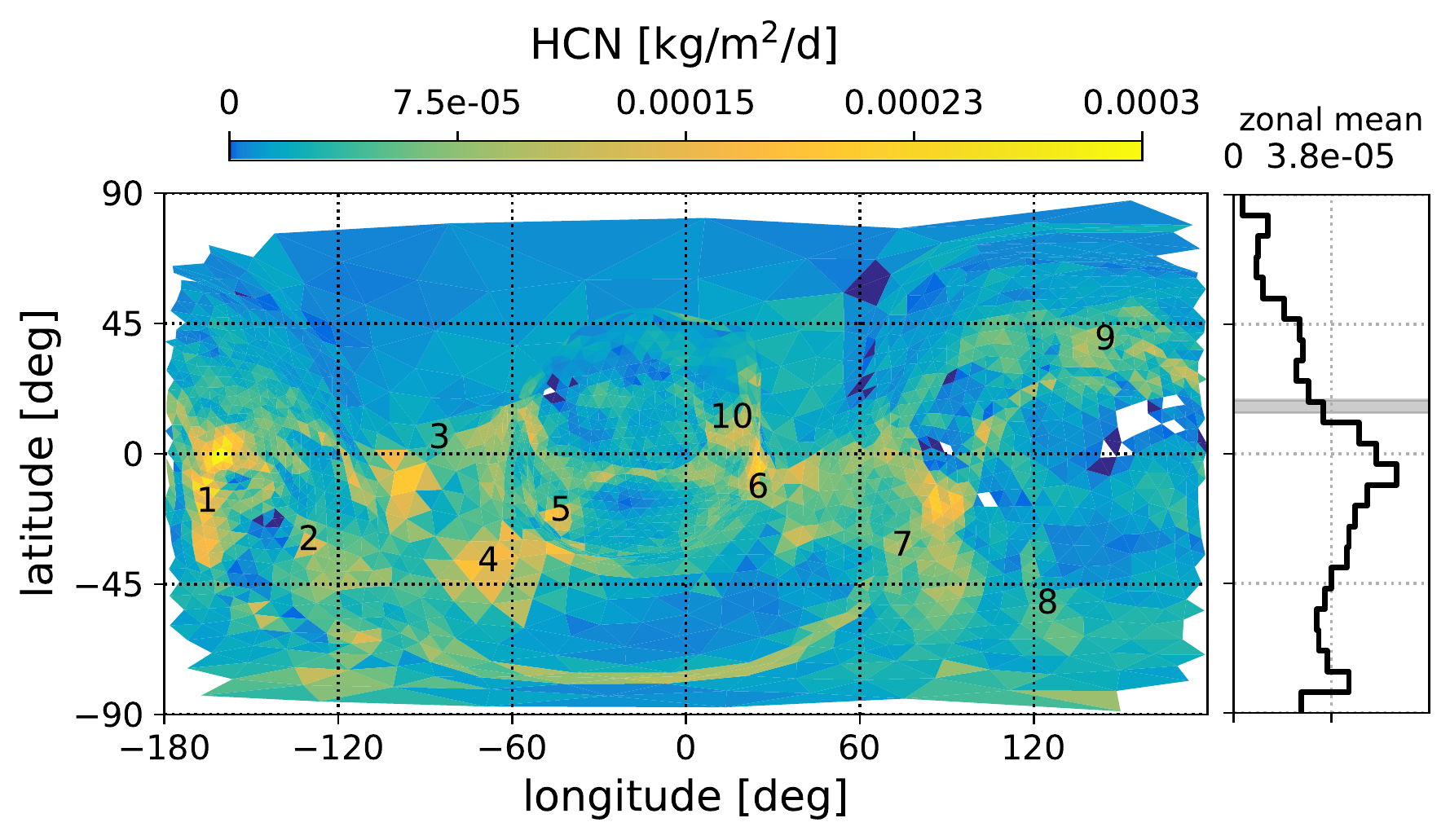}%
\hfill%
\includegraphics[width=0.5\textwidth]{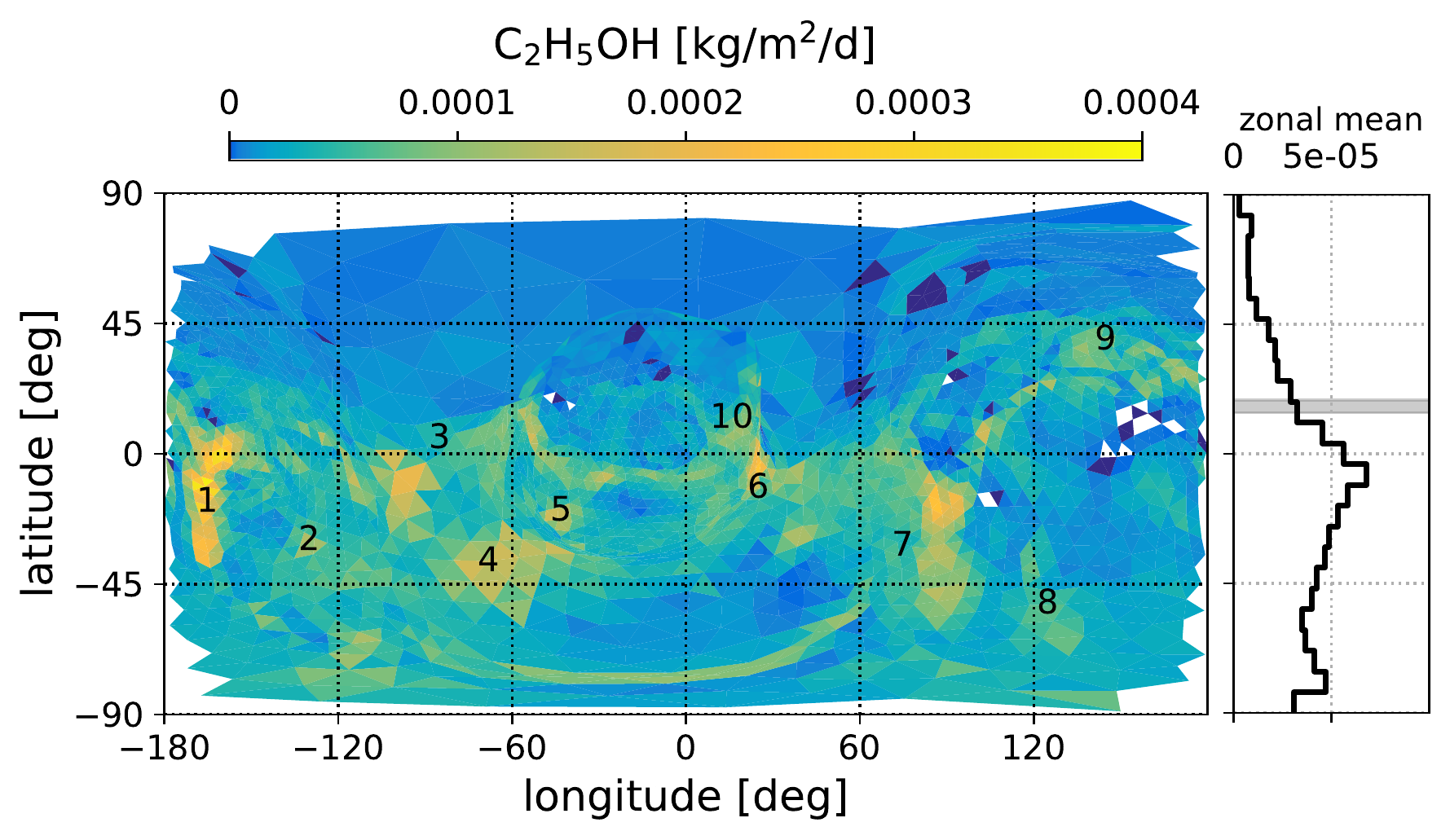}%
\\
\includegraphics[width=0.5\textwidth]{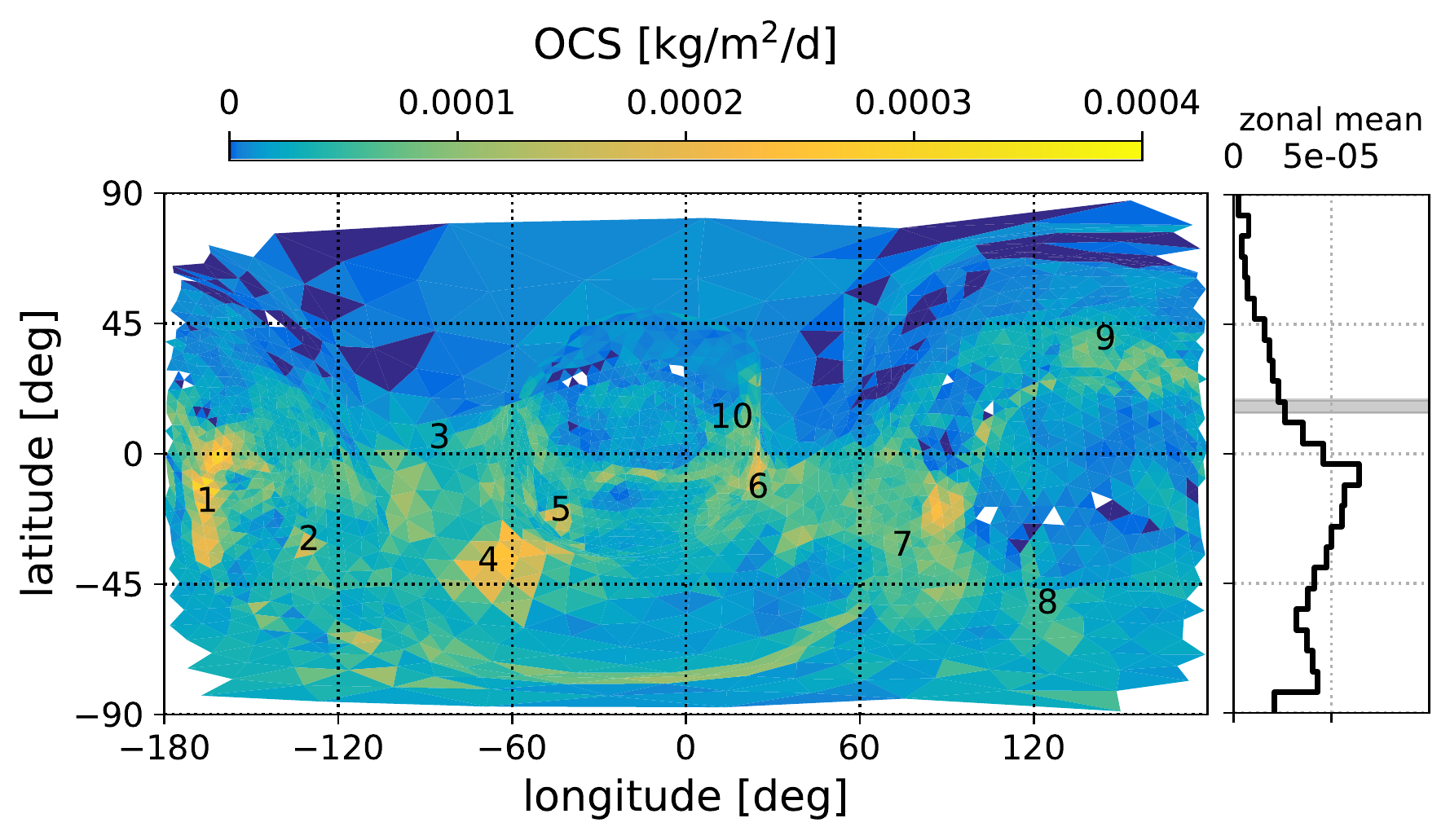}%
\hfill%
\includegraphics[width=0.5\textwidth]{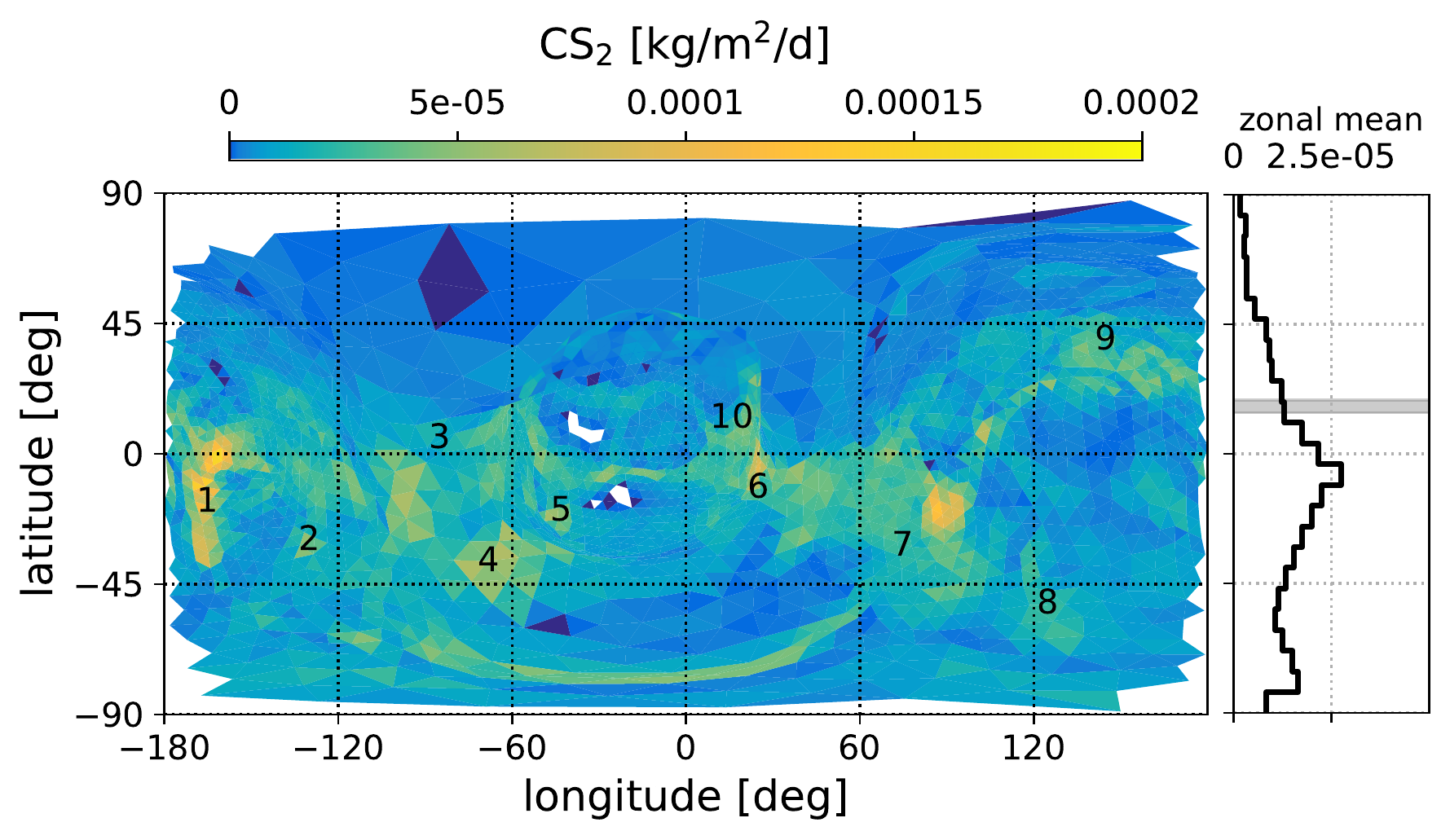}%
\caption{Surface emission rates for the minor gas species C$_2$H$_6$, NH$_3$, HCN, C$_2$H$_5$OH, OCS, and CS$_2$, averaged from 340~days to 390~days after perihelion, see $I_\mathrm{outbound}$ in Table~\ref{tab:intervals}.%
}
\label{fig:restoutbound}
\end{figure}

The Appendix contains maps of the surface emission rates for additional species not shown in the main text or in Läuter et al. \cite{Lauter2019}.
Läuter et al. \cite{Lauter2019} displays surface emissions for the major species H$_2$O, CO$_2$, CO, O$_2$.
Section~\ref{sec:globalgas} describes the surface emissions of H$_2$S, CH$_4$, CH$_3$OH, and H$_2$CO (Figures~\ref{fig:minorperihel}, \ref{fig:minorinbound}, and \ref{fig:minoroutbound}). 
Emission maps for all remaining species C$_2$H$_6$, NH$_3$, HCN, C$_2$H$_5$OH, OCS, and CS$_2$ in all three time intervals $I_\mathrm{inbound}$, $I_\mathrm{ph}$, and $I_\mathrm{outbound}$ are shown in the Figures~\ref{fig:restinbound}, \ref{fig:restperi}, and \ref{fig:restoutbound}.

\clearpage

\providecommand{\latin}[1]{#1}
\makeatletter
\providecommand{\doi}
  {\begingroup\let\do\@makeother\dospecials
  \catcode`\{=1 \catcode`\}=2 \doi@aux}
\providecommand{\doi@aux}[1]{\endgroup\texttt{#1}}
\makeatother
\providecommand*\mcitethebibliography{\thebibliography}
\csname @ifundefined\endcsname{endmcitethebibliography}
  {\let\endmcitethebibliography\endthebibliography}{}

\end{document}